\documentclass[11pt]{article}

\usepackage{booktabs}
\usepackage{placeins}
\usepackage{natbib}
\usepackage{amsmath}
\usepackage{graphicx}
\usepackage{authblk}
\usepackage{wasysym}
\usepackage{units}
\usepackage{url}

\usepackage[
  a4paper,
  left=2.0cm,
  right=2.0cm,
  top=2.5cm,
  bottom=3cm
]{geometry}

\newcommand{\Dmax}{{D}_\text{max}}
\newcommand{\Dmin}{{D}_\text{min}}
\newcommand{\Dopt}{{D}_\text{opt}}

\newcommand{\Tmin}{{T}_\text{min}}
\newcommand{\Topt}{{T}_\text{opt}}

\newcommand{\cke}{K_\text{c}}

\newcommand{\Celsius}{\;^\circ\mathrm{C}}

\newcommand{\mrime}{m_r}
\newcommand{\vrime}{V_r}
\newcommand{\rhorime}{\rho_\text{rime}}
\newcommand{\mliq}{m_\ell}
\newcommand{\mfrz}{m_f}
\newcommand{\vice}{V_i}

\title{Lagrangian particle simulations of an idealized deep \\ convective squall line: Aggregation, fragmentation, \\ and their polarimetric signatures}

\author{
  Axel Seifert\thanks{Corresponding author: axel.seifert@dwd.de}, Fabian Jakub, Christoph Siewert \\
  \vspace{-4mm}
  {\small Deutscher Wetterdienst, Offenbach, Germany}
  \vspace{-1mm}
  \and
  {Miklós} {Szakáll}\\
  \vspace{-4mm}
  {\small Johannes Gutenberg University, Mainz, Germany}
  \vspace{-1mm}
  \and
  Leonie von Terzi, Stefan Kneifel\\
  \vspace{-4mm}
  {\small Ludwig-Maximilians University of Munich, Munich, Germany}
  \vspace{-8mm}
}

\date{}

\begin{document}

\maketitle

\begin{abstract}

  Secondary ice production (SIP) in an idealized deep convective
  squall line is investigated using the Lagrangian particle model
  McSnow coupled to ICON. The model explicitly represents ice particle
  properties, including habit, degree of riming, and mass
  composition, allowing for a process-level analysis of ice formation
  pathways. We introduce a novel implementation of collision-induced
  ice--ice fragmentation that links fragmentation to prognostic particle
  properties through apparent density as a measure of particle
  fragility. We investigate three SIP mechanisms---rime splintering
  (RS), freezing fragmentation (FF), and collision-induced ice--ice
  fragmentation (CF)---and assess their relative importance together
  with the role of turbulence-enhanced collision rates.
  
  The simulations show that collision-induced fragmentation is the
  dominant SIP mechanism in the simulated squall line, producing a
  pronounced increase in ice particle number concentrations within the
  dendritic growth zone. In contrast, freezing fragmentation is limited
  by the availability of sufficiently large raindrops and contributes
  little in this convective regime. Rime splintering produces secondary
  ice at lower altitudes and indirectly promotes the formation of large
  aggregate snowflakes.
  
  Although turbulence has little influence on collision-induced
  fragmentation itself, it substantially enhances the aggregation of
  the generated fragments. Fragmentation alone shifts the particle
  size distribution toward smaller sizes and reduces precipitation,
  whereas turbulence-enhanced aggregation partially restores the
  large-particle tail and compensates for much of the
  fragmentation-induced reduction in precipitation. These results
  demonstrate that fragmentation and aggregation interact nonlinearly
  and cannot be considered independently in deep convective clouds.

  Forward simulations with the polarimetric radar model McRadar show
  that SIP modifies both bulk radar moments and Doppler spectra. Among
  the conventional radar moments, specific differential phase KDP
  exhibits the clearest signature of fragmentation, while spectral
  differential reflectivity identifies the source regions of secondary
  ice production through enhanced signals at the slow edge of the
  Doppler spectrum.

  Overall, our results identify collision-induced fragmentation as a
  key source of secondary ice in deep convection and demonstrate that
  its impact depends critically on its interaction with aggregation
  and turbulence. These findings highlight the importance of
  representing coupled microphysical processes in cloud models and
  suggest that spectral polarimetric radar observations provide a
  promising means of constraining secondary ice production in nature.

\end{abstract}

\section{Introduction}

Many microphysical processes in deep convective clouds remain incompletely
understood. Major challenges include the formation and growth of ice
particles, the relative importance of warm-rain versus ice-phase
precipitation pathways, the role of turbulence in enhancing collision
rates, and the impact of different ice-nucleating processes \citep{Morrison-2020}.
Additional uncertainty arises from secondary ice production (SIP)
mechanisms, such as rime splintering, freezing fragmentation, and
collision-induced breakup \citep{Field-2017,Korolev-Leisner-2020,Phillips-2021}. The
complexity of these interacting processes, combined with the
difficulty of obtaining reliable in situ observations within deep
convective updrafts, helps explain why our understanding of cloud
microphysics in deep convection remains limited.

Numerical simulations provide an essential tool for testing our
current process-level understanding. In particular, particle-based
models that explicitly represent microphysical processes offer a way
to evaluate whether our theoretical description is sufficient to
reproduce key features of deep convective systems \citep{Morrison-2020}. For example, can
such models improve the representation of observed characteristics of a deep convective
squall line? Do they require SIP to achieve realistic behavior, and can
they provide observable signatures of fragmentation processes
that help constrain the underlying mechanisms?

For many years, Hallett–Mossop rime splintering has been assumed to be
the dominant SIP mechanism in deep convective clouds. In situ
observations
\citep{Mossop-1985,Harris-Hobbs-Cooper-1987,Blyth-Latham-1993} and
modeling studies \citep{Beheng-1987,Blyth-Latham-1997,Sun-2010}
supported the existence and atmospheric relevance of this process, which
was originally postulated based on laboratory experiments
\citep{Hallett-Mossop-1974}.

In recent years, however, more detailed laboratory experiments on
freezing fragmentation
\citep{Lauber-2018,Keinert-2020,Kleinheins-2021}, corresponding in
situ observations
\citep{Korolev-2020,Pasquier-2022,Korolev-2022,Pfeifer-2025}, and
increasing evidence for collision-induced ice–ice fragmentation
\citep{Phillips-2017,Dedekind-2021,Grzegorczyk-2023,Gautam-2024,Yadav-2025}
have renewed the debate over which SIP mechanism dominates in deep
convective clouds. With new laboratory experiments challenging the
efficiency of the rime splintering mechanism \citep{Seidel-2024}, this
issue has become even more pressing.

Some studies suggest that ice–ice fragmentation may in fact be the
dominant SIP mechanism, or at least a viable alternative, for
secondary ice production in convective clouds
\citep{Gao-2023,Han-2024,Grzegorczyk-2025a,Grzegorczyk-2025b,Grzegorczyk2025c,Qu-2025,Waman-2026}. Other
studies favor freezing fragmentation
\citep{Lachapelle-2024,Cholette-2024}, rime splintering
\citep{Sun-2025}, or a combination of multiple SIP mechanisms
\citep{Sullivan-2018,Waman-2022} to reconcile model results with
observations.

These apparently conflicting results emphasize either the regime
dependence of SIP mechanisms or methodological difficulties, for
example limitations of bulk microphysics schemes in representing the
specific conditions required for SIP
\citep[e.g.][]{Blyth-Latham-1998}.

The growing interest in collision-induced fragmentation also
highlights the need to better understand turbulence-enhanced collision
rates, which remain poorly understood
\citep{Shaw-2003,Grabowski-Wang-2013}. For non-spherical ice particles
in particular, particle–turbulence interactions are complex due to
their irregular shapes and motions in turbulent flow
\citep{Siewert-2014a,Siewert-2014b,Voth-2017,Sheikh-2024}. While
parameterizations of turbulence effects on collision rates exist for
spherical particles \citep{Ayala-Wang-2008,Onishi-Seifert-2016},
corresponding formulations for non-spherical ice particles are still
lacking. This gap limits our ability to consistently represent the
interplay between turbulence, aggregation, and fragmentation in cloud
models.

In this study, we address these questions using the Lagrangian
particle model McSnow coupled to ICON to investigate secondary ice
production in deep convective clouds. We focus in particular on the
role of collision-induced fragmentation and its potential enhancement
by turbulence, and assess whether such processes can be identified by
observable polarimetric radar signatures.

The remainder of the paper is organized as follows: Section~2
describes the ICON/McSnow model, and Section~3 outlines the
experimental design. Section~4 analyzes the microphysical structure of
the control simulation, while Section~5 examines the roles of
stochastic snowflake geometry and turbulence in aggregation. Section~6
investigates secondary ice production, and Section~7 explores the
resulting polarimetric radar signatures. Finally, Section~8 summarizes
the main conclusions. Details of the implementation of the SIP
processes and turbulence effects on aggregation are provided in the
Appendix.

\section{The ICON/McSnow Lagrangian particle model}

The Monte Carlo Lagrangian particle model McSnow applies the
super-droplet algorithm of \citet{Shima-2009} to simulate the
evolution and interaction of liquid and ice particles
\citep{Brdar-Seifert-2018}. In McSnow, each super-particle is
described by eight fully prognostic properties: the ice mass $m_i$,
the monomer number $N$ of aggregates, the rime mass $\mrime$, the rime
volume $\vrime$, the liquid mass $\mliq$, the frozen mass $\mfrz$, the
aspect ratio $\phi_m$, and the ice volume $\vice$.  The latter
describe the ice particle habit and are essential for characterizing
branching and hollowing of ice crystals. Each super-particle
represents many real particles. This is described by the multiplicity
$\xi$, which is an additional prognostic variable.

The terminal fall velocity and collision efficiency of all particles
are based on the parameterizations of \citet{Boehm-1992a,
Boehm-1992b, Boehm-1992c}, and a habit-dependent empirical
ventilation coefficient is applied \citep{Welss-2024}. Habit
prediction follows the spheroidal model of \citet{Chen-Lamb-1994},
with extensions by \citet{Jensen-Harrington-2015} and
\citet{Welss-2024}. The evolution of $\phi_m$ and $\vice$ is governed
by the inherent growth function, which, even in its updated form,
depends only on temperature. For rimed particles, we apply the
classical fill-in model of \citet{Heymsfield-1982} instead of the
similarity model of \citet{Seifert-2019}.

Rime mass $\mrime$ and rime volume $\vrime$ increase through
collisions with supercooled liquid droplets. The instantaneous rime
density $\rhorime$ controlling the evolution of rime volume is
parameterized following \citet{Cober-List-1993}. The monomer number
$N$ increases by aggregation. Primary ice crystals that grow only by
deposition or riming retain $N=1$.

Melting and wet growth are treated using a quasi-stationary heat
budget for each super-particle, based on the parameterizations of
\citet{Rasmussen-Heymsfield-1987a, Rasmussen-Heymsfield-1987b} and
\citet{Rasmussen-1984a,Rasmussen-1984b}. Melting increases the liquid
mass $\mliq$, which is prognostic for each particle. Freezing of
liquid drops is represented using a probabilistic, time-dependent
interpretation of \citet{Barklie-Gokhale-1959}, leading to an
increase in frozen mass $\mfrz$.

In the present model configuration, depositional growth affects only
the ice mass $m_i$ and does not contribute to $\mrime$ or $\mfrz$.
This strict separation of growth pathways is important for the
representation of fragmentation processes. The newly implemented
secondary ice production (SIP) mechanisms, including freezing
fragmentation and collision-induced ice--ice fragmentation, are
described in the Appendix.

Optionally, McSnow includes a
stochastic model for aggregate geometry \citep{Seifert-2026}. In this
case, geometric properties such as maximum dimension, aspect ratio,
and cross-sectional area are sampled from prescribed probability
distributions after aggregation events and evolve between them.
This extends the microphysical state space to 9 dimensions.

McSnow is fully coupled to the ICOsahedral Non-hydrostatic ICON model
\citep{Zaengl-2015}. In this framework, only ice particles and
raindrops are represented by super-particles, while cloud droplets
remain described by the bulk scheme. Thus, McSnow replaces the cloud
ice, snow, graupel, hail and rain categories of the two-moment bulk
microphysics scheme of \citet{Seifert-Beheng-2006} by super-particles
without the need for any of these arbitrary hydrometeor categories. For
example, McSnow has only one parameterization for terminal fall
velocity of hydrometeors, similarly for collision efficiency,
ventilation coefficient and mass-size relation. All those
are free from arbitrary thresholds or categories. This design, which
largely relies on the parameterization of \citet{Boehm-1992a,Boehm-1992b,Boehm-1992c}, originates from the
initial goal of improving ice-phase representation through
high-fidelity benchmark simulations, and has since evolved into a
flexible research tool for detailed process studies.

Due to the coupling with the two-moment bulk scheme, raindrops are
initialized via autoconversion following
\citet{Seifert-Beheng-2001}. Ice-nucleating processes represented in
the bulk scheme such as heterogeneous nucleation, homogeneous freezing
of cloud droplets, and freezing of sulfate aerosols generate new
super-particles in McSnow. In contrast to many other Lagrangian
particle models, super-particles therefore exist only within cloud and
precipitation regions.

Because super-particles occupy only part of the domain, efficient
load balancing is essential. McSnow leverages the ICON
infrastructure, which employs an unstructured grid and allows
flexible mapping between grid cells and MPI tasks. This enables a
fine-grained, though static, domain decomposition in which each MPI
task samples a broad spatial distribution of particles. However,
this flexibility increases MPI communication and can reduce
efficiency in the dynamical core. In practice, a compromise must be
found between optimizing load balancing for microphysics and
maintaining dynamical performance.

\section{Baseline simulation and sensitivity experiments}

We simulate an idealized three-dimensional squall line using a setup similar to
\cite{Weisman-1997} and \citet{Rotunno-1988}, as used in previous
model studies like \citet{Bryan-2003} and \citet{Tijhuis-2025}.
The initial sounding follows \citet{Weisman-Klemp-1982} with a linear wind
profile from the surface to 2500 m height and constant wind speed of
10 m/s above. The water vapor mixing ratio near the surface is 12
g/kg. The ICON simulations apply a double-periodic triangular grid
corresponding to an equivalent grid spacing of 300 m with 1024 $\times$ 128
grid cells. The vertical grid has 128 levels with a domain top at 25
km and a damping layer starting at 22.5 km height.

The TKE-based Mellor-Yamada level 2.5 boundary layer scheme is applied
for vertical diffusion in combination with a 2D Smagorinsky closure in
the horizontal. The \citet{Phillips-2008} ice nucleation
parameterization is used with constant number densities for dust and
soot given as $n_{\text{dust}}=3.2\times10^6$ m$^{-3}$ and
$n_{\text{soot}}=60\times10^6$ m$^{-3}$, similar to
\citet{Seifert-2012}. The CCN activation is parameterized based on
\citet{Segal-Khain-2006} with $N_{\text{CN}}=250 \times 10^6$
m$^{-3}$.

ICON uses a time step of 2 s. McSnow is integrated with a 1 s
sub-step, while diffusional growth, riming, and melting are further
sub-stepped at 0.25 s to ensure numerical stability. Following
\cite{Weisman-1997} and similar to \citet{Tijhuis-2025}, a cold pool
initialization is used with a temperature reduction of 5 K at the
surface linearly reducing to zero at 2500 m height. Random temperature
perturbations with a maximum of 0.1 K are applied to trigger turbulent
motions. As in \citet{Bryan-2003}, we simulate 180 min, which is
sufficient to reach a quasi-stationary state of the squall line.

\begin{table}[t]
  \small
  \setlength{\tabcolsep}{4pt}
  \renewcommand{\arraystretch}{1.1}
  \caption{List of ICON/McSnow experiments and selected
    domain-averaged diagnostics.  $\xi_0$ is the base multiplicity. Runtime denotes
    wall-clock time in hours using 2048 cores.  $N_\text{sp}$ is the
    number of super-particles per grid cell (dimensionless); in-cloud mean and
    90th percentile values are shown.  The maximum vertical velocity
    $w_\text{max}$ is given in m s$^{-1}$, the upward mass flux
    $M_\text{up}$ in kg m$^{-2}$ s$^{-1}$, and cloud water path (CWP), rain water path (RWP), and ice water path (IWP) in kg
    m$^{-2}$.  The precipitation rate is given in mm h$^{-1}$.  All quantities are averaged
    over 120--180 min simulation time. For the domain-averaged
    precipitation rate the standard error is provided to allow a
    statistically meaningful comparison between experiments.}
  \label{tab:icon_timeseries}
  \vspace{3mm}
\begin{tabular}{llrrrrrrrrrr}
\toprule
\# & experiment & runtime & $N_\text{sp}$ & $N_\text{sp,p90}$ & $w_\text{max}$ & $M_\text{up}$ & CWP & RWP & IWP & precip. rate \\
\midrule
\midrule
1 & $\xi_0 = 3\times10^9$        & 45.7 & 154.9 & 333.1 & 31.12 & 3.69 & 0.130 & 0.184 & 1.003 & 1.628 $\pm$ 0.019 \\
2 & $\xi_0 = 7\times10^9$ (CTL)  & 36.2 & 108.9 & 232.7 & 29.82 & 3.64 & 0.129 & 0.183 & 1.032 & 1.667 $\pm$ 0.022 \\
3 & $\xi_0 = 3\times10^{10}$     & 20.5 & 51.7 & 108.8 & 32.03 & 3.62 & 0.129 & 0.187 & 1.058 & 1.758 $\pm$ 0.021 \\
4 & $\xi_0 = 7\times10^{10}$     & 14.8 & 31.5 & 64.6 & 28.65 & 3.56 & 0.135 & 0.205 & 1.047 & 1.878 $\pm$ 0.030 \\
5 & power-law agg. (MAgg)        & 34.5 & 110.9 & 236.2 & 31.05 & 3.67 & 0.132 & 0.193 & 1.049 & 1.650 $\pm$ 0.026 \\
6 & pow. agg. + turb. (MAggT)    & 42.2 & 109.8 & 242.5 & 30.31 & 3.44 & 0.124 & 0.155 & 0.960 & 1.873 $\pm$ 0.029 \\
7 & stoch. agg. + turb. (SAggT)  & 43.7 & 107.5 & 239.2 & 30.41 & 3.56 & 0.123 & 0.157 & 0.968 & 1.895 $\pm$ 0.028 \\
8 & rime splintering (RS)        & 34.0 & 115.6 & 249.5 & 29.64 & 3.54 & 0.136 & 0.182 & 1.113 & 1.683 $\pm$ 0.036 \\
9 & freezing fragmentation (FF)  & 36.5 & 108.6 & 233.0 & 31.95 & 3.52 & 0.135 & 0.185 & 1.038 & 1.666 $\pm$ 0.033 \\
10 & ice-ice fragmentation (CF)  & 38.2 & 161.9 & 364.3 & 27.61 & 3.68 & 0.130 & 0.193 & 1.134 & 1.627 $\pm$ 0.039 \\
11 & ice-ice frag. + turb. (CFT) & 47.3 & 159.0 & 377.8 & 27.75 & 3.65 & 0.108 & 0.127 & 0.958 & 2.086 $\pm$ 0.052 \\
\bottomrule
\end{tabular}
\end{table}

An overview of all experiments and their bulk characteristics is given
in Table~\ref{tab:icon_timeseries}. First, we performed a series of
sensitivity simulations varying the super-particle multiplicity
$\xi_0$. Note that in ICON/McSnow the (fractional) number of new
super-particles per process, time step, and grid point is limited to
1.25 (with stochastic sampling to either 1 or 2).  Consequently,
changes in the prescribed base multiplicity $\xi_0$ do not translate
linearly into changes in the simulated super-particle number.

Compared to the control simulation ($\xi_0 = 7\times10^9$), increasing
$\xi_0$ to $3\times10^{10}$ reduces the number of super-particles by
more than a factor of two, while for $\xi_0 = 7\times10^{10}$ the mean
number of super-particles drops below 40 per grid box. The number of
super-particle per grid point follows an approximately exponential
distribution (not shown), with some grid points reaching high values,
particularly in the convective core.  The 90th percentile is about
twice the mean, slightly below the value expected for a pure
exponential distribution. Despite this variability, the model runtime
scales approximately linearly with the number of super-particles for experiments 1-4.

With a reduced number of super-particles, the overall simulation
characteristics remain qualitatively similar, even for the largest
$\xi_0$, although averaged quantities such as the precipitation rate
show significant differences. In contrast, reducing $\xi_0$ to
$3\times10^9$ substantially increases computational cost while
producing only minor changes in physical quantities. The absence of a
statistically significant difference between $\xi_0 = 3\times10^9$ and
the control simulation
suggests that the domain-averaged precipitation rate is effectively converged
with respect to super-particle multiplicity. Therefore, all subsequent
experiments use $\xi_0 = 7\times10^9$ as the control
configuration.

The present sensitivity experiments indicate that a mean of
approximately 100 super-particles per grid cell provides a suitable
compromise between computational cost and convergence of the bulk
diagnostics considered here. This is consistent with previous studies
of complex particle system with many particle properties
\citep{Jaruga-2018,Shima-2020}. This does not ensure that the system
is converged in higher-order moments or particle size distributions.
For this, many more super-particles would be necessary. In 1D McSnow
simulations, we find true convergence only for more than 1000
particles per grid cell.

The remaining experiments are analyzed in detail in the following sections.
Here, we briefly summarize the response of the domain-averaged precipitation rate.

Experiments 1-4 including CTL make use of the stochastic model of snowflake aggregation of \citep{Seifert-2026}. Using a deterministic description of aggregation that predicts only
the mean geometry of aggregate snowflakes (MAgg) leads to no
significant change in precipitation rate.  Turbulence effects on
aggregation consistently increase precipitation
rate. This increase is observed for both the mean aggregation geometry
(MAgg vs.\ MAggT) and stochastic aggregation (CTL vs.\ SAggT).

Rime splintering (RS) shows a marginal decrease in surface precipitation, whereas
collision-induced ice–ice fragmentation (CF) causes a pronounced and
statistically robust reduction in precipitation rate. Freezing
fragmentation (FF), on the other hand, shows no statistically
significant impact on precipitation.  When combining collision-induced
fragmentation with turbulence effects on collisions (CFT),
the turbulence-induced enhancement more than offsets the
reduction in precipitation due to fragmentation, resulting in precipitation
rates slightly above CTL.

Variations in maximum updraft velocity are small and show no
systematic response across experiments. Updraft mass flux also varies
only weakly and does not explain the differences in
precipitation. Cloud water path increases with precipitation (with the
exception of CFT), whereas among simulations with the same base
multiplicity, ice water path decreases with precipitation. Rain water
paths exhibit no consistent relationship with precipitation
changes. This indicates that the precipitation differences primarily
arise from changes in microphysical process efficiencies rather than
from dynamical feedbacks. This is consistent with the experimental
design, as all modifications affect primarily ice-phase processes and thus
couple only weakly to latent heating and dynamics.

The static load balancing strategy has proven reasonably effective for
these simulations. All simulations were performed on 16 AMD EPYC nodes
with a total of 2048 CPU cores, using 64 MPI tasks with 32 OpenMP
threads per task. The relatively large number of OpenMP threads helps
to further improve load balancing for the Lagrangian particles
(maximum of $272 \times 10^6$ super-particles). The total wall-clock
time for CTL was 36.2~h, corresponding to 74,137 core hours.  A
comparable simulation using the two-moment bulk scheme requires 1,212
core hours, implying that McSnow is approximately 60 times more
expensive in this configuration.

\section{Microphysical structure of the control experiment}

\begin{figure}[p]
  \centering
  \def\mywidth{0.495}
  \begin{minipage}{\mywidth\textwidth}
    a) total mass density \\[3mm]
    \includegraphics[width=\textwidth, viewport= 0 0 630 400, clip=]{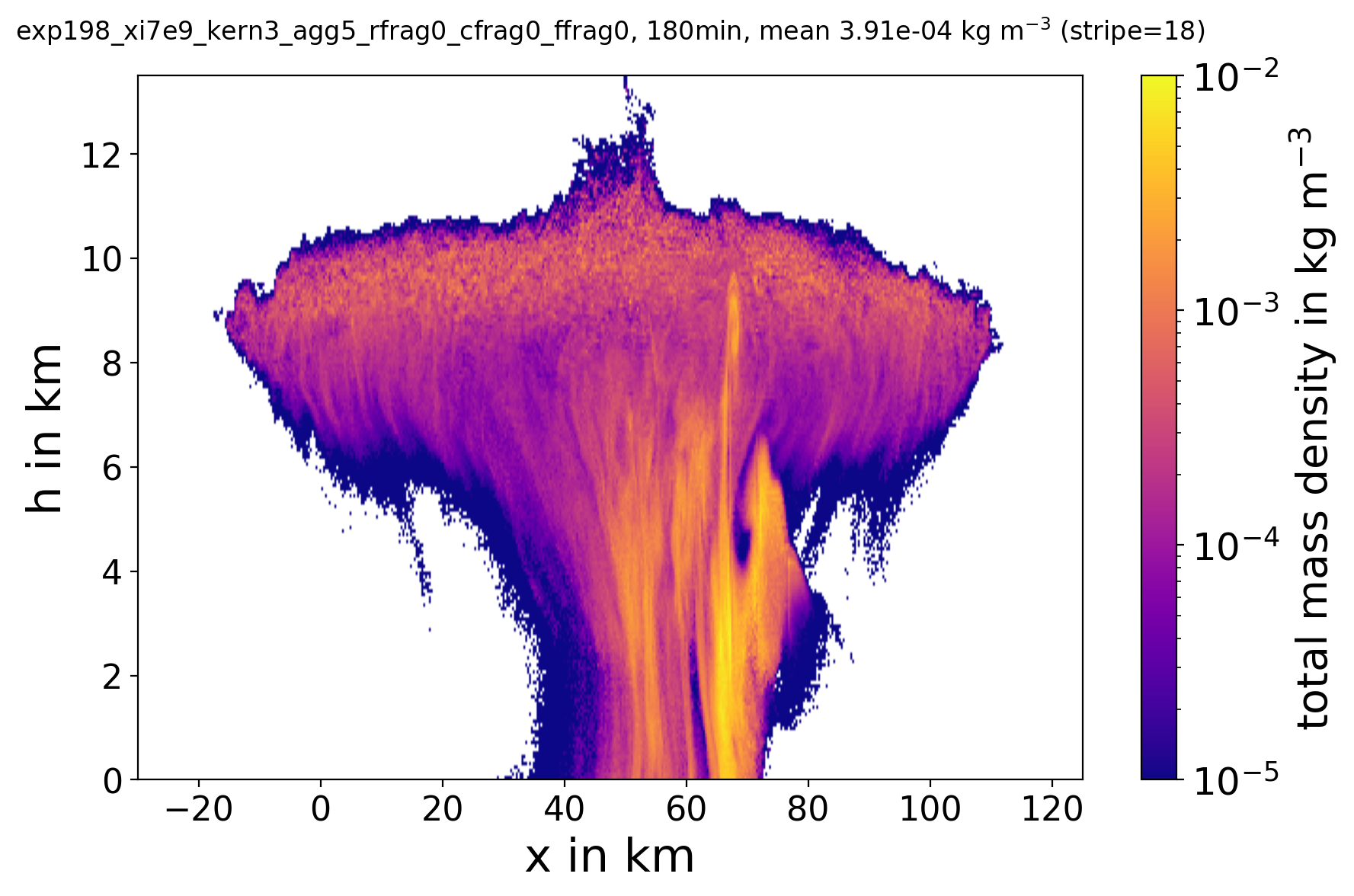}
  \end{minipage}
  \begin{minipage}{\mywidth\textwidth}
    b) total number density \\[3mm]
    \includegraphics[width=\textwidth, viewport= 0 0 630 400, clip=]{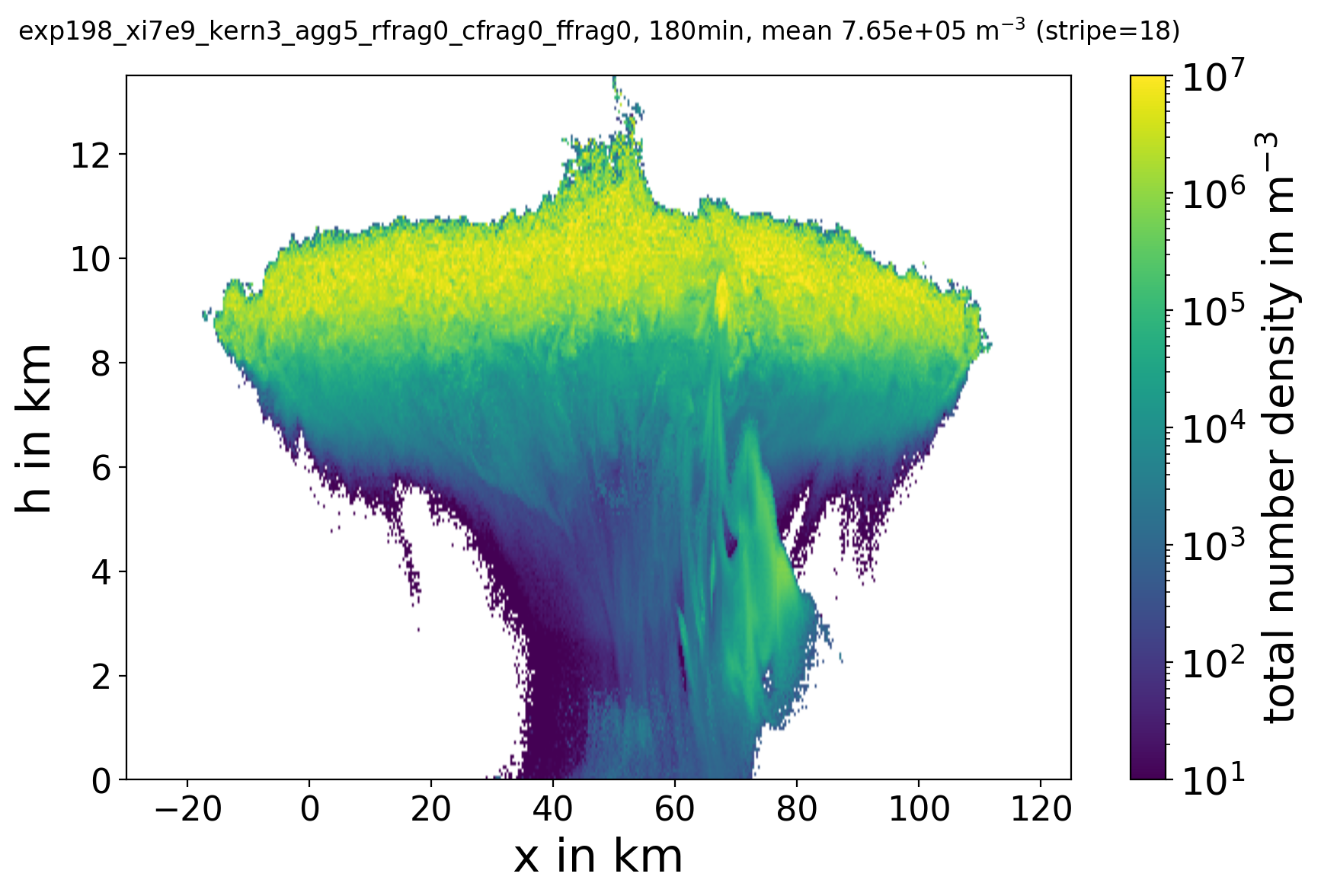}
  \end{minipage}

  \begin{minipage}{\mywidth\textwidth}
    c) rime mass fraction \\[3mm]
    \includegraphics[width=\textwidth, viewport= 0 0 630 400, clip=]{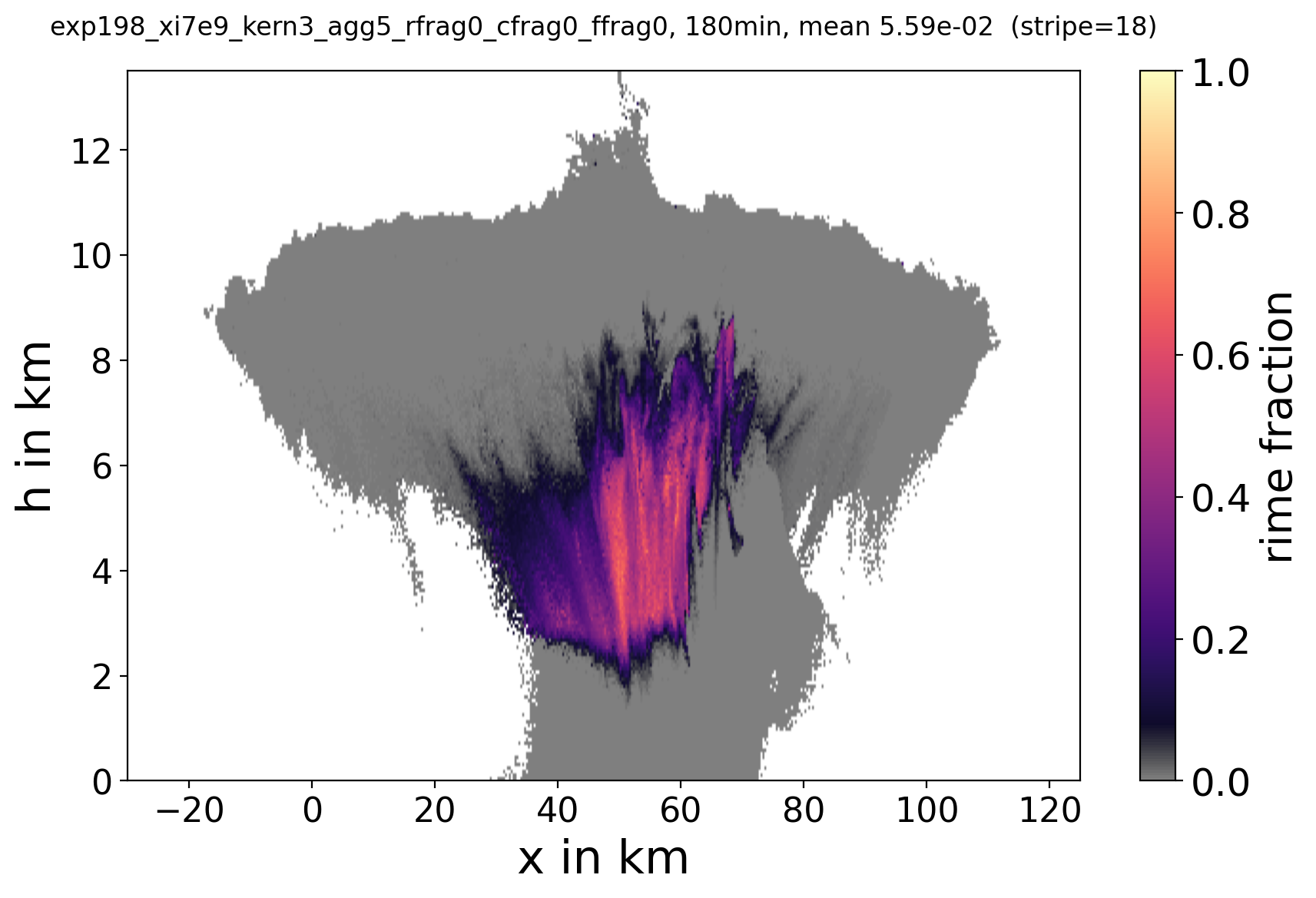}
  \end{minipage}
  \begin{minipage}{\mywidth\textwidth}
    d) frozen mass fraction \\[3mm]
    \includegraphics[width=\textwidth, viewport= 0 0 630 400, clip=]{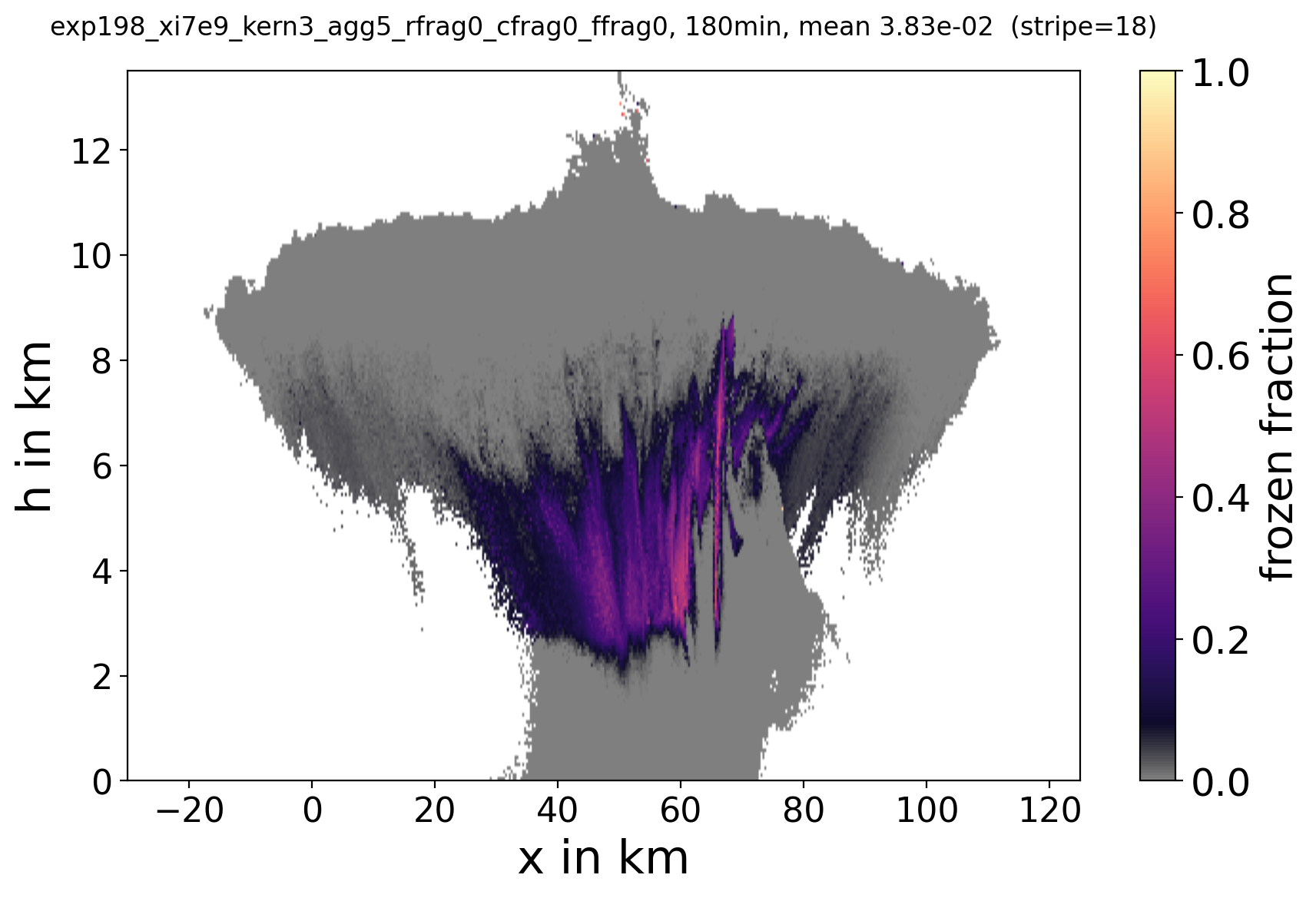}
  \end{minipage}

  \begin{minipage}{\mywidth\textwidth}
    e) liquid mass fraction \\[3mm]
    \includegraphics[width=\textwidth, viewport= 0 0 630 400, clip=]{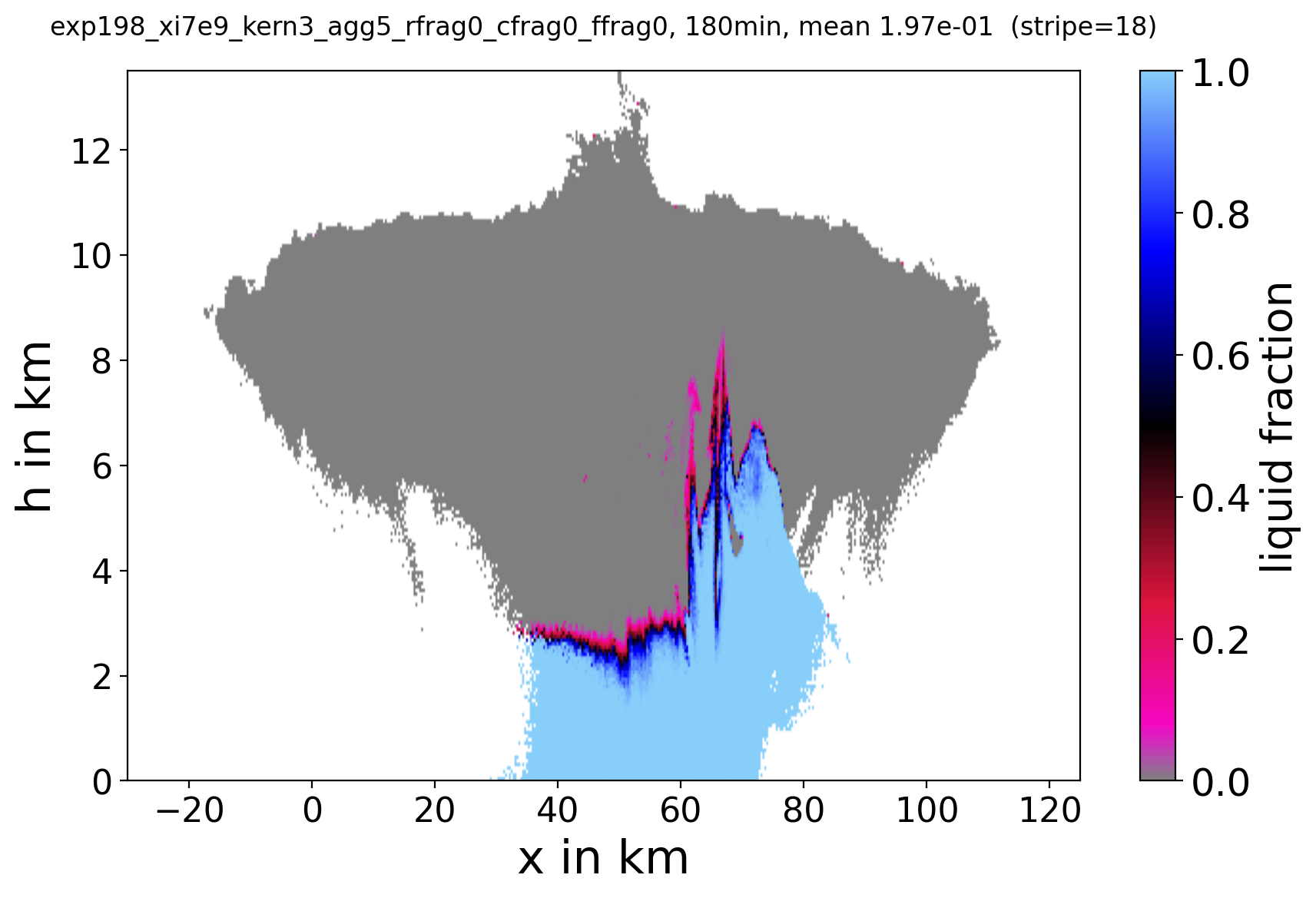}
  \end{minipage}
  \begin{minipage}{\mywidth\textwidth}
    f) maximum dimension \\[3mm]
    \includegraphics[width=\textwidth, viewport= 0 0 630 400, clip=]{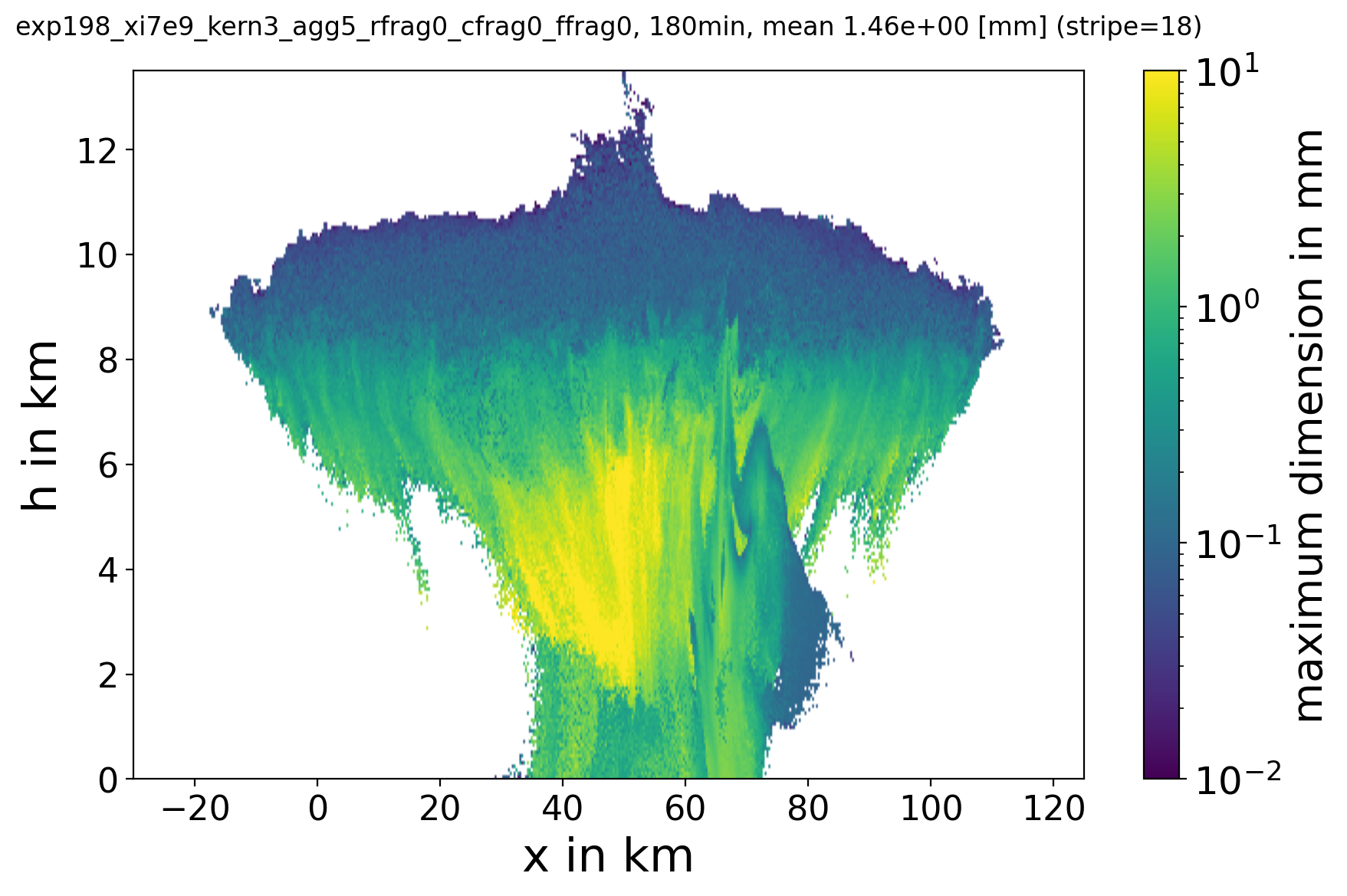}
  \end{minipage}

  \caption{\label{fig:xz-control}
    Line-averaged vertical cross sections of the CTL
    simulation at 180 min of (a) total mass density, (b) total number
    density, (c) rime mass fraction, (d) frozen mass fraction, (e)
    liquid mass fraction and (f) maximum dimension. The line average is
    for 1/20 of the model domain corresponding to a length of 1
    km. For the intensive variables (c,d,e,f) a mass-weighted average is applied.}
\end{figure}

\begin{figure}[t]
  \centering
  \def\mywidth{0.495}
  \begin{minipage}{\mywidth\textwidth}
    a) monomer count \\[3mm]
    \includegraphics[width=\textwidth, viewport= 0 0 630 400, clip=]{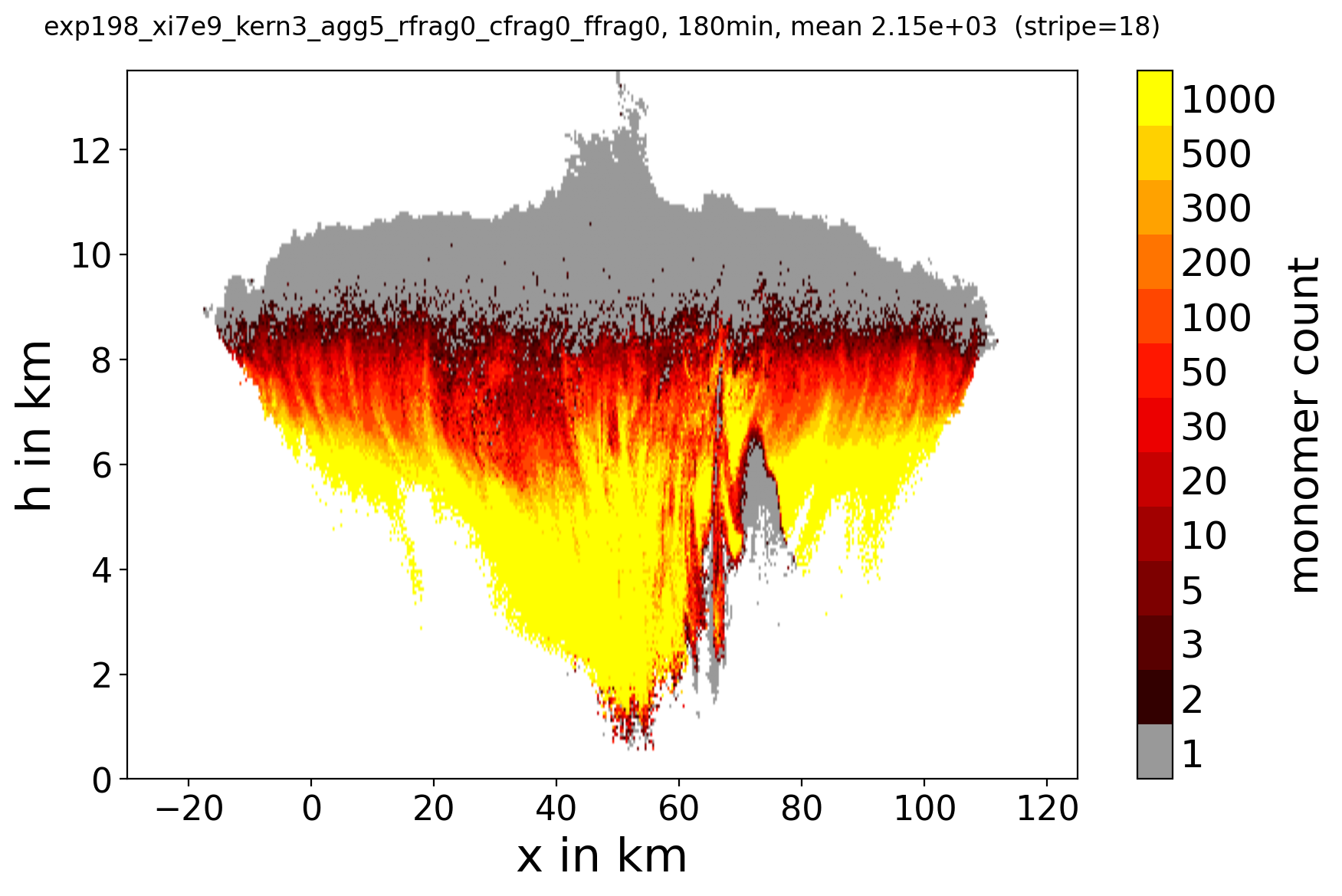}
  \end{minipage}
  \begin{minipage}{\mywidth\textwidth}
    b) apparent density \\[3mm]
    \includegraphics[width=\textwidth, viewport= 0 0 630 400, clip=]{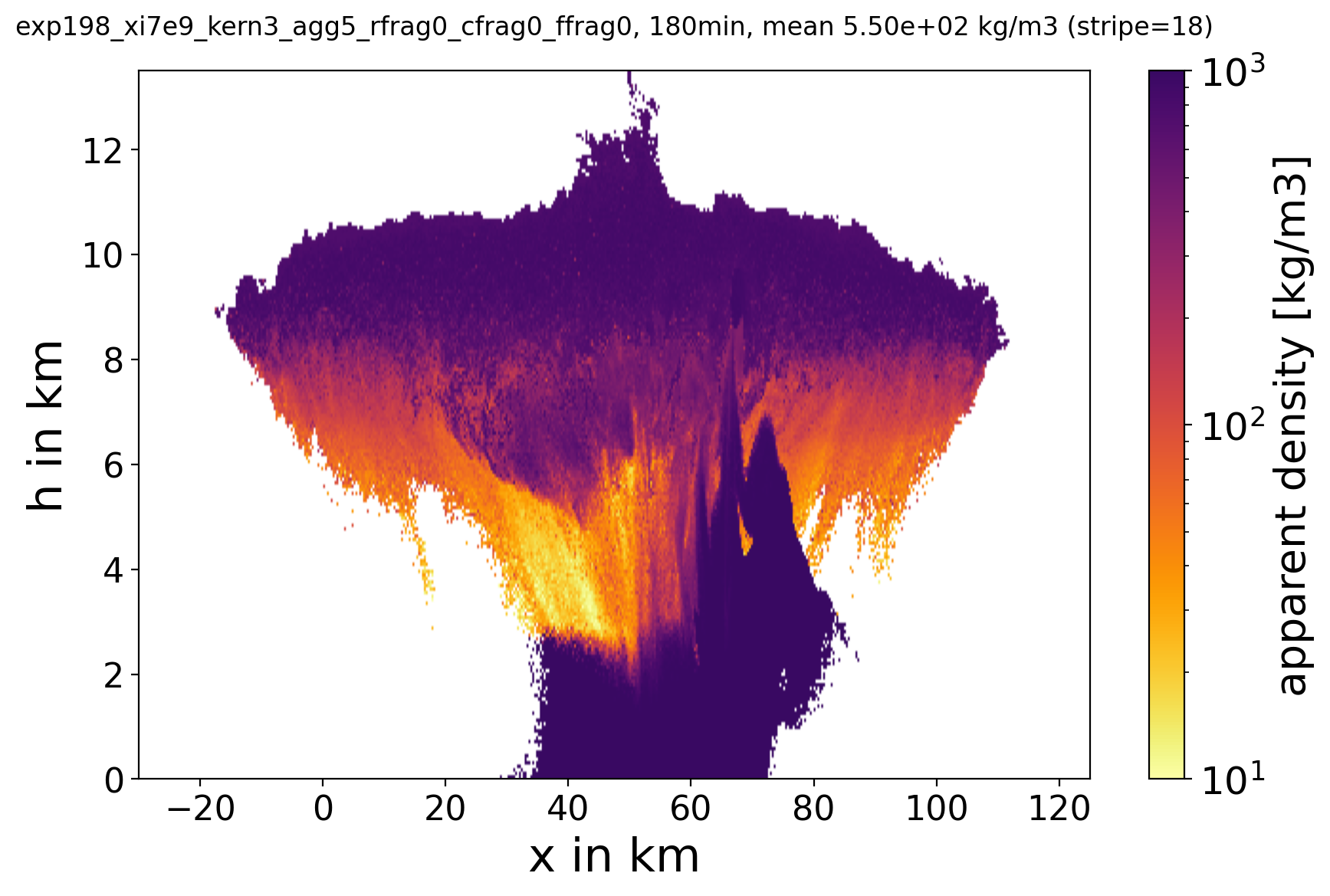}
  \end{minipage}

  \caption{\label{fig:xz-control2} As Figure \ref{fig:xz-control} but
    showing (a) monomer count and (b) apparent density. }
\end{figure}

Figures \ref{fig:xz-control} and \ref{fig:xz-control2} show vertical
cross sections of total mass density, total number density, rime mass
fraction, frozen mass fraction, liquid mass fraction, maximum
dimension, apparent density, and monomer count for the CTL
simulation. Cloud liquid water is not included, as it is not
represented by super-particles.  The fields are averaged over one
twentieth of the domain in the $y$-direction (approximately 1 km), and
the slice with the highest total water content is shown.  The convective
core is characterized by high total water density and a liquid mass
fraction close to unity.
Within the convective core, approximately from $x=60$ km to $x=80$ km, supercooled raindrops freeze and
subsequently grow rapidly by riming, producing the largest particles
in the simulation. As a result, small and
large particles coexist in close proximity, while most particles have
a monomer count of one, indicating that aggregation remains limited.

In contrast, the trailing stratiform region, approximately from $x=30$ km to $x=60$ km, shows decreasing rime and
frozen mass fractions, increasing monomer count, and decreasing
apparent density in mid-levels behind the convective core. This
indicates a dominance of lightly rimed and unrimed aggregate
snowflakes.  In the convective anvil above 8 km height, the highest
number densities of ice particles are found, while rimed particles and
frozen drops are largely absent. The monomer count remains close to
one, consistent with unrimed primary ice crystals. This is consistent
with the well-known structure of squall lines
\citep[e.g.][]{Xue-2017}. While the ice water content in the anvil is
high, the trailing stratiform region is relatively weak compared to
many observed systems. On one hand, this lower precipitation
efficiency is consistent with the weak low-level shear of
$10~\text{m/s}$ and moderate low-level moisture of $12~\text{g/kg}$,
which favors a more tilted, less dynamically 'optimal' convective
organization.  Alternatively, the relatively weak stratiform region
may indicate that particle growth in McSnow may be too slow to
efficiently populate the intermediate particle sizes that sustain
stratiform precipitation. Quantifying this possibility would require
dedicated comparison with observations or alternative microphysical
schemes and is beyond the scope of the present study.

Together, these patterns indicate a transition from riming-dominated
growth in the convective core to aggregation-dominated growth in the
trailing stratiform region.

\def\expname{exp198}

\begin{figure}[p]
  \centering
  \begin{minipage}{0.95\textwidth}
    \def\mywidth{0.45}
    \begin{minipage}{\mywidth\textwidth}
      a) particle mass \\[1mm]
      \includegraphics[width=\textwidth, viewport= 0 319 710 833, clip=]{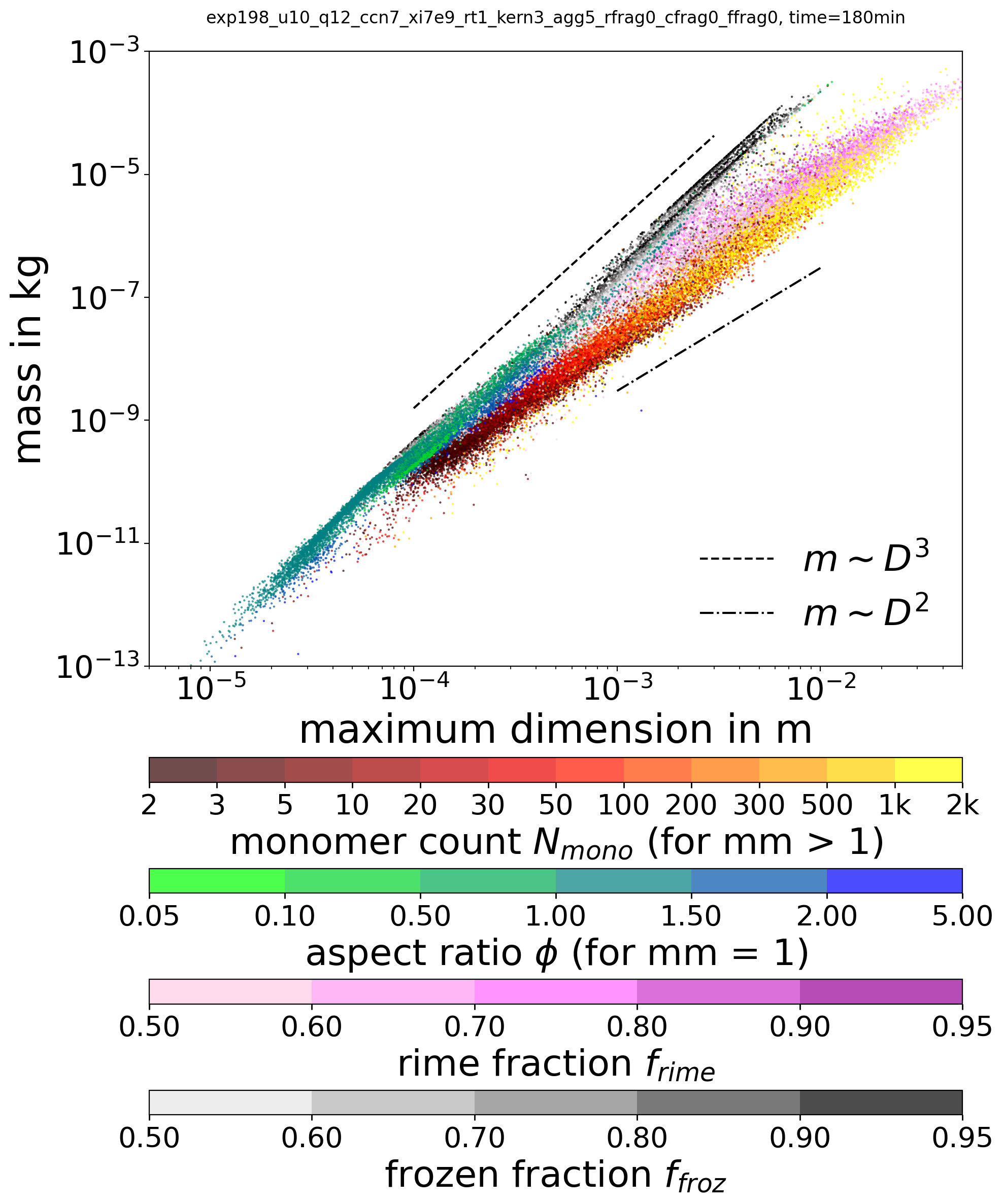}
    \end{minipage}
    \hfill
    \begin{minipage}{\mywidth\textwidth}
      b) ellipsoidal area fraction \\[1mm]
      \includegraphics[width=\textwidth, viewport= 0 319 710 833, clip=]{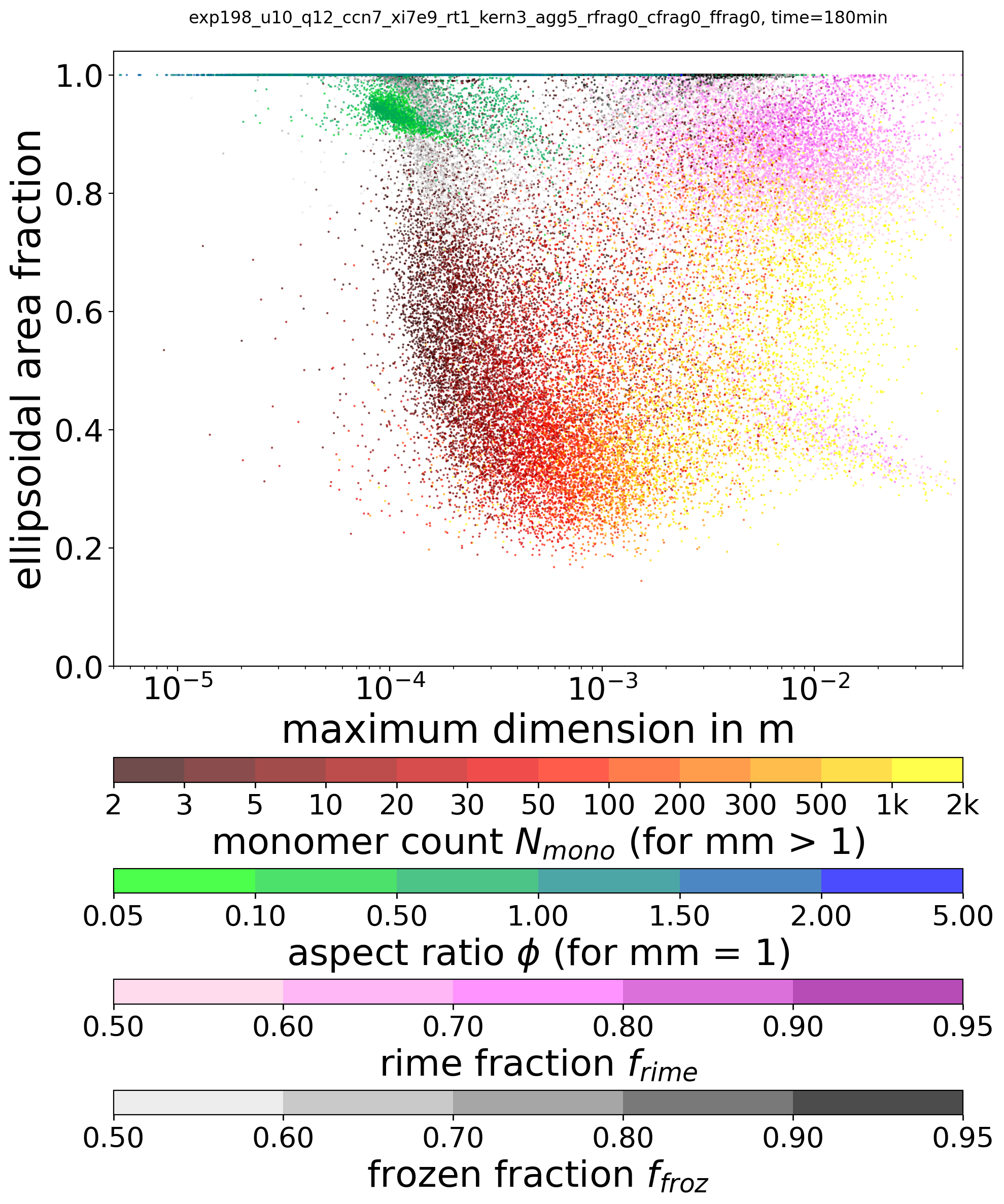}
    \end{minipage}
    \vspace{2mm}    
    
    \begin{minipage}{\mywidth\textwidth}
      c) aspect ratio \\[1mm]
      \includegraphics[width=\textwidth, viewport= 0 319 710 833, clip=]{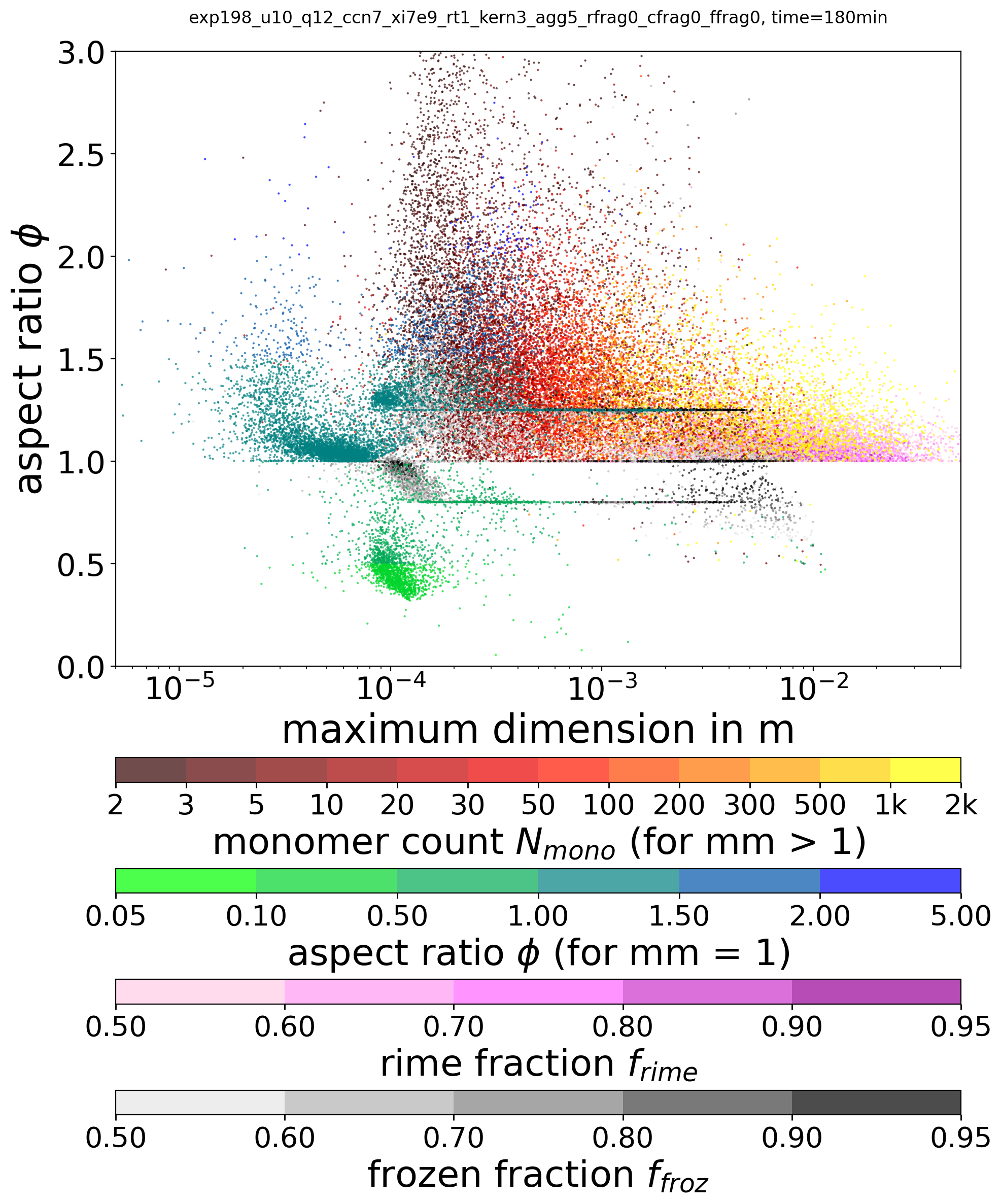}
    \end{minipage}
    \hfill
    \begin{minipage}{\mywidth\textwidth}
      d) apparent density \\[1mm]
      \includegraphics[width=\textwidth, viewport= 0 319 710 833, clip=]{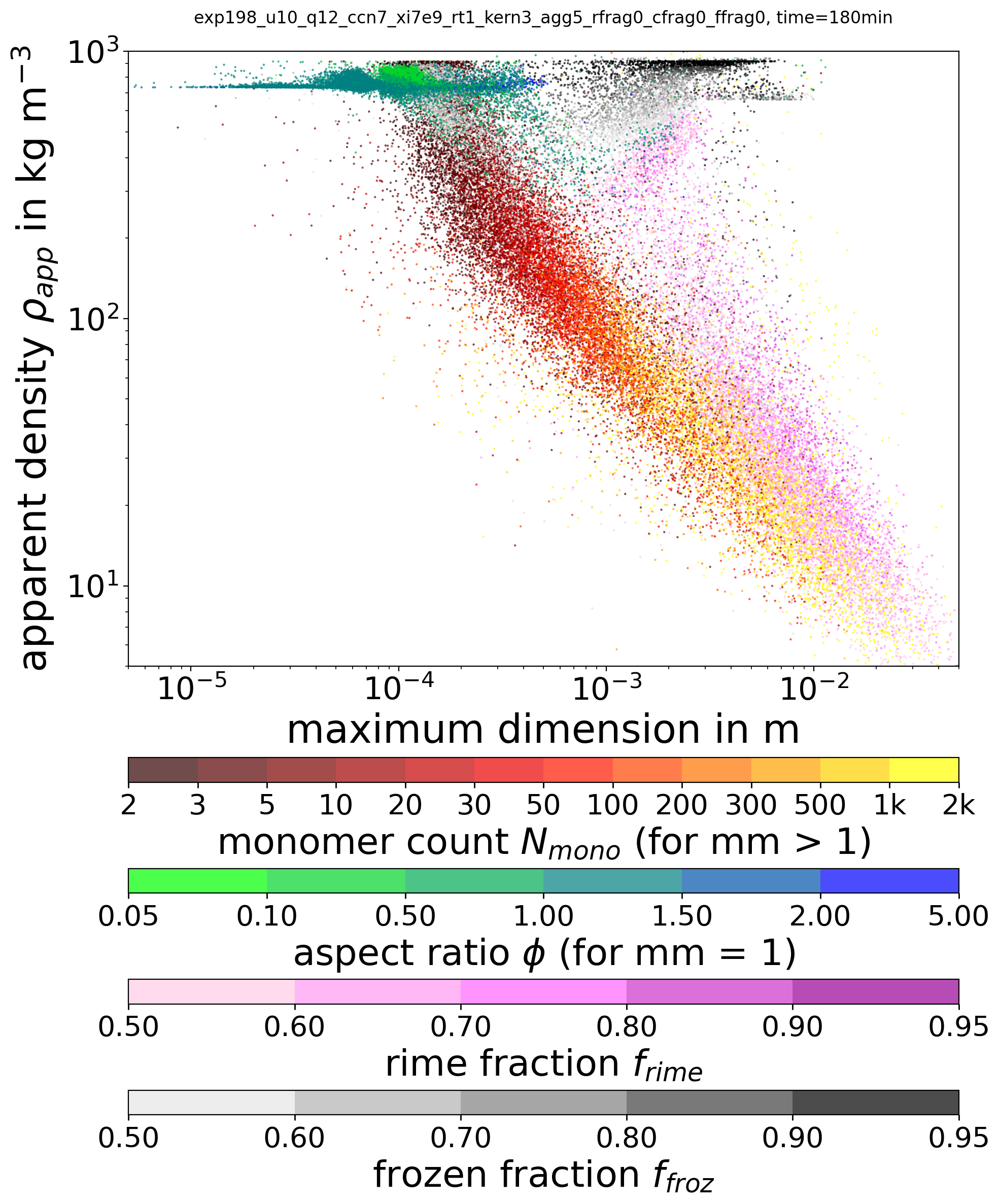}
    \end{minipage}
    \vspace{2mm}    

    \begin{minipage}{\mywidth\textwidth}
      e) velocity (log-log) \\[1mm]
      \includegraphics[width=\textwidth, viewport= 0 319 710 833, clip=]{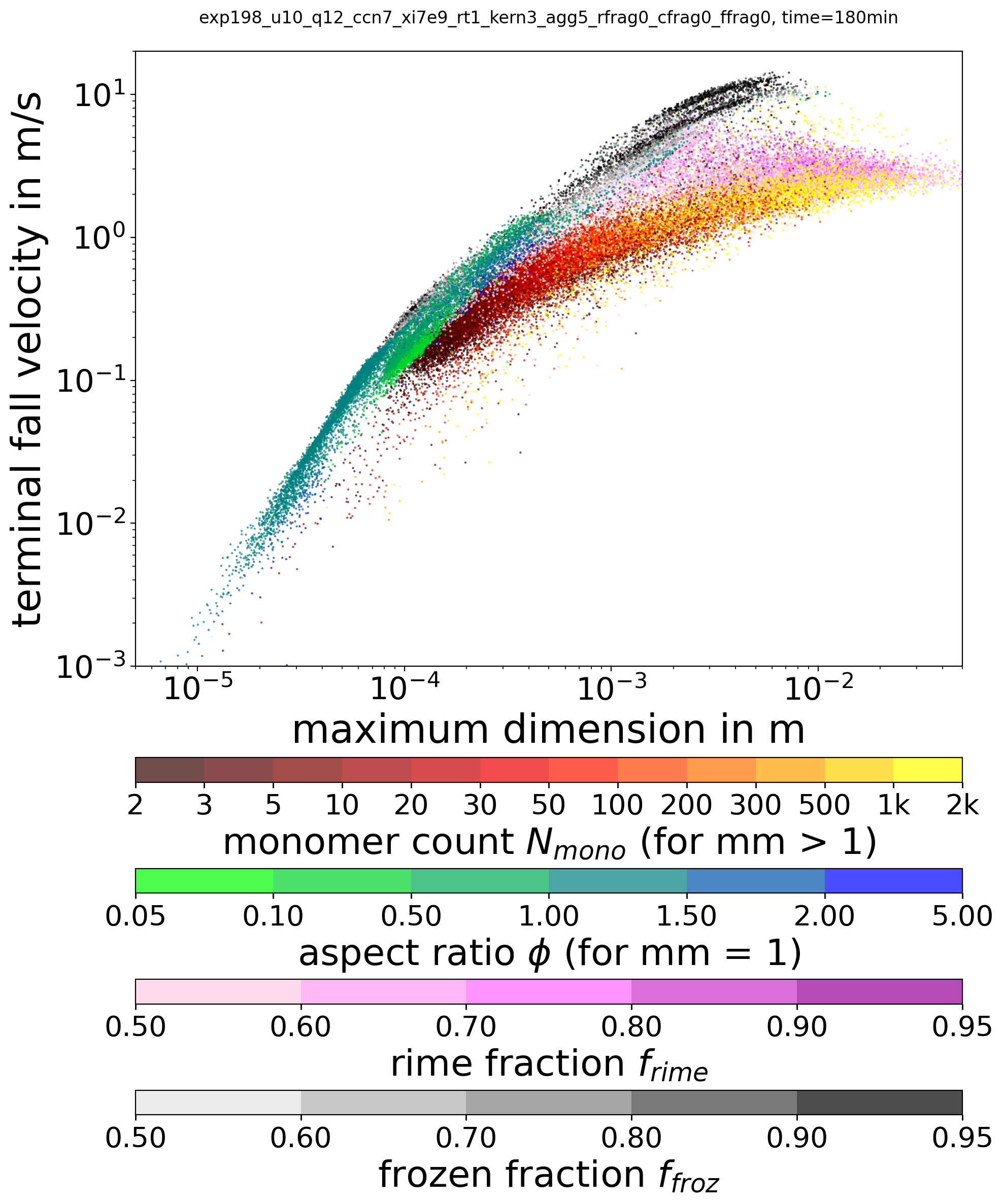}
    \end{minipage}
    \hfill
    \begin{minipage}{\mywidth\textwidth}
      f) velocity (log-linear) \\[1mm]
      \includegraphics[width=\textwidth, viewport= 0 319 710 833, clip=]{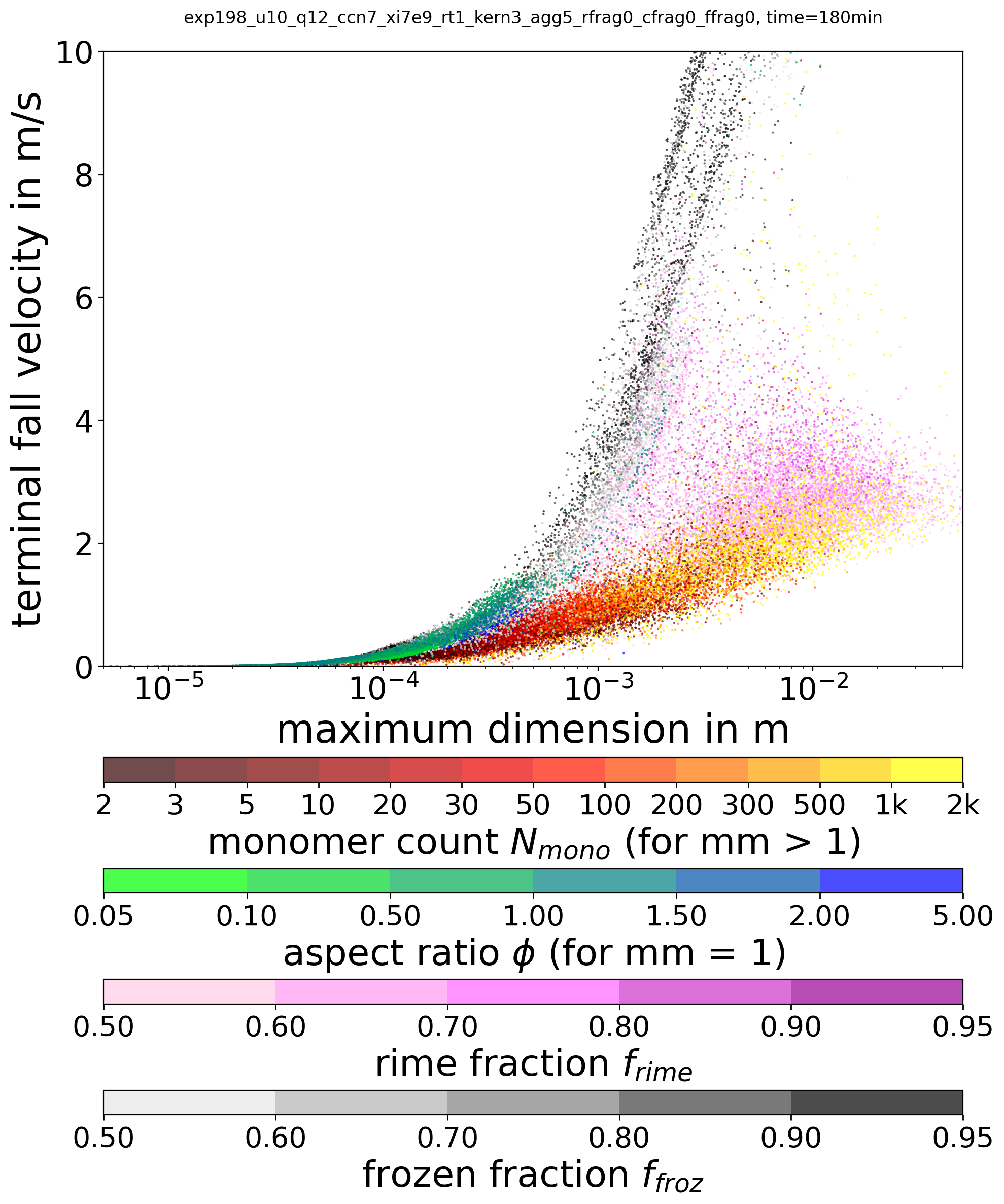}      
    \end{minipage}
    \vspace{3mm}

    \begin{minipage}{\mywidth\textwidth}
      \includegraphics[width=\textwidth,viewport= 0 162 710 240, clip=]{mcsnow_logvel_size_scatter_subset_j18.png}
    \end{minipage}
    \hfill
    \begin{minipage}{\mywidth\textwidth}
      \includegraphics[width=\textwidth,viewport= 0 82 710 160, clip=]{mcsnow_vel_size_scatter_subset_j18.png}
    \end{minipage}
    \vspace{2mm}    

    \begin{minipage}{\mywidth\textwidth}
      \includegraphics[width=\textwidth,viewport= 0 242 710 320, clip=]{mcsnow_logvel_size_scatter_subset_j18.png}
    \end{minipage}
    \hfill
    \begin{minipage}{\mywidth\textwidth}
      \includegraphics[width=\textwidth,viewport= 0 5 710 85, clip=]{mcsnow_vel_size_scatter_subset_j18.png}
    \end{minipage}
    \vspace{4mm}    
  \end{minipage}

  \caption{\label{fig:particlescatter} Scatter plots illustrating how
    particle geometry and fall speed evolve with maximum dimension in
    the CTL simulation.  Shown are a) particle mass, b) ellipsoidal
    area fraction, c) aspect ratio, d) apparent density and the
    terminal fall velocity with log axis (e) and linear axis (f). Note
    that the terminal fall velocity is not at reference pressure. The
    data are from the same cross section as the previous figures, but
    only a very small subset of the super-particles are shown. The
    sub-sampling uses a logarithmic  binning in particle mass to show also some
    particles from the tail of the distribution.}
\end{figure}

Figure \ref{fig:particlescatter} shows scatter plots of particle
properties in the same cross section as Figs.~\ref{fig:xz-control} and
\ref{fig:xz-control2}. A subset of approximately 50,000
super-particles is shown, sampled from all particles in the cross
section using logarithmic binning in total mass. This sampling helps
ensure that the tails of the distribution are adequately represented.
The color coding in the scatter plots provides additional information
on particle properties. For monomers with $N=1$ the aspect ratio is
color-coded using the green-to-blue colorbar. For unrimed and lightly
rimed aggregates with rime fraction $f_\text{rime} < 0.5$ and frozen
fraction $f_\text{frz} < 0.5$ the monomer count $N$ is shown using the
brown-red-yellow colorbar. For rimed particles with $f_\text{rime} \ge
0.5$ the rime fraction is shown using the magenta colorbar and for
particles with $f_\text{frz} \ge 0.5$ the frozen fraction is shown
using the grey colorbar.  The mass–size relation in
Fig.~\ref{fig:particlescatter}a shows cubic scaling for primary
crystals and spherical particles, and quadratic scaling for unrimed
aggregates. Rimed particles populate the transition between these two
regimes.

The area fraction (Fig.~\ref{fig:particlescatter}b) ranges from about
0.2 to 0.7 for unrimed aggregates, but increases rapidly with riming.
Monomers are predominantly plate-like with high area fraction, though
some prolate, needle-like crystals also occur, as shown in
Fig.~\ref{fig:particlescatter}c.  Unrimed aggregates are generally
prolate, with aspect ratios between 1 and 3 consistent with
\citet{Seifert-2026}. The two horizontal lines in
Fig.~\ref{fig:particlescatter}c at $\phi=0.8$ and $\phi=1.25$ indicate
the range of quasi-spherical tumbling graupel particles
\citep{Jensen-Harrington-2015}; see also Section~2.1.3 of
\citet{Welss-2024} for details. The apparent density
(Fig.~\ref{fig:particlescatter}d) is high for primary crystals and
frozen drops, and can become very low for aggregate snowflakes. As
expected, riming increases the apparent density.

The terminal fall velocity spans four orders in magnitude from
small primary crystals with negligible fall speeds to large graupel
exceeding 10 m/s (Fig.~\ref{fig:particlescatter}e).
For unrimed
aggregates the terminal fall velocity shows a linear increase in the
log-linear plot (Fig.~\ref{fig:particlescatter}f) indicating an
approximately logarithmic growth of terminal velocity with maximum
dimension. Rimed snowflakes fall faster due to their higher
density. The highest terminal fall velocity is reached by particles
that contain a large fraction of frozen mass. Overall, the particle
properties are consistent with Figs.~4--7 of \citet{Shima-2020}.

\begin{figure}[t]
  \centering
  \includegraphics[width=\textwidth, viewport= 0 0 1070 350, clip=]{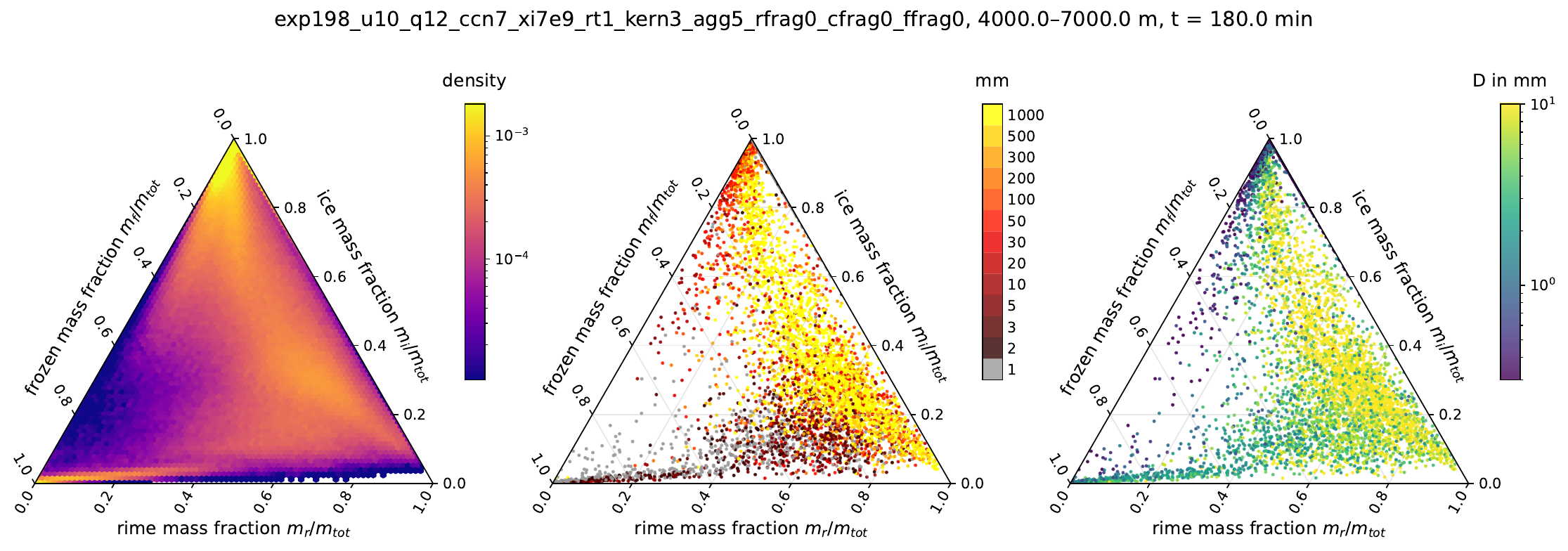}

  \caption{\label{fig:ternary} Ternary plots of the mass composition
    of ice particles in the mixed-phase region of the squall line
    between 4000 m and 7000 m height illustrating the dominant
    particle growth pathways. Probability density (left), and scatter
    plots of the monomer count (center) and the maximum dimension
    (right).}
\end{figure}

The ternary representation provides a compact view of the relative
contributions of deposition, freezing, and riming to particle growth.
A ternary plot helps identify the dominant particle growth pathways
corresponding to distinct trajectories in composition space.  Ternary
plots are based on the relation $m_i + m_f + m_r = m_{\text{tot}}$,
i.e., ice mass, frozen mass, and rime mass sum to the total mass. The
corresponding mass fractions $m_k/m_{\text{tot}}$ with $k \in
\{i,f,r\}$ sum to unity and define the coordinates within the
triangle.  Ternary representations are widely used to visualize
three-component compositions, for example in chemistry (flammability
diagram), pedology (soil texture triangle), petrology (classification
diagrams for rocks), and population genetics (de Finetti
diagram). Figure \ref{fig:ternary} shows ternary diagrams for
particles from the mixed-phase region between 3 and 7 km height,
displaying the probability density, monomer count, and maximum
dimension. The probability density indicates that a wide range of mass
compositions occur in the simulation, although some compositions occur
much less frequently. A purely rime-dominated state
($m_r/m_{\text{tot}} \approx 1$) is essentially absent, since riming
is typically accompanied by depositional growth under mixed-phase
conditions.

Particles formed by heterogeneous ice nucleation originate near the
$m_i/m_{\text{tot}} = 1$ corner and grow by deposition, aggregation,
and riming. Frozen drops, particles undergoing wet growth, and
refrozen particles originate near the $m_f/m_{\text{tot}} = 1$ corner
and typically grow rapidly by riming. Only after reaching sufficient
size do they participate in aggregation.  Most particles in the
millimeter size range are either large aggregates or heavily rimed
graupel or hail particles. Two dominant pathways emerge. Primary ice
crystals evolve from the deposition-dominated corner toward aggregates
through deposition and aggregation, whereas frozen drops originate
near the frozen-mass corner and evolve toward heavily rimed graupel
through riming before eventually entering the aggregate regime.
The absence of large hail with
$m_f/m_{\text{tot}} \approx 1$ and $D_{\max} > 5$ mm in the figure
does not imply that such particles are absent in the simulation;
rather, they are too rare to be captured by the sampling (logarithmic binning
in particle mass) used in the figure.

Overall, the CTL simulation exhibits three distinct microphysical
regimes: an anvil dominated by primary ice crystals, a convective core
characterized by riming and frozen drops, and a trailing stratiform
region dominated by aggregate snowflakes. The particle-property and
ternary analyses reveal two principal growth pathways, one originating
from primary ice crystals and one from frozen drops, which converge
toward particles that have experienced both aggregation and
substantial riming.

\section{Aggregation, stochastic snowflakes and turbulence}

Aggregation in clouds is dominated by the differential sedimentation
of ice particles. The corresponding gravitational collection kernel
is:
\begin{equation}
  K_{\text{grav}} = \pi (r_1 + r_2)^2 \, |v_1 - v_2| \, \eta_g  
\end{equation}
where $r_1$ and $r_2$ are the radii perpendicular to the flow, $v_1$
and $v_2$ are the terminal fall velocities, and $\eta_g$ is the
collection efficiency. For non-spherical particles, we adopt a spheroid
model with the maximum dimension oriented perpendicular to the flow:
\begin{equation}
  r_n = \frac{D_n}{2} \min \! \left ( 1, \, \phi^{-\frac{1}{2}} \right )
\end{equation}
where $\phi$ is the aspect ratio and $D_n$ is the maximum
dimension. The term $\phi^{-1/2}$ describes the reduced cross-sectional
area of prolate particles like columns or needles, whereas for oblate
particles the cross-sectional area does not depend on $\phi$.

In turbulent flows, the dynamical formulation of the collection kernel
is used:
\begin{equation}
  K_{\text{dyn}} = 2 \pi (r_1 + r_2)^2 \, \langle |w_r| \rangle \, g_{12} \, \eta_g \, \eta_e  
\end{equation}
This formulation is based on the mean radial relative velocity at
contact $\langle |w_r| \rangle$ and the corresponding radial
distribution function $g_{12}$. The factor $\eta_e$ accounts for the
turbulent enhancement of collision efficiency
\citep{Ayala-Wang-2008}. In McSnow, turbulence is
parameterized following \citet{Onishi-Seifert-2016}, see Appendix C,
and the total collection kernel combines gravitational and turbulent
contributions as:
\begin{equation}
  K_{\text{coll}} = \sqrt {K_{\text{grav}}^2 + K_{\text{dyn}}^2 }.
\end{equation}
The stochastic snowflake model of \citet{Seifert-2026} additionally
accounts for the natural geometric variability of aggregates.

Standard power-law descriptions predict identical terminal velocities
and collision cross sections for particles of equal
mass. Consequently, the gravitational collision kernel vanishes for
equal-mass particles.  However, the stochastic model introduces
variability in $v$ and $r$ even among particles of equal mass,
enabling collisions that a purely gravitational kernel would
neglect. This makes the stochastic snowflake model a compelling
geometric counterpart to the physical effects of turbulence, although
stochastic geometry modifies the particle properties entering the
kernel, whereas turbulence acts directly on the collision rate.

\def\expname{exp209_u10_q12_ccn7_xi7e9_rt1_kern4_agg5_rfrag0_cfrag0_ffrag0}
\begin{figure}[t]
  \centering
  \begin{minipage}{0.7\textwidth}
    \includegraphics[width=\textwidth]{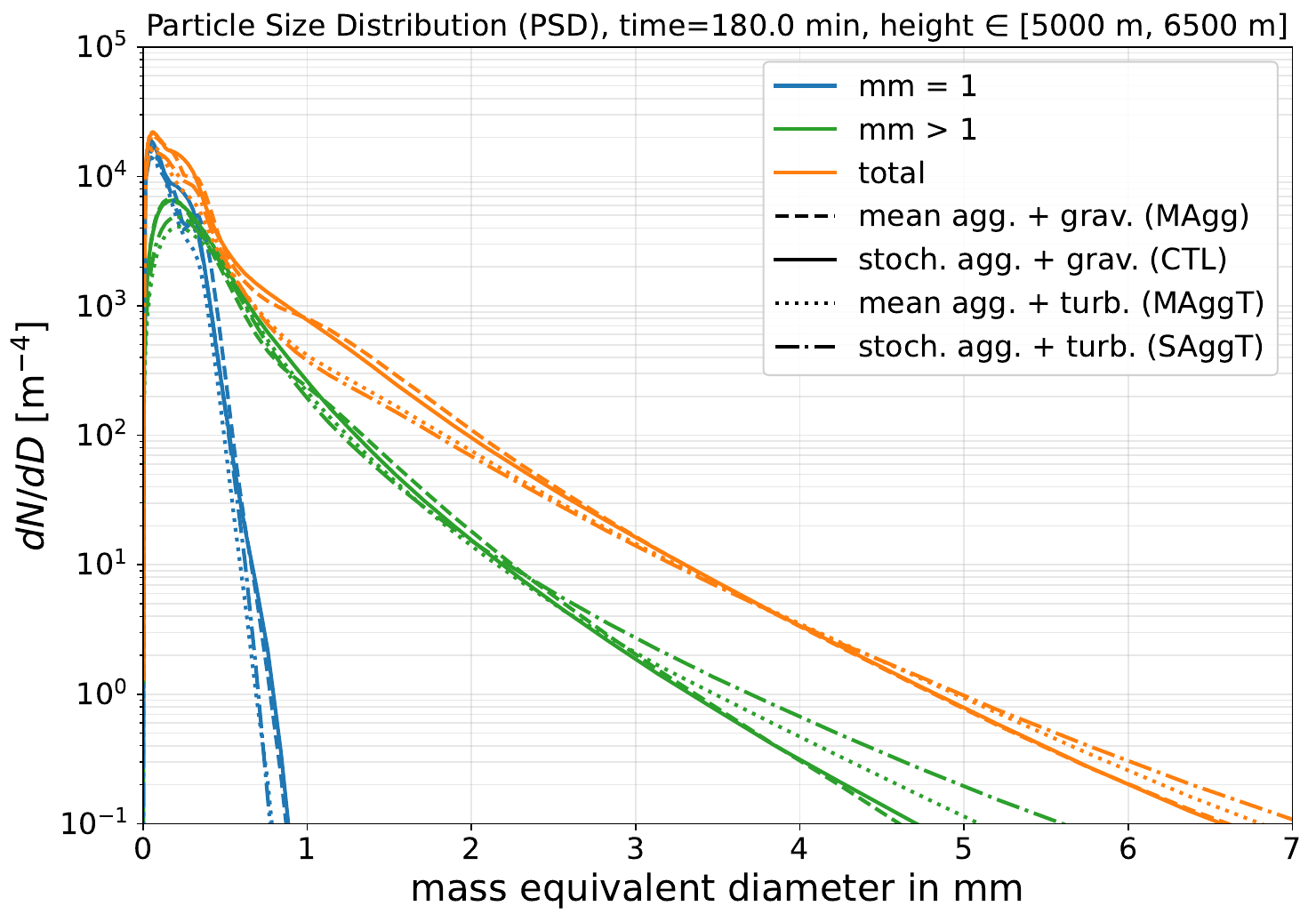}
  \end{minipage}
  
  \caption{\label{fig:psd_agg_and_turb} Particle size distributions
    for monomer crystals (mm=1, blue), unrimed and lightly rimed
    aggregates (mm $>$ 1, green), and the total (orange). Shown is
    experiment MAgg assuming mean aggregate snowflake and
    gravitational kernel (solid lines), CTL with stochastic aggregates
    and gravitational kernel (dashed), MAggT with mean-aggregate and
    turbulence effects on collision (dotted), and SAggT with
    stochastic aggregates and turbulence effects on collision
    (dashed-dotted). All PSDs are for 180 min simulation time for the
    whole layer between 5000 m and 6500 m height.}
\end{figure}

The four experiments shown below disentangle the effects of stochastic
particle geometry and turbulence on aggregation. Figure
\ref{fig:psd_agg_and_turb} shows particle size distributions as a
function of mass-equivalent diameter in the mixed-phase region of the
squall line. Four experiments are shown: a deterministic aggregation
case without turbulence (MAgg), a stochastic baseline (CTL),
deterministic aggregation with turbulence (MAggT), and stochastic
aggregation with turbulence (SAggT).

Turbulence enhances aggregation as expected, producing more large
aggregates and graupel. In contrast, stochastic snowflake geometry
reduces the abundance of large snowflakes. At first sight, this appears
counterintuitive because stochastic geometry introduces variability in
fall speed and projected area among particles of similar mass, thereby
allowing differential sedimentation between particles that would
otherwise have identical terminal velocities.  However, it also
produces more compact aggregates with smaller projected cross
sections. Although these particles fall faster, the reduction in
collision cross section outweighs the increase in differential fall
velocity. Aggregation in the present simulations is therefore
controlled more strongly by particle geometry than by relative
velocity.  The dominance of projected cross section is a consequence
of the high Reynolds numbers of snowflakes, for which the drag law—and
thus the terminal fall velocity—becomes only weakly dependent on
particle size and shape. Consequently, the decrease in collision cross
section more than compensates for the increase in fall speed of more
compact snowflakes.

Consequently, in the present squall-line simulation, stochastic
snowflake geometry slows aggregation rather than accelerating it. This
may vary across different cloud regimes depending on the local
particle size distribution of snowflakes. Regarding turbulence, we
find a moderate but significant enhancement of aggregation and,
subsequently, precipitation formation. Precipitation increases by
approximately 8--10\% when turbulent effects are included. As
discussed in Appendix C, the application of the Onishi kernel to
non-spherical particles should be understood as a sensitivity study
rather than a fully consistent physically-based parameterization.

\section{Fragmentation of ice particles}

We investigate three secondary ice production (SIP) mechanisms: rime
splintering (RS), freezing fragmentation (FF), and collision-induced
ice--ice fragmentation (CF). While their implementation is described
in detail in the Appendix, this section focuses on their occurrence,
efficiency, and relative importance in the simulated squall line.

\def\expname{exp209_u10_q12_ccn7_xi7e9_rt1_kern3_agg5_rfrag0_cfrag6_ffrag0}
\begin{figure}[t]
  \centering
  \begin{minipage}{\textwidth}
    \def\mywidth{0.495}
    \begin{minipage}{\mywidth\textwidth}
      a) CKE dependence \\[0mm]
      \includegraphics[width=\textwidth]{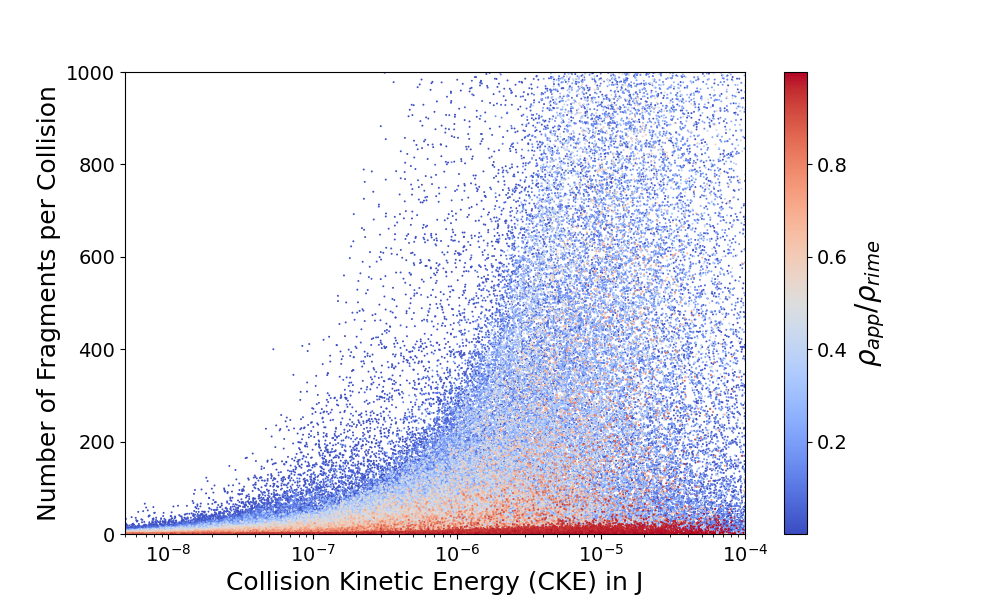}
    \end{minipage}
    \hfill
    \begin{minipage}{\mywidth\textwidth}
      b) temperature dependence \\[0mm]
      \includegraphics[width=\textwidth]{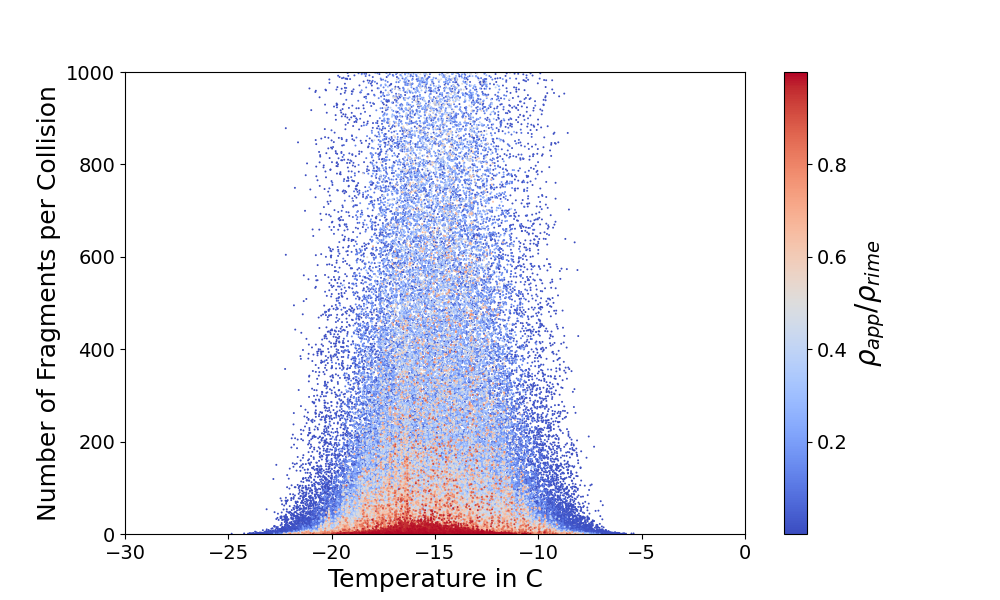}
    \end{minipage}
  \end{minipage}
  
  \caption{\label{fig:iceice_frag} Scatter plots of the number of
    fragments per fragmentation event for collision-induced ice-ice
    fragmentation as a function of the collision kinetic energy CKE (a)
    and temperature (b) for experiment CF. Color indicates the ratio of the apparent
    density over rime density (Eq.~\eqref{eq:densityratio}). Hence, low density particles like
    unrimed aggregates are represented by blue dots, intermediate
    particles like graupel with dendrites by grey dots, and bare
    graupel by red dots. The density characterizes the more fragile
    particle of both collision partners.}
\end{figure}

Figure \ref{fig:iceice_frag} shows for the collision-induced
fragmentation experiment (CF) that collisions with high collision
kinetic energy $\cke$ do occur in the squall line simulation for a
range of particle types, including low-density unrimed aggregates,
graupel with dendrites, and bare graupel. Among these collision types,
low-density particles and graupel with dendrites produce the largest
number of fragments.

Although this behavior is partly prescribed by the parameterization,
it is not obvious that the required collisions occur frequently enough
in the simulated cloud to generate substantial secondary ice
production. In particular, large unrimed aggregates and large graupel
particles could remain sufficiently separated in space and time that
such fragmentation events would be rare. Instead, the simulations
indicate that collisions between fragile aggregate snowflakes -- or
graupel with dendritic surface structures -- and fast-falling
high-density particles such as graupel or hail occur frequently enough
to sustain substantial fragmentation.
The temperature dependence is largely determined by the explicit
temperature function of Eq.~\eqref{eq:final-Nf}, and restricted to the
dendritic growth zone.

\begin{figure}[t]
  \centering
  \begin{minipage}{\textwidth}
    \centering
    \def\mywidth{0.7}
    \begin{minipage}{\mywidth\textwidth}
      \includegraphics[width=\textwidth]{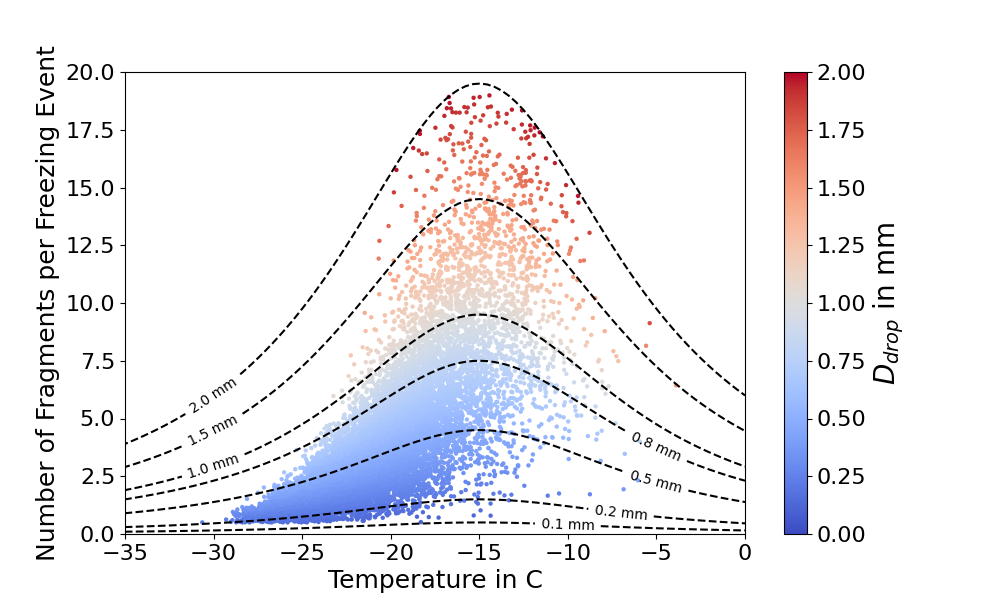}
    \end{minipage}
  \end{minipage}
  
  \caption{\label{fig:freeze_frag} Scatter plot of the number of
    fragments per freezing fragmentation event of raindrops as a
    function of temperature for experiment FF. Color indicates the raindrop diameter. The dashed lines show the prescribed number of fragments per event from the FF parameterization for six fixed raindrop diameters. Note that freezing fragmentation only occurs
    where sufficiently large raindrops exist and freeze in the squall line
    simulation.}
\end{figure}

As shown in Figure~\ref{fig:freeze_frag}, freezing fragmentation is most efficient near $-15^\circ$C,
consistent with the temperature dependence prescribed in the
parameterization (Appendix B). For raindrops smaller than 2 mm diameter, the number of fragments per freezing event increases linearly with drop diameter. Hence, large drops of 1 mm diameter and larger produce many more fragments than the more frequent drops of 0.2-0.5 mm diameter.  Freezing fragmentation events occur over a broad
temperature range, extending from close to $0^\circ$C to about
$-30^\circ$C, but only where raindrops are present and undergo
freezing. 
Large raindrops tend to freeze at relatively high sub-zero temperatures, whereas smaller drizzle drops can remain supercooled to substantially lower temperatures. Consequently, raindrops with diameters larger than
about 1.5~mm are found only above approximately $-20^\circ$C, strongly
limiting the contribution of large drops to freezing fragmentation at
lower temperatures.

\begin{figure}[p]
  \centering
  \begin{minipage}{\textwidth}
    \def\mywidth{0.49}
    \begin{minipage}{\mywidth\textwidth}
      a) ice particle number density \\[1mm]
      \includegraphics[width=\textwidth]{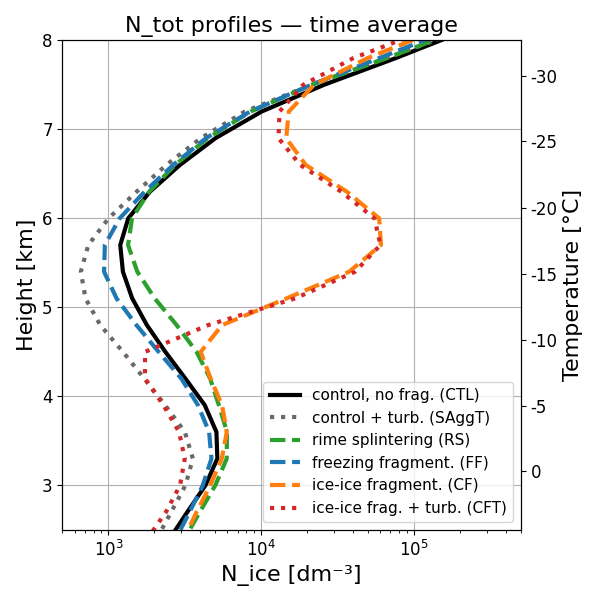}
    \end{minipage}
    \hfill
    \begin{minipage}{\mywidth\textwidth}
      b) ice water content \\[1mm]
      \includegraphics[width=\textwidth]{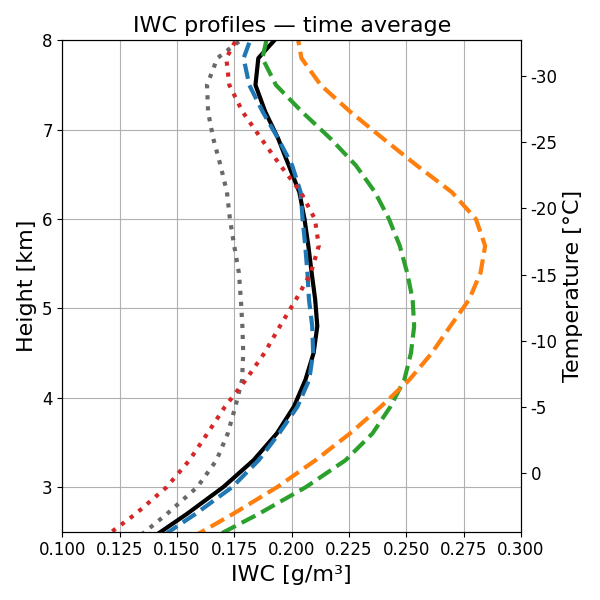}
    \end{minipage}

    \vspace{2mm}
    \begin{minipage}{\mywidth\textwidth}
      c) ice multiplication factor \\[1mm]
      \includegraphics[width=\textwidth]{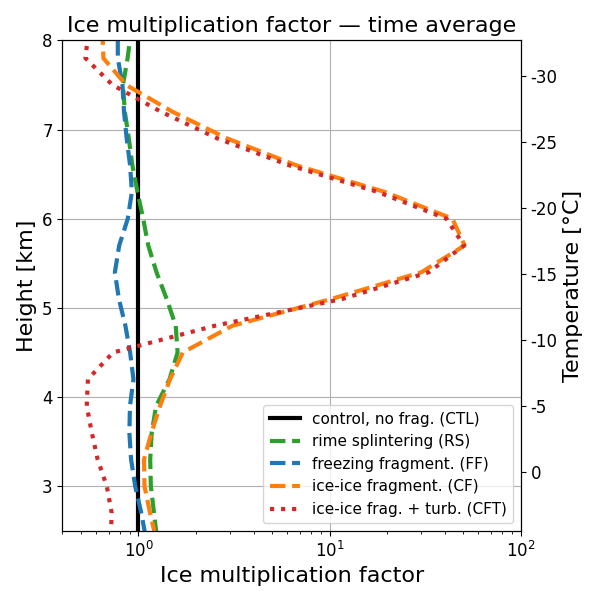}
    \end{minipage}
    \hfill
    \begin{minipage}{\mywidth\textwidth}
      d) number of SIP particles\\[1mm]
      \includegraphics[width=\textwidth]{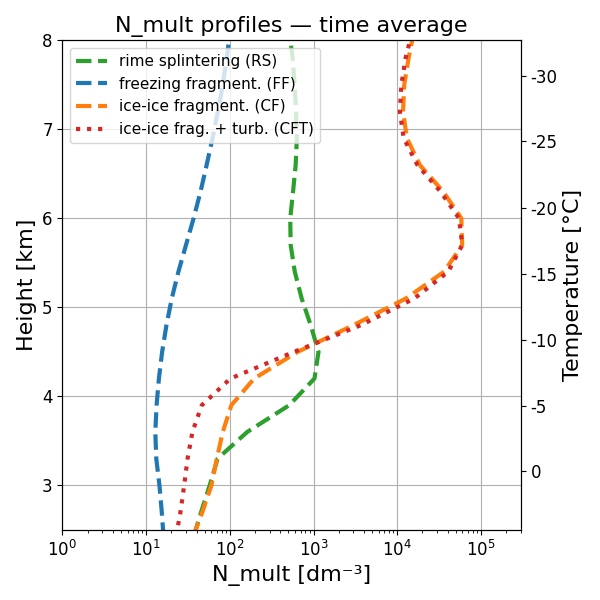}
    \end{minipage}
  \end{minipage}

  \caption{\label{fig:profiles_frag} Vertical profiles of (a) ice
    particle number density, (b) ice water content (IWC), (c) ice
    multiplication factor of ice particle number (relative to the
    control simulation CTL), and (d) number density of secondary ice
    production (SIP) particles for the experiments CTL, SAggT, RS, FF, CF,
    and CFT. Profiles are averaged over the time period 120--180~min
    and over the entire domain using 5 min sampling. Shown is only the mixed-phase region
    between 2500 m and 8000 m height.  The ice multiplication factor
    in (c) quantifies the net effect of SIP on the ice population,
    while (d) shows the contribution of SIP particles tracked
    explicitly in the Lagrangian framework. The control simulation CTL
    is shown as black solid line, fragmentation simulations without
    turbulence as dashed lines (RS, FF, CF), and simulations with turbulence effects
    on collisions as dotted lines (SAggT and CFT). The temperature profile on the
    right y-axis is based on the environmental WK82 profile.}
\end{figure}

\begin{figure}[t]
  \centering
  \begin{minipage}{\textwidth}
    \centering
    \def\mywidth{0.49}
    \begin{minipage}{\mywidth\textwidth}
      \includegraphics[width=\textwidth]{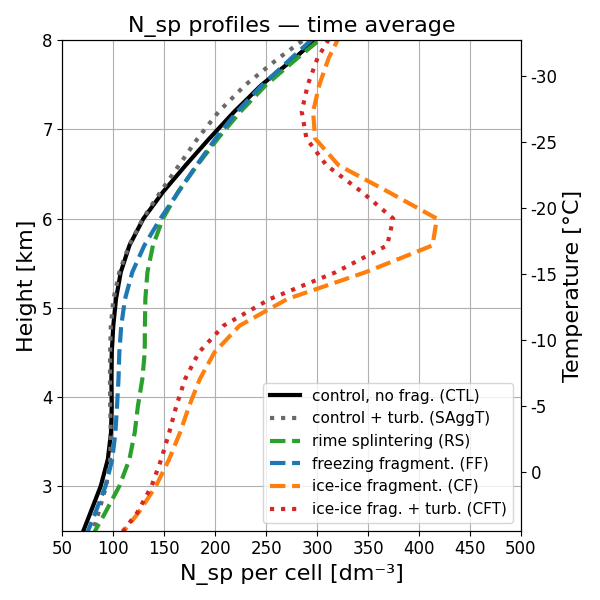}
    \end{minipage}
  \end{minipage}

  \caption{\label{fig:profiles_nsp} As Fig.~\ref{fig:profiles_frag},
    but showing vertical profiles of the number of super-particles per
    grid cell for the experiments CTL, SAggT, RS, FF, CF, and
    CFT. Differences between the experiments reflect both direct
    changes in particle number due to SIP and indirect effects through
    modified sedimentation and residence times.}
\end{figure}

The total effect of the secondary ice production processes is
revealed by the vertical profiles for the experiments RS, FF, CF,
CFT, and CTL shown in Figure \ref{fig:profiles_frag}. For CTL the
basic vertical structure exhibits a double peak, with a local maximum
around 3.5~km height for ice particle number density and around 5~km
for ice water content (IWC), followed by a minimum at about 6~km and
7.5~km, respectively, and an increase further aloft.  This structure
reflects the combined effects of heterogeneous ice nucleation,
freezing of drizzle drops, subsequent particle growth, aggregation,
and sedimentation. As hydrometeors grow, their terminal fall
velocities increase, reducing both particle number density and ice
water content at a given height. The local minimum in particle number
near 6 km coincides with the dendritic growth zone (DGZ), where
aggregation is particularly efficient because sticking efficiencies
are highest near $-15 \Celsius$. The profiles shown here are averaged
over space and time, which complicates a direct process-level
interpretation.

However, they are sufficient to assess the impact of secondary ice
production (SIP).  RS, CF, and CFT show a clear increase in ice
particle number compared to CTL, particularly between 4 and 7~km
height. For CF and CFT the increase in ice particle number happens
mostly in the DGZ, since the process is restricted to this temperature
range in our simulations.  Thus, collision-induced fragmentation
reverses the depletion of particle number seen in the DGZ in CTL,
producing a local maximum instead.  Turbulence effects in CFT do not
increase fragmentation but accelerate the aggregation of newly generated fragments
leading to a reduced ice particle number density outside of the DGZ
relative to CTL. Compared to SAggT, which includes turbulence effects
but no fragmentation, CFT shows an increase in ice particle number
throughout all height levels. RS shows only a moderate increase in
ice particle number density between 4-6 km height. 

A corresponding increase in IWC is evident for RS and CF, whereas the
response is much weaker for FF and CFT, consistent with partial
compensation between fragmentation and turbulence-enhanced aggregation
in CFT. The increased IWC does not directly imply an enhanced
precipitation flux, but primarily reflects the slow fall speeds and longer
residence times of newly generated fragments. Although turbulence
offsets part of the IWC increase relative to CF, CFT still exhibits
enhanced IWC in the DGZ compared to SAggT, confirming that
collision-induced fragmentation robustly increases the ice mass
present in this layer.

The impact of SIP is further illustrated by the ice multiplication
factor, defined here as the ratio of the ice particle number density
in a given experiment to that of the control simulation (CTL), which
does not include SIP processes. This is directly analogous to the
conventional ice multiplication factor, which is defined as the ratio
of the ice particle concentration to the concentration of primary ice
particles (or activated INPs). The CF and CFT experiments show an
increase by more than a factor 50 at around 6~km height, while RS
reaches a factor of about 1.5 from 4.5-5.5~km. In contrast, FF yields
amplification factors around and even below 1.

While the ice multiplication factor quantifies the net effect of SIP
on the ice population, the Lagrangian particle framework additionally
allows us to directly track SIP particles. Each SIP particle is
tagged, any hydrometeor containing at least one SIP particle is
counted.  The resulting vertical profiles show that RS has the highest
number of SIP particles around 4 km height. However, rime splintering
decreases above 4.5~km, whereas collision-induced fragmentation has
its source region in the DGZ and therefore exhibits a maximum around
6~km. 

The particle tracking demonstrates that turbulence has little
influence on the production of fragments themselves. Instead, its
primary effect is to accelerate the subsequent aggregation of those
fragments.  This suggests that fragmentation-producing collisions
occupy a Stokes-number regime outside that for which turbulent
collision enhancement is strongest (approximately $St=1$). Turbulence
therefore has little influence on the high-CKE collisions responsible
for fragmentation, but substantially affects the subsequent
aggregation of the generated fragments.  

Freezing fragmentation contributes only about 10--100 SIP particles L$^{-1}$ in the present simulation. Both the fragment yield per FF event and the number of FF events are small compared with collision-induced fragmentation. The latter reflects the limited number of sufficiently large supercooled raindrops available for freezing fragmentation. The net contribution of FF 
depends on both the fragment yield per event and the occurrence of
freezing fragmentation events. In the present squall-line simulation,
the resulting contribution to the total ice population remains small,
as indicated by the weak response of the ice particle number density
and ice multiplication factor in the FF experiment
(Fig.~\ref{fig:profiles_frag}). Thus, although the FF parameterization
allows fragmentation over a broad temperature range, the availability
of sufficiently large supercooled raindrops limits its overall
importance in this deep convective regime. Accordingly, FF has only a small effect on IWC and ice particle number concentration in the present simulation.

Although the number of super-particles is a numerical rather than a
physical quantity, it provides additional insight into how
fragmentation is represented and the corresponding physical
mechanisms. RS exhibits a moderate increase between 3 and 6~km height,
whereas FF shows no significant difference. In the CF and CFT
simulation, the number of super-particles increases strongly in the
DGZ but also throughout the whole domain. This shows that many
fragmentation events are in the regime where the fragment mass is much
smaller than the total particle mass, e.g. graupel-graupel collisions
within the DGZ with dendritic ice on the surface of graupel particles,
which in our implementation leads to the creation of new
super-particles (Appendix C).

\def\expname{exp209_u10_q12_ccn7_xi7e9_rt1_kern3_agg5_rfrag0_cfrag6_ffrag0}
\begin{figure}[t]
  \centering
  \begin{minipage}{0.99\textwidth}
    \def\mywidth{0.49}
    \begin{minipage}{\mywidth\textwidth}
      a) 3500 m -- 5000 m \\[2mm]
      \includegraphics[width=\textwidth]{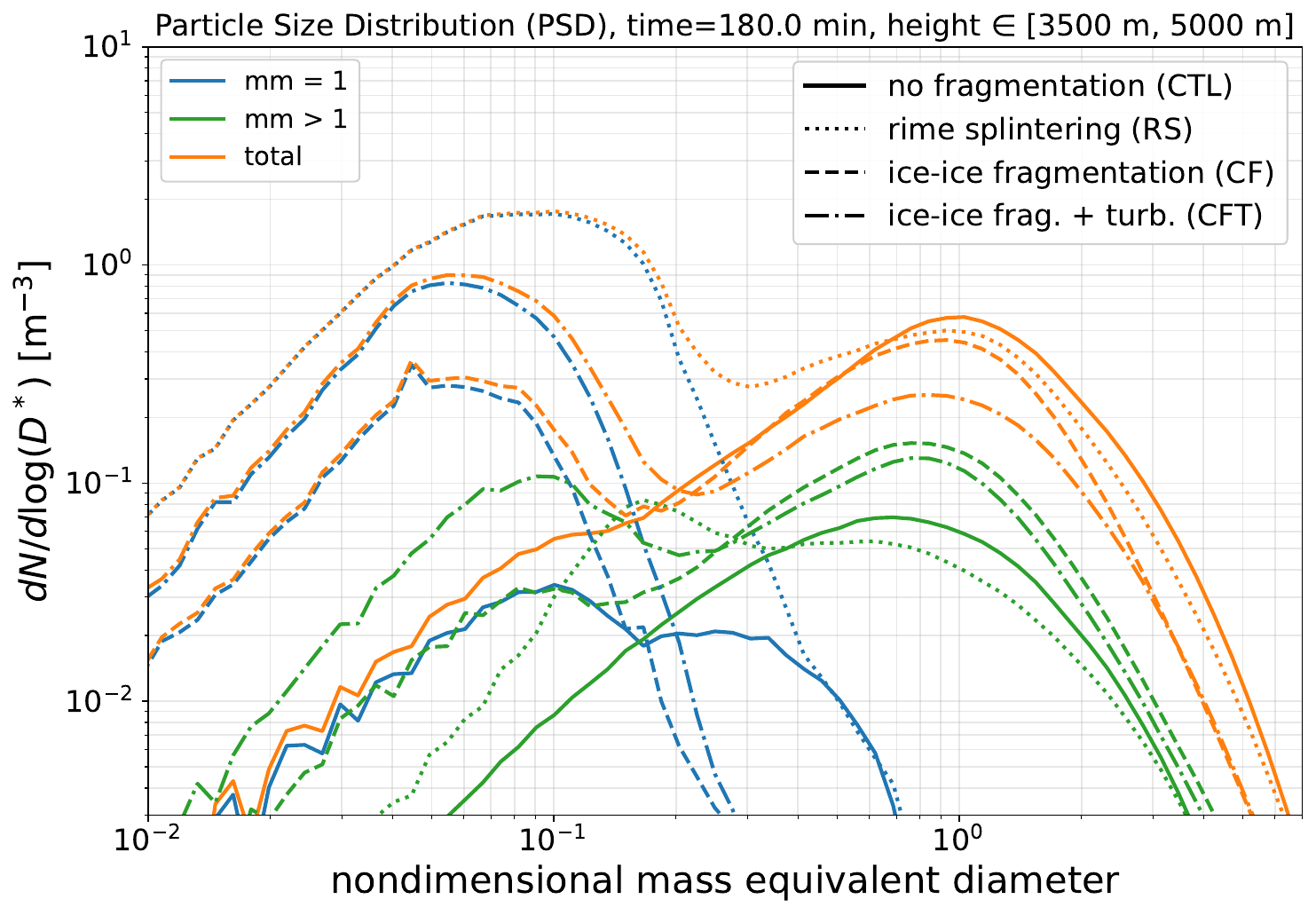}
    \end{minipage}
    \hfill
    \begin{minipage}{\mywidth\textwidth}
      b) 5000 m -- 6500 m \\[2mm]
      \includegraphics[width=\textwidth]{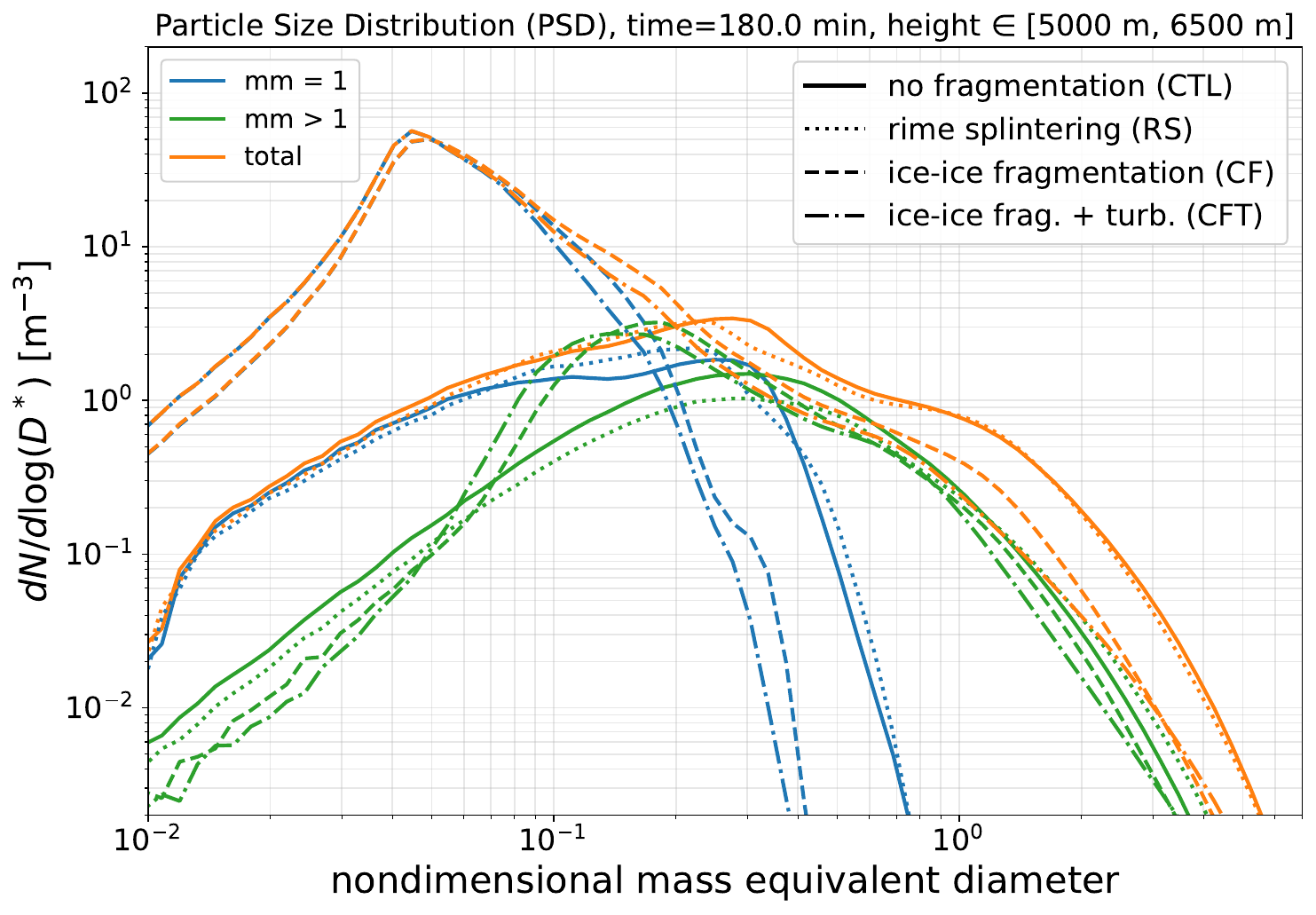}
    \end{minipage}
  \end{minipage}
  
  \caption{\label{fig:psd_frag} Particle size distributions for
    monomer crystals (mm=1, blue), unrimed and lightly rimed
    aggregates (mm $>$ 1, green), and the total (orange) using mass
    equivalent diameter.  Shown are the PSDs of CTL, RS, CF, and CFT
    for 180 min simulation. Here $D^*$ is the non-dimensional mass
    equivalent diameter normalized with $\ell=1$ m.}
\end{figure}

The particle size distributions (PSDs) shown in
Fig.~\ref{fig:psd_frag} confirm the interpretation derived from the
bulk profiles. Between 3500 m and 5000 m height, rime splintering (RS)
leads to a pronounced increase in small ice crystals compared to CTL
as this is the source region of rime splintering. CF and CFT show an
increase of small particles in this height, but have lower
concentrations of small particles compared to RS in this layer
consistent with the bulk profiles discussed above. In this layer, CFT
has a more pronounced small particle mode than CF, which is not
obvious from the bulk profiles shown in
Fig.~\ref{fig:profiles_frag}. The aggregate mode with a maximum around
1 mm equivalent diameter is similar for CF and CFT.

Between 5000 m and 6500 m, CF and CFT show a strong increase of small
ice particles consistent with the DGZ being the source region for
collision fragmentation. The RS simulation has no significant small
particle mode in this height.  In both, CF and CFT we find a shift of
the aggregate mode toward smaller sizes with a bimodality in
aggregates in 3500-5000 m height. Hence, the fragments are the source
of an additional aggregate mode with an equivalent diameter around 0.1
mm, which does not exist without collision-induced
fragmentation. Interestingly, larger aggregates are not less frequent
in the CF and CFT simulations, although they can be depleted by
collision fragmentation. In both simulations the tail of the aggregate
PSD is similar or even more pronounce than in CTL.

Interestingly, the abundance of large aggregates is not reduced in the
CF and CFT simulations, despite ongoing fragmentation. Instead, the
broadening of the PSD promotes aggregation by increasing the
availability of small and intermediate-sized collision partners. Thus,
fragmentation continuously replenishes the part of the size spectrum
that is most efficiently collected by larger snowflakes. Rather than
suppressing aggregation, collision-induced fragmentation redistributes
ice mass toward smaller particles while simultaneously sustaining the
growth of large aggregates.

The total PSD is shifted toward smaller particles, which
shows that heavily rimed snowflakes and graupel lose mass due to
collision fragmentation. In many cases this might be the 'graupel with
dendrites' regime in which graupel growing by vapor deposition in the
DGZ can lose these dendrites in graupel-graupel collisions. This
fragmentation mechanism limits the growth of graupel and provides more
mass, in form of fragments, for the growth of aggregates.  Hence,
collision fragmentation affects the whole size distribution with a
shift toward smaller particles and from graupel to
aggregates. Overall, the PSDs show the pronounced effects of both, RS
and CF/CFT, for the occurence of small ice particles, especially in
those height levels where the SIP is active.

\begin{figure}[t]
  \centering
  \def\mywidth{0.49}
  \begin{minipage}{\mywidth\textwidth}
    a) total mass density \\[3mm]
    \includegraphics[width=\textwidth, viewport= 0 0 630 400, clip=]{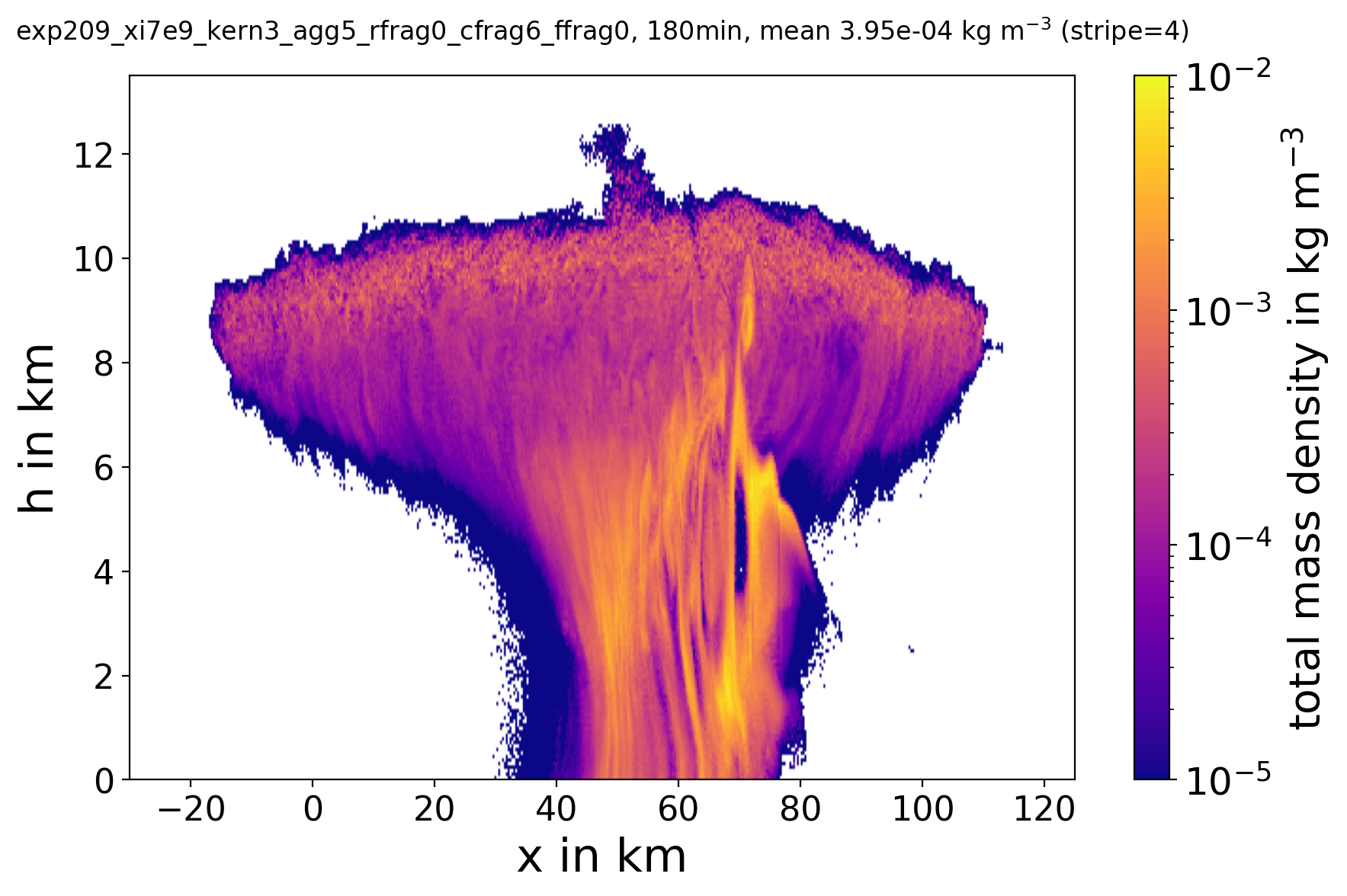}
  \end{minipage}
  \begin{minipage}{\mywidth\textwidth}
    b) total number density \\[3mm]
    \includegraphics[width=\textwidth, viewport= 0 0 630 400, clip=]{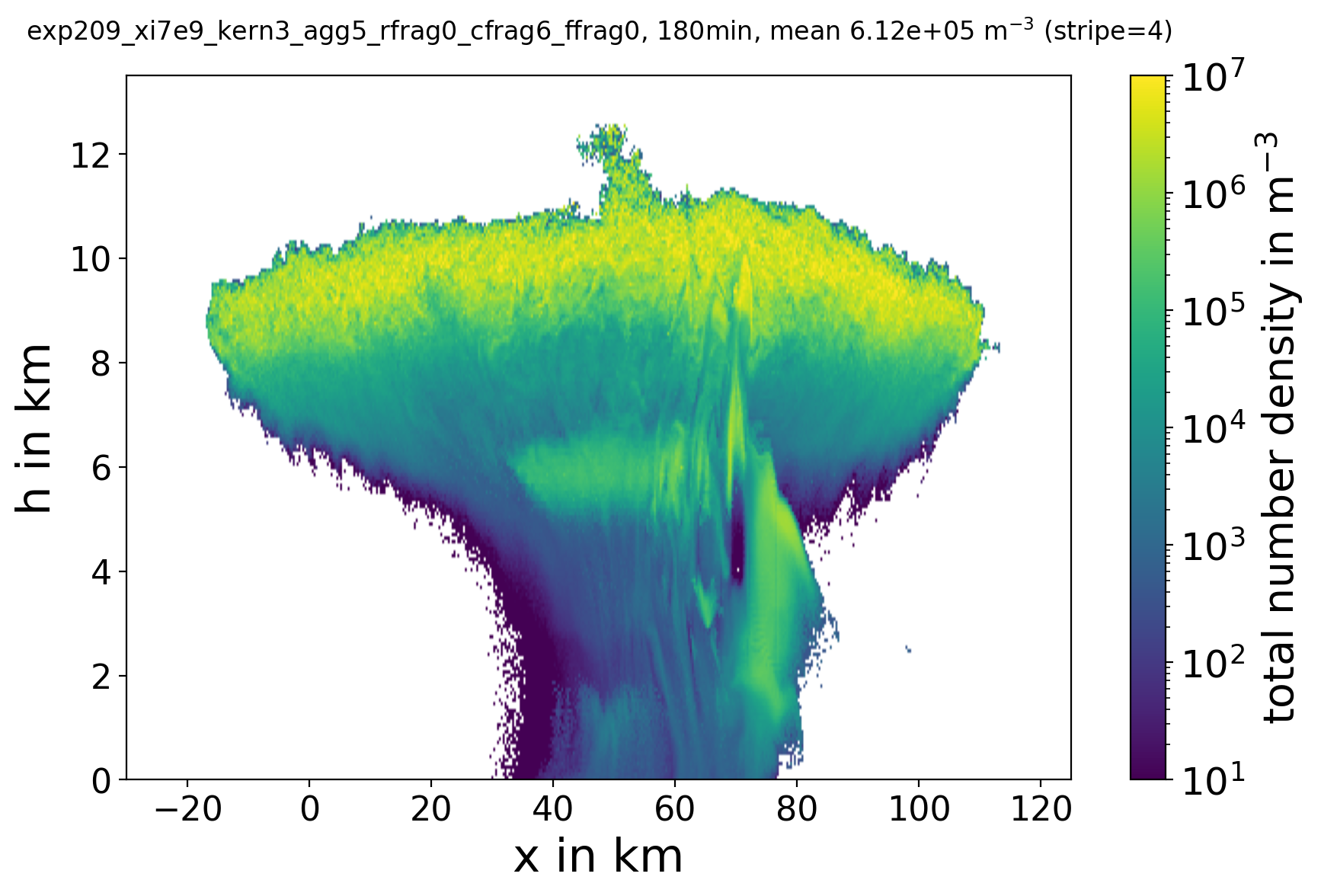}
  \end{minipage}

  \caption{\label{fig:xz-fragmentation} Line-averaged vertical cross
    sections of the CF simulation at 180 min of (a) total mass
    density, (b) total number density, The line average is for 1/20 of
    the model domain corresponding to a length of 1 km.}
\end{figure}

Finally, the vertical cross sections of Figure
\ref{fig:xz-fragmentation} show that the increased particle
concentrations are located behind the convective core and extend into
the trailing stratiform region around 6 km height. In contrast, the
total mass density shows no pronounced increase in that
region and looks, apart from random convective and turbulent
structures, very similar to CTL shown in Figure
\ref{fig:xz-control}. The significant increase in IWC, which we have
seen in Figure 8, is not clearly visible in Figure
\ref{fig:xz-fragmentation} because the color scale spans four orders of magnitude.

\def\expname{exp209_u10_q12_ccn7_xi7e9_rt1_kern3_agg5_rfrag0_cfrag6_ffrag0}

\section{Simulated radar signatures}

Remote sensing by radar provides valuable information about the
microphysics of deep convective systems and clouds in
general. Especially for deep convection, obtaining in-situ
observations can be challenging and measurements of vertical profiles
within convective clouds are rare.  Polarimetric radar measurements,
which are sensitive to particle number, size, and shape, are more readily
available and offer the potential to validate or falsify model
assumptions.

Cloud radars commonly operate at higher frequencies, such as Ka- or
W-band, than weather radars operating at X-, C-, or S-band. These
higher frequencies provide greater sensitivity to small ice particles
but can suffer from substantial attenuation and multiple scattering in
deep convection, limiting their ability to observe the full depth of a
squall line. X-band systems are often used for research applications
because they offer a good compromise between sensitivity and
attenuation for observing convective clouds
\citep{DOW_Wurman_1997}. Similar to conventional Ka- or W-band cloud
radars, X-band systems have also been used in vertical mode.

Recent advances in radar technology allow X-band systems to achieve
sensitivities comparable to those of conventional cloud radars, while
also providing access to Doppler spectra
\citep{kneifel_tripleSpec_2016,von2022ice}.  We use McRadar to
simulate polarimetric radar moments and Doppler spectra from the
ICON/McSnow simulations for a virtual X-band radar. This allows us to
quantify the sensitivity of radar observables to SIP mechanisms like
rime splintering and collision-induced ice-ice fragmentation.

McRadar \citep{vonTerzi-2026} is specifically designed for forward
modeling of super-particle simulations using a microwave scattering
database based on the discrete dipole approximation. Because it
operates directly on the McSnow super-particle representation without
an intermediate spectral grid \citep{Seifert-2026}, it provides a
physically consistent simulation of polarimetric radar variables and
Doppler spectra.

McRadar simulations were performed for six cross sections through the
ICON/McSnow domain. Along each cross section, virtual radar locations
were sampled between $x=40$ and $65$ km at 1 km intervals. Radar
observables were calculated at elevation angles of $30^\circ$ and
$90^\circ$. At each virtual location, ten beams with small spatial
offsets were simulated and subsequently averaged to reduce sampling
variability. We assume a radar beamwidth of $1^\circ$ and a noise
level of -50 dBZ.

\begin{figure}[t]
  \centering
  \begin{minipage}{\textwidth}
    \def\mywidth{0.325}
    \begin{minipage}{\mywidth\textwidth}
      a) Ze \\[1mm]
      \includegraphics[width=\textwidth]{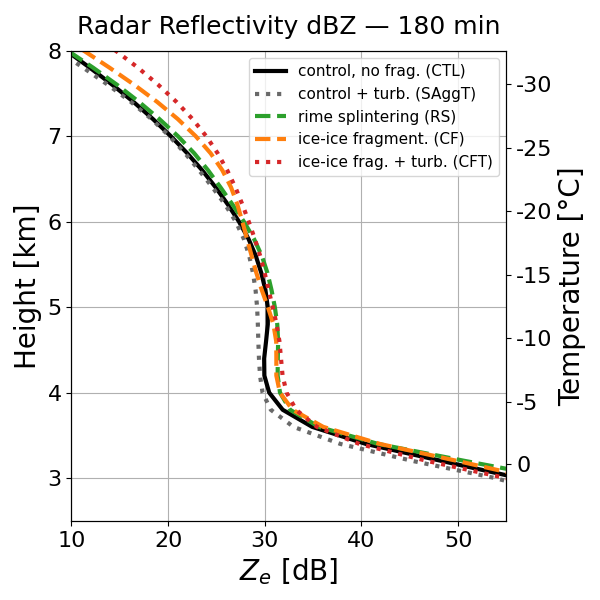}
    \end{minipage}
    \hfill
    \begin{minipage}{\mywidth\textwidth}
      b) ZDR \\[1mm]
      \includegraphics[width=\textwidth]{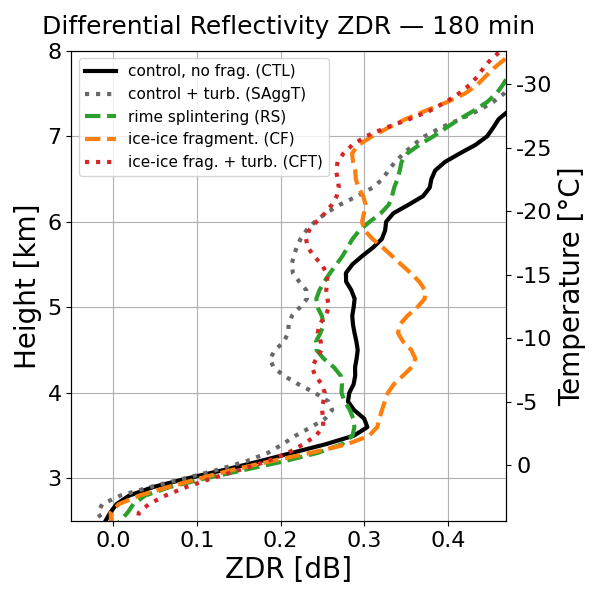}
    \end{minipage}
    \hfill
    \begin{minipage}{\mywidth\textwidth}
      c) KDP \\[1mm]
      \includegraphics[width=\textwidth]{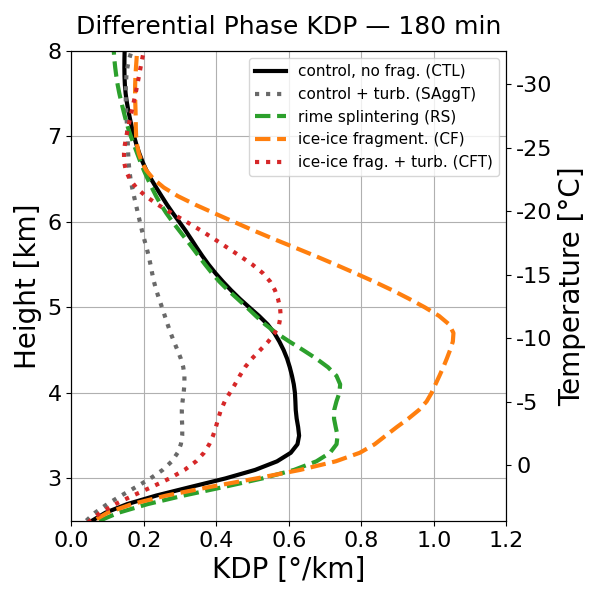}
    \end{minipage}
  \end{minipage}

  \caption{\label{fig:profiles_mcradar} Vertical profiles of radar
    reflectivity at 90° elevation (Ze), differential reflectivity
    (ZDR), and specific differential phase (KDP) at 30° elevation at
    180 min simulation time. Shown are the simulations CTL, SAggT, RS,
    CF, and CFT.}
\end{figure}

In the equivalent radar reflectivity $Z_e$, expressed in dBZ
(Fig.~\ref{fig:profiles_mcradar}a), we find an increase by 2 dB for RS,
CF, and CFT above the melting layer in 4-5 km height, no significant
difference between 5-6 km height, and an increase further aloft for CF
and CFT compared to CTL.  Overall, $Z_e$ is only weakly sensitive to
SIP because changes in the particle size distribution and ice water
content can compensate each other in the radar reflectivity.

For differential reflectivity (ZDR, Fig.~\ref{fig:profiles_mcradar}b),
the average differences between simulations are small (on the order of
0.05 dB). As the differences are similar to the ZDR noise in the
simulation, the results should be interpreted with care.
In general, an increased contribution from small, non-spherical
monomers tends to increase ZDR, whereas the formation of larger, more
isotropic aggregates tends to decrease it.
Because SIP simultaneously modifies the abundance of small monomers
and larger aggregates, these opposing effects can largely compensate,
resulting in only subtle changes in ZDR.

Among the conventional polarimetric radar moments, KDP exhibits the
clearest signature of SIP (Fig.~\ref{fig:profiles_mcradar}c). For RS
we find a moderate increase in KDP in 3-4.5 km height, where rime
splintering is active. CF exhibits a much stronger increase compared
to RS and over a deeper layer from 3-6.5 km height. The initial
increase in KDP at 6500 m is consistent with the strong increase in
number concentration Figure~\ref{fig:profiles_frag}a), and increase in
aggregate size (not shown). Despite the decrease in number
concentration between 5800 and 4500m, KDP shows a distinct peak at
4800m. Since KDP scales with the number concentration and the mass of
the individual particles, the KDP maximum reflects the competition
between these two factors. Above the peak, the increase in mean
aggregate mass more than compensates for the decreasing number
concentration; below the peak, the continued decrease in number
concentration dominates, leading to the observed reduction in
KDP. Turbulence effects on collisions enhance aggregation, reduce
particle number density, and, hence, reduce KDP, which is seen clearly
in SAggT compared to CTL and CFT vs CF.  In CFT an increase in KDP due
to fragmentation is seen when using SAggT as the reference instead of
CTL.  Hence, among the bulk radar moments, KDP is the most promising
candidate for identifying SIP, although its interpretation can still
be confounded by other microphysical processes.  To obtain a more
process-specific signature of SIP, we therefore examine polarimetric
Doppler spectra, which provide more detailed information than the bulk
radar moments.

To isolate the microphysical signatures from the highly variable
convective dynamics of the squall line, the synthetic Doppler spectra
are calculated solely from the hydrometeor terminal fall
velocities. Dynamical broadening and spectral shifts due to background
winds are intentionally neglected, allowing us to assess the maximum
potential radar fingerprint of the microphysical processes involved.

Doppler spectra for vertical observation - as commonly performed by
cloud radars - are shown in the upper panels of Figure
\ref{fig:spectra_mcradar}. The lower panels include Doppler spectra at
30° elevation (slant view necessary for polarimetric signatures) on an
individual cross section through the Squall Line. Please note that the
slant spectra are narrower because the vertical component only
contributes 50\% to the radial Doppler velocity at this angle. The
reflectivity Doppler spectra (Figure \ref{fig:spectra_mcradar}a-c)
show the broad range of particles with fall velocities from 0.5 m/s to
above 3 m/s.  For velocities larger than 1 m/s the differences between
CTL, RS and CF are small. For small velocities we find a signal
between 4 and 5 km height in RS and around 6 km in CF. These are the
source regions of the SIP processes where small particles are
generated -- and this is visible in the Doppler spectra. In actual
measurements, this weak spectral signal might be difficult to separate
from the noise floor. In contrast, the spectral differential
reflectivity (sZDR) exhibits a pronounced enhancement at the slow edge
of the spectrum in the source regions of both rime splintering and
collision-induced fragmentation.
Because spectral differential reflectivity resolves the polarimetric
signature as a function of terminal fall velocity, it can link newly
generated ice particles to both their Doppler velocity and their
altitude. Hence, sZDR preserves microphysical signatures that are
largely lost in bulk radar moments.

Polarimetric Doppler spectra therefore provide a promising means of
identifying SIP mechanisms because sZDR has the potential to locate
the source regions of newly generated ice particles.
This may help distinguish rime splintering, which primarily occurs
between $-3\,^\circ$C and $-8\,^\circ$C, from collision-induced
fragmentation, whose source region is centered on the dendritic growth
zone near $-15\,^\circ$C.

Overall, we find that among the conventional radar moments, KDP is the
most sensitive indicator of SIP, whereas ZDR responds primarily
through changes in aggregate abundance and reflectivity is least
diagnostic because of compensating effects of particle size and ice
water content. None of these bulk radar moments uniquely identifies
the underlying SIP mechanism.  In contrast, spectral polarimetric
observables such as sZDR provide information on both the altitude —
which can often be related to the local temperature regime — and the
terminal fall velocities of newly generated ice particles, making them
a promising observational diagnostic of secondary ice production.
Analysis of real radar observations is needed in order to investigate
whether these spectral features can be identified in real observations
or are obscured by dynamical broadening and other sources of spectral
variability.

\def\expname{exp209_u10_q12_ccn7_xi7e9_rt1_kern3_agg5_rfrag0_cfrag6_ffrag0}

\begin{figure}[t]
  \centering
  \begin{minipage}{\textwidth}
    \def\mywidth{0.325}
    \begin{minipage}{\mywidth\textwidth}
      a) CTL - sZe \\[1mm]
      \def\expname{exp198}
      \includegraphics[width=\textwidth, viewport= 750 460 1220 900, clip=]{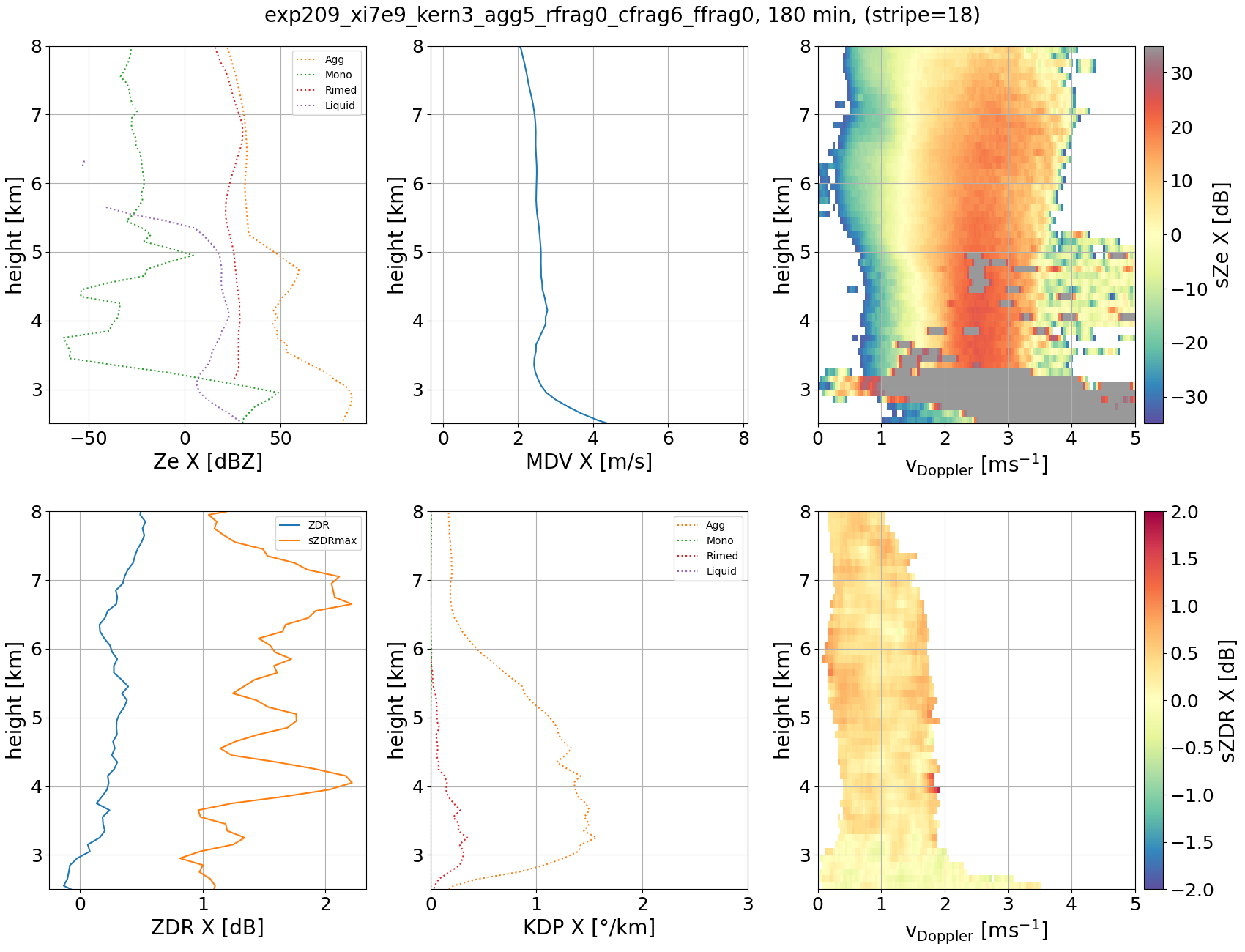}
    \end{minipage}
    \hfill
    \begin{minipage}{\mywidth\textwidth}
      b) RS - sZe \\[1mm]
      \def\expname{exp209_u10_q12_ccn7_xi7e9_rt1_kern3_agg5_rfrag1_cfrag0_ffrag0}
      \includegraphics[width=\textwidth, viewport= 750 460 1220 900, clip=]{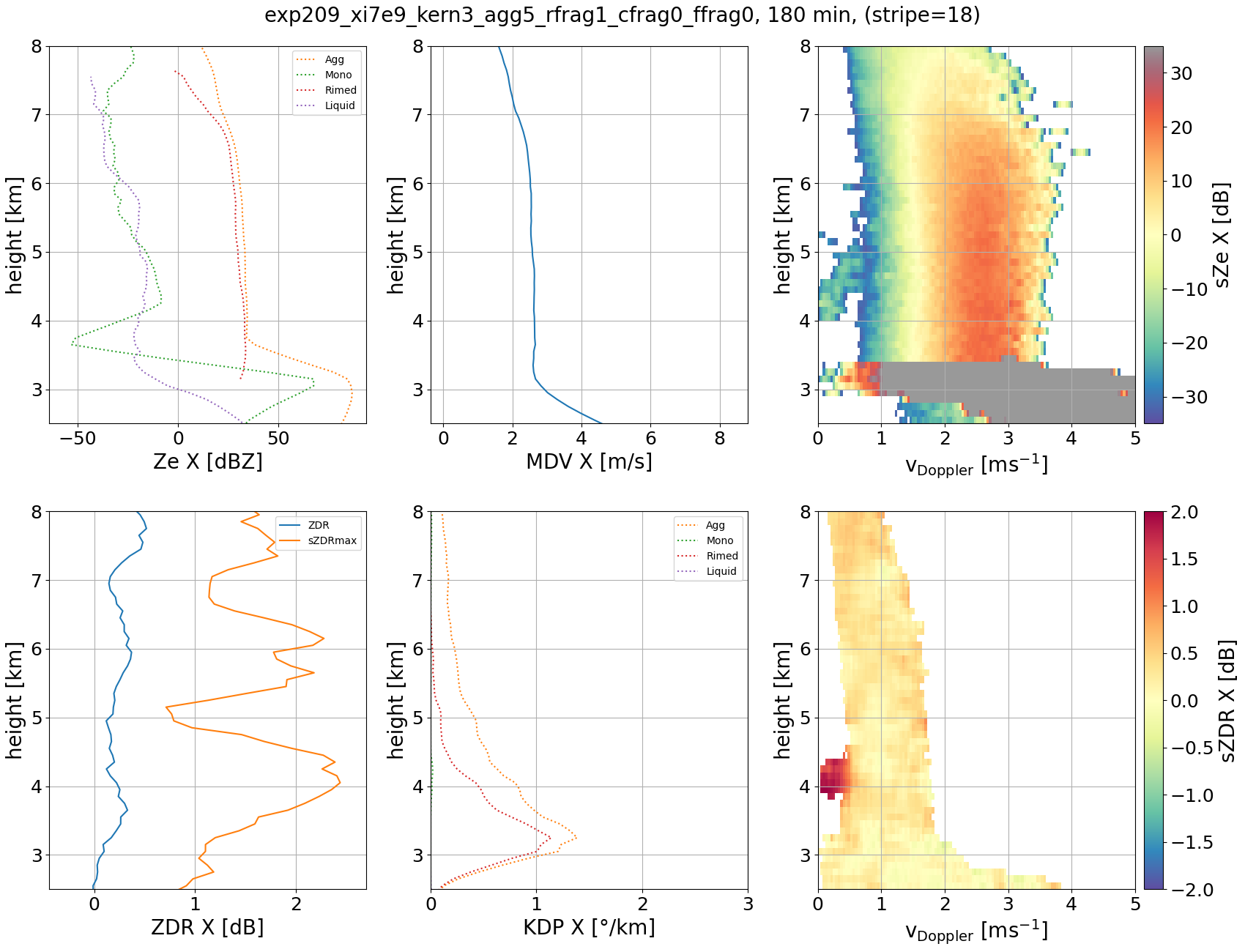}
    \end{minipage}
    \hfill
    \begin{minipage}{\mywidth\textwidth}
      c) CF - sZe \\[1mm]
      \def\expname{exp209_u10_q12_ccn7_xi7e9_rt1_kern3_agg5_rfrag0_cfrag6_ffrag0}
      \includegraphics[width=\textwidth, viewport= 750 460 1220 900, clip=]{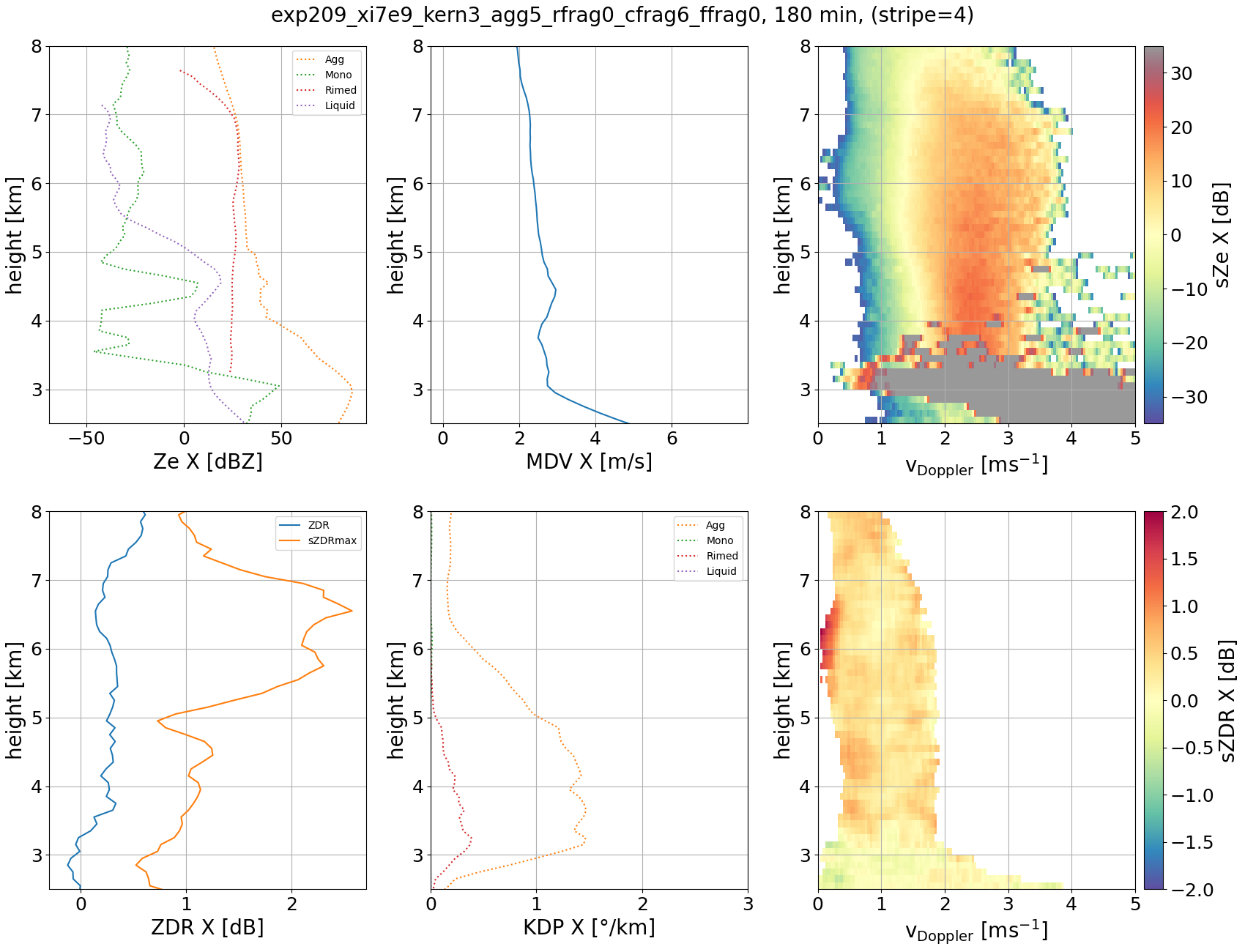}
    \end{minipage}
    \vspace{3mm}

    \begin{minipage}{\mywidth\textwidth}
      d) CTL - sZDR \\[1mm]
      \def\expname{exp198}
      \includegraphics[width=\textwidth, viewport= 750 0 1220 450, clip=]{Radar_Spectra_10800s_Xband_CTL_18.png}
    \end{minipage}
    \hfill
    \begin{minipage}{\mywidth\textwidth}
      e) RS - sZDR \\[1mm]
      \def\expname{exp209_u10_q12_ccn7_xi7e9_rt1_kern3_agg5_rfrag1_cfrag0_ffrag0}
      \includegraphics[width=\textwidth, viewport= 750 0 1220 450, clip=]{Radar_Spectra_10800s_Xband_RS_18.png}
    \end{minipage}
    \hfill
    \begin{minipage}{\mywidth\textwidth}
      f) CF - sZDR \\[1mm]
      \def\expname{exp209_u10_q12_ccn7_xi7e9_rt1_kern3_agg5_rfrag0_cfrag6_ffrag0}
      \includegraphics[width=\textwidth, viewport= 750 0 1220 450, clip=]{Radar_Spectra_10800s_Xband_CF_04.png}
    \end{minipage}
  \end{minipage}

  \caption{\label{fig:spectra_mcradar} Simulated Doppler spectra of
    equivalent spectral reflectivity sZe (top row, a--c, 90° elevation), and spectral
    differential reflectivity sZDR (bottom row, d--f, 30° elevation), for the simulations CTL, RS, and CF.}
\end{figure}

\section{Summary and Conclusions}

This study investigated how aggregation, fragmentation, stochastic
snowflake geometry, and turbulence jointly control the evolution of
ice particles in an idealized squall line using the Lagrangian
particle model ICON/McSnow. The explicit representation of particle
properties and growth pathways allows for a detailed process-level
analysis that avoids many of the limitations of bulk microphysics
schemes.

The control simulation shows a separation between convective
and stratiform region, with rimed particles and frozen drops
dominating in the convective core, and unrimed or lightly rimed
aggregates prevailing in the stratiform region. Particle growth is
governed by a combination of deposition, riming, and aggregation, with
substantial variability in particle properties emerging naturally from
the Lagrangian particle framework.

A key finding is that stochastic variability in snowflake geometry
reduces aggregation despite increasing variability in fall
velocity. In the simulated squall line, aggregation is controlled
primarily by particle cross section rather than differential
sedimentation because terminal fall velocities vary only weakly at the
high Reynolds numbers of snowflakes. Turbulence effects on the
collision rate enhance aggregation and increase precipitation rates
by about 8–10 \%.

Among the SIP mechanisms, collision-induced fragmentation (CF) emerges
as the most effective source of secondary ice in the simulated squall
line. It produces large numbers of fragments in the dendritic growth
zone and significantly increases ice particle number
concentrations. Rime splintering and collision fragmentation increase
KDP and both show a signature of SIP in spectral differential
reflectivity sZDR in the source region of the SIP mechanism. Hence,
polarimetric Doppler spectra are potentially instrumental for the
identification of SIP.

The main result of this study is that convective updrafts promote the
coexistence of low-density aggregates and fast-falling rimed
particles, creating favorable conditions for efficient
collision-induced fragmentation (CF).  Rime splintering (RS) contributes
mainly at lower altitudes and enhances the population of small
crystals, thereby indirectly promoting aggregation but with little
effect on precipitation. Freezing fragmentation (FF), in contrast, is
strongly limited by the availability of sufficiently large raindrops
and plays only a minor role in this regime. In environments with
different drop size distributions, weaker vertical transport, or
re-freezing after melting, freezing fragmentation may play a larger
role.

Although the parameterization of collision-induced fragmentation is
based on laboratory measurements of snowflakes, graupel with
dendrites, and graupel, the results may depend on the assumed
relationship between particle fragility and apparent density in the
Lagrangian framework. While this formulation is physically motivated,
it remains uncertain and may influence the quantitative strength of
the simulated SIP. At the same time, the Lagrangian particle approach
provides detailed information on particle properties and collision
characteristics, such as collision kinetic energy, and the variability
of these particle properties allow for a physically consistent
representation of fragmentation processes.

Another key result of this study is the interaction between
fragmentation and turbulence. While collision-induced fragmentation
alone shifts the particle size distribution toward smaller sizes and
reduces precipitation efficiency, the inclusion of turbulence enhances
collision rates, thereby increasing aggregation.  The increase in
aggregation due to turbulence compensates for the
fragmentation-induced depletion of large particles and restores the
formation of precipitation-sized hydrometeors. As a result, the
combination of collision-induced fragmentation and turbulence yields
precipitation rates comparable to the control simulation without
fragmentation, though still lower than in simulations with
turbulence-enhanced aggregation alone. Although turbulence increases
collision rates, it does not significantly enhance collision-induced
fragmentation, because collision-induced fragmentation requires
sufficient collision kinetic energy, which is only present in the high
Stokes number regime. Hence, turbulence effects are less relevant for
collision-induced fragmentation itself, but act primarily to enhance
the aggregation of fragments produced by collision-induced
fragmentation.

The applied turbulence parameterization is based on formulations for
spherical particles, and its extension to non-spherical ice particles
introduces additional uncertainty in the magnitude of the turbulence
effects. This gap should be addressed in future work by connecting
existing results on orientation statistics \citep{Gustavsson-2017},
analyses of swept volumes \citep{Gavze-Khain-2022}, direct numerical
simulations \citep{Sheikh-2024}, and theoretical models
\citep{Zaichik-2009,Luo-2026}.

Forward simulations with McRadar further show that conventional
polarimetric radar moments provide only limited discrimination between
SIP mechanisms. Among these, KDP exhibits the clearest response,
whereas spectral polarimetric observables such as sZDR retain
information on the altitude and fall-speed distribution of newly
generated ice particles and therefore provide a promising
observational diagnostic of SIP.

The results of this study are specific to the simulated squall line
regime, characterized by strong updrafts and significant graupel
production. The relative importance of SIP mechanisms may differ in
other cloud types, such as stratiform or shallow convective clouds.

Overall, our results demonstrate that fragmentation, aggregation,
stochastic snowflake geometry, and turbulence cannot be considered
independently in deep convective clouds. Their coupled interaction
controls the evolution of the ice particle size distribution,
secondary ice production, and ultimately precipitation
formation. Capturing these interactions therefore appears essential
for physically realistic simulations of deep convective cloud
microphysics.

\appendix

\section{Implementation of rime splintering}

The implementation of Hallett–Mossop rime splintering in McSnow
follows well-established formulations
\citep[hereafter PK98]{Pruppacher-Klett-1998}. For cloud droplets
larger than 24~$\mu$m, it is assumed that 350 splinters are produced
per $10^{-6}$~kg of rime accreted, corresponding to
$C_\text{RS} = 3.5 \times 10^{8}$~kg$^{-1}$.
Rime splintering occurs within a narrow temperature range between
$-3^\circ$C and $-8^\circ$C, with a maximum at $-5^\circ$C.

In McSnow, this process is represented using a stochastic formulation with
\begin{equation}
  p_\text{RS} = \frac{N_\text{RS}}{\xi_0},
\end{equation}
where $\xi_0$ is the base multiplicity of super-particles, as used for
other processes like ice nucleation. The number of splinters per grid
box and time step is
\begin{equation}
  N_\text{RS} = C_\text{RS} \, f_\text{RS}(T)
  \sum_{i=1}^{N_\text{SP}} \Delta m_{\text{rime},i}.
\end{equation}
Here, $\Delta m_{\text{rime},i}$ denotes the rime mass accreted by the $i$-th
super-particle during one time step, and $N_\text{SP}$ is the number
of super-particles in the grid box.
The temperature-dependent factor is given by
\begin{equation}
  f_\text{RS}(T) =
  \begin{cases}
    0,                          & T > 270.16 \; \unit{K}, \\
    (270.16 - T)/2,             & 270.16 \ge T > 268.16 \; \unit{K}, \\
    (T - 265.16)/3,             & 268.16 \ge T \ge 265.16 \; \unit{K}, \\
    0,                          & T < 265.16 \; \unit{K}.
  \end{cases}
\end{equation}
This corresponds to Eq.~(16-71) in PK98.

To limit the growth in super-particle number, at most one splinter
super-particle is created per grid box and time step. The quantity
$p_\text{RS}$ is used as a control parameter for rime splintering
events. For $p_\text{RS} \ge 1$, a deterministic event is triggered
and a single super-particle is generated that represents the total
number of produced splinters, with multiplicity $\xi_\text{RS} =
N_\text{RS}$.  For $p_\text{RS} < 1$, Bernoulli sampling is applied:
a random number $r \in [0,1]$ is drawn, and a new super-particle with
multiplicity $\xi_\text{RS} = \xi_0$ is created only if $r <
p_\text{RS}$. For $r \ge p_\text{RS}$, no rime splintering occurs
during the time step. This can be written compactly as an equation for
$\xi_\text{RS}$:
\begin{equation}
  \xi_\text{RS} =
  \begin{cases}
    N_\text{RS}, & p_\text{RS} \ge 1, \\
    \xi_0,       & p_\text{RS} < 1 \ \text{and} \ r < p_\text{RS}, \\
    0,           & p_\text{RS} < 1 \ \text{and} \ r \ge p_\text{RS},
  \end{cases}
\end{equation}
This approach preserves the expected number of splinters while
limiting the number of super-particles.

Hallett–Mossop splinters are typically assumed to be a few micrometers
in diameter, and thus primarily enhance ice number concentration
rather than directly contributing to ice mass. In McSnow, we assume
that the splinters have a diameter of 5~$\mu$m and they are treated as
solid spherical ice particles. Splinters are assigned only ice mass $m_i$
and no rime mass $m_r$. Hence, in McSnow, the splinters are considered
small unrimed ice crystals.

\section{Implementation of freezing fragmentation}

The implementation of freezing fragmentation follows \citet[hereafter
Pf25]{Pfeifer-2025} and \citet[hereafter Ph18]{Phillips-2018}. The size
dependence is based on the empirical in-situ data of Pf25 (their Fig.~5). We
assume a piecewise linear function with a maximum fragmentation probability of
$p_{\max} = 0.30$ at an optimum diameter $\Dopt = 0.5\:$mm. No freezing
fragmentation occurs for droplets smaller than
$D_{\min} = 0.05 \: \mathrm{mm} = 50\:\mu\mathrm{m}$. For droplets larger than
$\Dopt$, the probability decreases linearly to a minimum value of
$p_{\min} = 0.05$ at $\Dmax = 2\:$mm and remains constant thereafter.
\begin{equation}
  p_1(D) =
  \begin{cases}
    0, & D \le D_{\min}, \\[0pt]
    p_{\max} \, \frac{D - \Dmin}{\Dopt - \Dmin}, & \Dmin < D \le \Dopt, \\[0pt]
    \max \! \left( p_{\max} - (p_{\max} - p_{\min})\,\frac{D - \Dopt}{\Dmax - \Dopt}, \, p_{\min} \right), & D > \Dopt.
  \end{cases}
\end{equation}
Based on Ph18 and \citet{Keinert-2020}, we assume an optimum
temperature $\Topt = -15 \, \Celsius$ and no freezing fragmentation
below $\Tmin = -30\,\Celsius$. The temperature dependence is represented by a Gaussian probability:
\begin{equation}
  p_2(T_c) =
  \begin{cases}
    0, & T_c \le \Tmin, \\[0pt]
    \exp \left ( - \left ( \frac{T_c - {\Topt}}{\sigma_\text{FF}} \right )^2 \right ), & {\Tmin} < T_c \le 0 \Celsius, \\[0pt]
    0, & T_c > 0 \Celsius.
  \end{cases}
\end{equation}
with $\sigma_\text{FF}=10\:\text{K}$. Temperature and size effects
are assumed to be statistically independent; hence, the total freezing
fragmentation probability is given by
\begin{equation}
  p_\text{FF}(D, T_c) = p_1(D)\, p_2(T_c)
\end{equation}
and is bounded by $0 \le p_\text{FF} \le p_{\max}$. 
The number of fragments per fragmentation event $n_\text{FF}$ follows Ph18 (their Mode 1):
\begin{equation}
  n_\text{FF} = \alpha_\text{FF} \, \min \left ( \frac{D}{\Dmin}, 40 \right ) \,
                \left ( {1 + \left ( \frac{T_c-\Topt}{\sigma_\text{FF}} \right )^2 } \right )^{-1}
\end{equation}
with $\alpha_\text{FF}=0.5$. The fragment mass is sampled from a gamma
distribution with shape parameter $k=1$ and mean mass corresponding to
a diameter of $5\:\mu$m. For computational efficiency, we assume that
all fragments produced in a single fragmentation event of a
super-particle have the same size.  In addition, we assume that the
frozen drop breaks into two half spheres. In McSnow the half spheres
are approximated by spheroids with an aspect ratio of 0.5. For each
freezing fragmentation event a new super-particle is created in McSnow
that represents the fragments, whereas the existing super-particle
represents the broken or unbroken residual.

\section{Implementation of collision-induced ice-ice fragmentation}

The implementation of collision-induced ice-ice fragmentation in McSnow is
based on the empirical parameterizations of \citet[G23 hereafter]{Grzegorczyk-2023} and
\citet[Y25 hereafter]{Yadav-2025}. The formulation follows the methodology of \citet{Phillips-2017} and
\citet{Gautam-2024}. The primary variable that determines the fragmentation
event is collision kinetic energy
\begin{equation}
  \label{eq:cke}
  \cke = \frac{1}{2} \frac{m_1 m_2}{m_1 + m_2} |v_1 - v_2|^2
\end{equation}
and the number of fragments follows
\begin{equation} \label{eq:nfrag-cke}
  N_{f} = \alpha_f  A_f  \left [ 1 - \exp \left (-  \left( 
      \frac{C_f \cke}{\alpha_f A_f} \right )^{\gamma} \, \right ) \right ]
\end{equation}
where $\alpha_f$ is the surface area of the more fragile particle and $A_f$ is
a measure of the areal number density of breakable asperities. $C_f$ is the
asperity fragility coefficient, and $\gamma=0.78$ is a dimensionless exponent.

First, the more fragile particle of the two collision partners is determined
based on apparent density. The particle with lower apparent density is assumed
to fragment. In the following all McSnow variables refer to this more fragile
particle and we assume that this is particle 1 with mass $m_1$, surface area
$\alpha_f =\pi D_1^2$, and apparent density
\begin{equation}
  \label{eq:rhoapp}
  \rho_\text{app} =
  \begin{cases}
    \frac{6 m_1}{\pi D_1^3 \phi},   & \phi \le 1, \\
    \frac{6 m_1 \phi^2}{\pi D_1^3}, & \phi > 1. \\
  \end{cases}
\end{equation}
with the aspect ratio $\phi$. 

G23 provide relations for snow-graupel and graupel-graupel collisions, the
latter including both dendritic and non-dendritic (bare) graupel. Y25 provide
parameterizations for bare graupel and for ice spheres of different apparent
density. Hence, we have three regimes that we want to connect in a continuous
way: snow-graupel with coefficients $C_{f,sg}$ and $A_{f,sg}$, graupel with
dendrites with $C_{f,gg}$ and $A_{f,gg}$, and bare graupel including solid
spheres or frozen drops with $C_{f,bg}$ and $A_{f,bg}$. The latter are
interpolated from Table 4 of Y25. The method described
below ensures a continuous transition between all three regimes by making use of the
particle properties predicted by McSnow.

Primary crystals like solid plates or columns will most likely not lead to
fragmentation. We use the ellipsoidal area fraction $q_A$ to exclude those particles with
\begin{equation}
  C^*_{f,sg} = C_{f,sg} \min(1, \max(0, (0.7 - q_A)/0.3))  
\end{equation}
Hence, for particles with $q_A \ge 0.7$ the asperity fragility coefficient
becomes zero, and only for $q_A \le 0.4$, which is clearly dendritic growth or
aggregates, we retain the original values of G23.

To make the transition from snow to graupel with dendrites we use the rime and frozen fraction
\begin{equation}
  \label{eq:frimefroz}
  f_\text{rf} = \frac{m_r+m_f}{m_1}
\end{equation}
and interpolate the areal number density and the asperity fragility
coefficient linearly:
\begin{align}
  C_{f,dg} &= C^*_{f,sg} (1-f_\text{rf}) + C_{f,gg} f_\text{rf} \\
  A_{f,dg} &= A_{f,sg} (1-f_\text{rf}) + A_{f,gg} f_\text{rf}
\end{align}
With this set of coefficients we get $N_{f,dg}$ for the dendritic growth
regime from Eq.~\eqref{eq:nfrag-cke}.

The number of fragments for the bare graupel regime $N_{f,bg}$ is based on Table 4 of Y25 using rime-frozen density
\begin{equation}
  \label{eq:rimefrozendensity}
  \rho_\text{rf} = \frac{m_r+m_f}{V_r + V_f}  
\end{equation}
where $V_r$ is the rime volume, which is a prognostic variable in McSnow, and $V_f = {m_f}/{\rho_i}$ is the
volume of the frozen mass of the particle. Here $\rho_i$ is the density of
material ice.

As a final step, we use the ratio of apparent density over rime-frozen density
\begin{equation}
  \label{eq:densityratio}
  f_\rho = \frac{\rho_\text{app}}{\rho_\text{rf}} 
\end{equation}
to decide whether the particle is in the dendritic regime
or in the bare graupel regime. For $f_\rho=1$ the particle is bare graupel or
a frozen drop, whereas for $f_\rho < 1$ it has undergone dendritic growth,
for $f_\rho \ll 1$ it is an unrimed snowflake. Hence, for $f_\rho \le
f^*_\rho = 0.5$
we apply $N_{f,dg}$, for $f_\rho=1$ the $N_{f,bg}$ is used. In between a linear
interpolation based on $f_\rho$ is used with:
\begin{equation}
  \label{eq:interpolation_rhoapp}
  N_f = N_{f,dg} (1-x_\rho) + N_{f,bg} x_\rho
\end{equation}
and
\begin{equation}
  \label{eq:interpolation_xrho}
  x_\rho = \min \left ( \max \left [ \frac{f_\rho-f^*_\rho}{1-f^*_\rho},
      0 \right ], 1 \right )
\end{equation}
Before this interpolation $N_{f,dg}$ and  $N_{f,bg}$ are limited to 1000 and
100 particles, respectively.

With this treatment of collision-induced fragmentation that makes use of the
habit prediction of McSnow, we may already capture different growth regimes:
dendritic snowflakes and aggregate snowflake, graupel with dendrites, bare
graupel, and frozen drops, and we have excluded primary crystals like plates
or columns.  Even without an explicit temperature constraint, this scheme
produces a physically plausible temperature dependence, with a fragmentation
maximum around $-20 \Celsius$ for the idealized squall line.

Nevertheless, we apply an additional temperature constraint following
\citet{Phillips-2017} and \citet{Gautam-2024} to suppress ice-ice
fragmentation more strongly outside of the dendritic growth regime.  Hence,
the final number of fragments is
\begin{equation}
  \label{eq:final-Nf}
  N_f^* = N_{f} \exp \left [ - \left ( \frac{T_c - \Topt}{\sigma_\text{cf}}
    \right )^2 \right ]
\end{equation}
with $\Topt = -15 \Celsius$ and $\sigma_\text{cf} = 2.5 \;\unit{K}$.

The fragment mass $m_\text{frag}$ is sampled from a gamma distribution
with shape parameter $k=1$ and mean mass corresponding to a diameter of
$75\,\mu$m. For computational efficiency, we assume that all fragments produced
in a single fragmentation event of a super-particle have the same size.

Collision-induced fragmentation occurs frequently in the squall line
simulation.  Creating a new super-particle for each individual
fragmentation event would lead to a rapid increase in super-particle
number. Here we apply a two-fold strategy: For events in which the
total mass of fragments is much smaller than the mass of the
fragmented ice particle, i.e.~$m_\text{frag} < 10^{-3} m_1$, we
collect all fragments from all fragmentation events at each grid point
during this time step and generate a single SP with the mean mass of
all fragments. This fragment SP is pure ice mass $m_i$. For cases in
which the fragment mass is not much smaller than the fragmented
particle, $m_\text{frag} \ge 10^{-3} m_1$, a mass-weighted
super-sampling strategy, also referred to as stochastic mode
selection, is applied as suggested by \citet{Seifert-Siewert-2023} and
adopted by \citet{deJong-2023} and \citet{Bayley-2025b} for
collisional breakup of raindrops. Hence, for each fragmentation event
the super-particle represents either the fragments or the residual
particle based on the probability
\begin{equation}
  \label{eq:supersampling}
  p_\text{res} = \frac{m_\text{res}}{m_1},  \quad \text{with} \quad m_\text{res} = m_1 - m_\text{frag} \, N_f^*
\end{equation}
For each fragmentation event, Bernoulli sampling is applied by drawing
a random number $r_p \in [0,1]$. For $r_p \le p_\text{res}$, the
residual mode is selected, and the multiplicity of the super-particle
is adjusted to conserve mass. For $r_p > p_\text{res}$, the
super-particle represents $m_1/m_\text{frag}$ fragments in the
fragment mode.  This stochastic mode selection conserves mass exactly
in each individual realization and yields the correct expected
contributions of the residual and fragment modes. Note that $N_f^*$ is
only used in the definition of $p_\text{res}$ and is not explicitly
realized, consistent with the mass-weighted sampling strategy. The
super-sampling conserves the microphysical structure/composition of
the ice particle, i.e., their ice, rime, and frozen mass etc., for
fragments and the residual particle.

This approach limits the number of super-particles created by
fragmentation, but requires a sufficiently large number of
super-particles to adequately sample the stochastic fragmentation
process. In contrast to raindrop breakup, where fragment populations
can be merged, this is not straightforward for ice fragmentation due
to the high dimensionality of fragment properties (e.g., size, shape,
density, and riming state), which prevents efficient averaging in
phase space.

\section{Implementation of turbulence effects on collisions}

The implementation of turbulence effects in McSnow follows
\citet[hereafter OS16]{Onishi-Seifert-2016} and the corresponding
Fortran code provided as an online supplement. The Onishi kernel
consists of three contributions: the radial distribution function
$g_{12}$, the mean radial relative velocity $\langle |w_r| \rangle$,
and the turbulent enhancement factor $\eta_e$.

For $g_{12}$, \citet{Onishi-2015} proposed a model for clustering in
monodisperse systems of non-sedimenting particles, primarily as a
function of the Stokes number
\begin{equation}
  \mathrm{St} = \frac{\tau_p}{\tau_k} = \frac{v_t}{g \, \tau_k} 
  \qquad \text{with} \qquad
  \tau_p = \frac{v_t}{g}, \qquad \tau_k = \left ( \frac{\nu}{\varepsilon} \right )^\frac{1}{2}.
\end{equation}
Here $\tau_p$ is the particle relaxation time with the terminal fall
velocity $v_t$ and gravitational acceleration $g$. The Kolmogorov time
scale $\tau_k$ is a function of the kinematic viscosity $\nu$ and
the turbulent dissipation rate $\varepsilon$ of the fluid.
In contrast to the original formulation, we estimate $\tau_p$ from the
terminal fall velocity as $\tau_p = v_t/g$, which implicitly accounts
for nonlinear drag effects. This should be interpreted as an effective
particle response time, extending the applicability of the
parameterization beyond the strict Stokes regime.

The radial distribution function for a monodisperse system with Stokes
number $\mathrm{St}_n$ is provided by OS16 as:
\begin{equation}
g_{nn} - 1 =
\begin{cases}
A_1 \, \mathrm{St}_n^2, & \text{for } \mathrm{St}_n < \mathrm{St}_a, \\
A_2 \, \mathrm{Re}_\lambda \, \mathrm{St}_n^{-2}, & \text{for } \mathrm{St}_n \geq \mathrm{St}_a,
\end{cases}
\end{equation}
with $A_1 = 110$ and $A_2 = 0.38$, and
\begin{equation}
\mathrm{St}_a = \left( \frac{A_2}{A_1} \right)^{1/4} \mathrm{Re}_\lambda^{1/4}.
\end{equation}
A $\tanh$-based smoothing is applied at $\mathrm{St}_a$ (see OS16 for
details). Based on this parameterization the bidisperse radial
distribution function $g_{12}$ is obtained from the model of
\citet{Zhou-2001}
\begin{equation}
g_{12} = 1 + \rho_{12} \, (g_{11} - 1)^{1/2} \, (g_{22} - 1)^{1/2},
\end{equation}
with
\begin{equation}
  \rho_{12} = a_1 \exp(-\mathrm{St}_{\max})
            + a_2 \exp(-a_3 \, \mathrm{St}_{\max}) \, \left[ 1 + \tanh(\mathrm{St}_{\max} - 3) \right],
\end{equation}
where $\mathrm{St}_{\max} = \max(\mathrm{St}_1, \mathrm{St}_2)$ and $a_1=2.6$, $a_2=0.1025$, and $a_3=0.0206$.

For $\langle |w_r| \rangle$, we follow \citet{Onishi-2015} and
\citet{Wang-2000} with
\begin{equation}
\langle |w_r| \rangle =
\left[
\frac{2}{\pi}
\left(
w_{\text{shear}}^2 + w_{\text{accel}}^2
\right)
\right]^{1/2},
\end{equation}
and
\begin{equation}
w_{\text{shear}}^2 = \frac{(r_1+r_2)^2 \, \varepsilon}{15 \, \nu},
\qquad
w_{\text{accel}}^2 =
\frac{1}{3}
\left[ 1 + 0.6 \exp\left( -(\mathrm{St}_{\max} - 1)^{1.5} \right) \right]
\, f_{\mathrm{KK}},
\end{equation}
where $\nu$ is the kinematic viscosity and $\varepsilon$ the turbulent
kinetic energy dissipation rate. The formulation of $f_{\mathrm{KK}}$
follows \citet{Kruis-1997} as given in OS16.

For the turbulent enhancement factor $\eta_e$, the tabulated values
from \citet{Wang-2008} and \citet{Wang-Grabowski-2009} are used as
described in detail in OS16. The tabulated values are part of the
online supplement to OS16.

Although gravitational settling is present in the underlying DNS of
\citet{Onishi-2015} and OS16, the parameterizations of $g_{12}$ and
$\langle |w_r| \rangle$ assume isotropy and do not explicitly account
for gravitational effects. In particular, the radial velocity does not
include a contribution from sedimentation and therefore does not
recover the differential terminal fall velocity in the limit of large
Stokes numbers.  To account for this, the gravitational
and turbulent contributions are combined to obtain the total
collection kernel,
\begin{equation}
  K_{\text{coll}} = \sqrt {K_{\text{grav}}^2 + K_{\text{dyn}}^2 }.
\end{equation}

No comparably mature turbulence collision kernel exists for
non-spherical ice particles.  We therefore apply the Onishi kernel to
ice particles using the mass-equivalent diameter for interpolation in
the collision efficiency lookup table. For $g_{12}$ and $\langle |w_r|
\rangle$, the Stokes number is computed using the actual terminal fall
velocity, consistent with McSnow. While this formulation of the
dynamical kernel provides a reasonable approximation for
near-spherical particles, the applicability of the Onishi model is not
valid for strongly non-spherical shapes, where particle tilting and
associated accelerations become important. This is expected to reduce
preferential concentration while potentially increasing $\langle |w_r|
\rangle$ (see, e.g., \citet{Gustavsson-2017} and references
therein). Unforunately, no quantitative parameterizations exist for
these effect. Here we apply the Onishi kernel to all ice particles,
but this should be understood as a sensitivity study rather than a
physically-based model.  Despite these limitations, we are confident
that the formulation captures the expected order of magnitude of
turbulence-induced enhancements and allows to assess their potential
impact in the present simulations.

In the current study, turbulence effects apply only to super-particle
interaction. Hence, in the current ICON/McSnow implementation this
excludes riming and accretion. This is an arbitrary choice but it
simplifies the interpretation of the sensitivity studies. Including
turbulence effects of riming would change the latent heat release and
consequently the dynamics significantly and confound micophysical and
dynamical effects. \\

{\textbf{Code availability:} McSnow is part of the ICON modeling
  framework and available under a BSD-3C license model at
  https://gitlab.dkrz.de/mcsnow/msnow. The implementation of McSnow in
  ICON has not yet been released to the public due to the additional
  infrastructure necessary for the coupling, which is not compatible
  with the current ICON Open Source Release version.
  The McRadar version used in this study is available on github at
  \url{https://github.com/lterzi/McRadar/tree/ICON_McSnow}. The
  scattering database of snowflake aggregates used in this study is
  archived on zenodo at \url{10.5281/zenodo.23240257}.} \\

\textbf{Author contributions:} {C.S.~developed the original ICON implementation of
  McSnow, which was updated and extended by F.J. for the current
  study. C.S., F.J, and A.S.~implemented the fragmentation
  schemes. A.S.~performed and analyzed the idealized squall line
  simulations together with C.S. M.S.~contributed to the
  implementation of the ice-ice fragmentation.  L.v.T.~developed
  McRadar and performed the DDA simulations together with
  S.K. A.S.~wrote the inital version of the manuscript. All authors
  discussed the results and contributed to the writing of the
  manuscript.} \\

\textbf{Funding statement:} The study contributes to and is partly
funded by the project FRAGILE of the SPP 2115 Fusion of Radar
Polarimetry and Numerical Atmospheric Modelling Towards an Improved
Understanding of Cloud and Precipitation Processes funded by the
Deutsche Forschungsgemeinschaft (DFG 492234709).


\begingroup
\small
\setlength{\bibsep}{2pt}
\bibliographystyle{apalike}
\bibliography{squall}

@string{ACP     = {Atmos.~Chem.~Phys.}}

@string{AR      = {Atmos.~Res.}}

@string{GRL     = {Geophys.~Res.~Lett.}}

@string{GMD     = {Geosci.~Model~Dev.}}

@string{ASL     = {Atmos.~Sci.~Lett.}}

@string{JAS     = {J.~Atmos.~Sci.}}

@string{JFM     = {J.~Fluid Mech.}}

@string{JGR     = {J.~Geophys.~Res.}}

@STRING{JTECH   = {J.~Atmos. Ocean. Tech. }}

@string{QJ      = {Quart.~J.~Roy.~Met.~Soc.}}

@string{MAP     = {Meteorol.~Atmos.~Phys.}}

@string{MWR     = {Mon.~Wea.~Rev.}}

@string{JAMES   = {J.~Adv. in Modeling Earth Systems}}

@article{kneifel_tripleSpec_2016,
    author = {Kneifel, S. and Kollias, P. and Battaglia, A. and Leinonen, J. and Maahn, M. and Kalesse, H. and Tridon, F.},
    title = {First observations of triple-frequency radar Doppler spectra in snowfall: Interpretation and applications},
    journal = grl,
    volume = {43},
    number = {5},
    pages = {2225-2233},
    doi = {https://doi.org/10.1002/2015GL067618},
    url = {https://agupubs.onlinelibrary.wiley.com/doi/abs/10.1002/2015GL067618},
    year = {2016}
}

@article {DOW_Wurman_1997,
    author = {Joshua  Wurman and Jerry Straka and Erik Rasmussen and Mitch Randall and Allen  Zahrai},
    title = {Design and Deployment of a Portable, Pencil-Beam, Pulsed, 3-cm Doppler Radar},
    journal = jtech,
    year = 1997,
    volume = {14},
    number = {6},
    doi = {10.1175/1520-0426(1997)014<1502:DADOAP>2.0.CO;2},
    pages = {1502 - 1512},
    url = {https://journals.ametsoc.org/view/journals/atot/14/6/1520-0426_1997_014_1502_dadoap_2_0_co_2.xml}
}

@article{vonTerzi-2026,
  title={A microwave scattering database of oriented ice and snow particles: supporting habit-dependent
         growth models and radar applications (McRadar 1.0.0)},
  author={von Terzi, Leonie and Ori, Davide and Kneifel, Stefan},
  journal=GMD,
  volume={19},
  number={2},
  pages={887--910},
  year={2026},
  publisher={Copernicus Publications G{\"o}ttingen, Germany}
}

@article{von2022ice,
  title={Ice microphysical processes in the dendritic growth layer: a statistical analysis combining multi-frequency and polarimetric Doppler cloud radar observations},
  author={von Terzi, L. and Dias Neto, J. and Ori, D. and Myagkov, A. and Kneifel, S.},
  journal=ACP,
  volume={22},
  pages={11795-11821},
  year={2022},
  publisher={Copernicus GmbH},
  doi={10.5194/acp-22-11795-2022}
}

@article{Seifert-Beheng-2001,
  author = 	 {Axel Seifert and Klaus D. Beheng},
  title = 	 {A double-moment parameterization for simulating
		  autoconversion, accretion and selfcollection},
  journal  =     AR,
  year     =     2001,
  volume = 	 {59-60},
  pages = 	 {265--281}
}

@article{Shima-2009,
  title={The super-droplet method for the numerical simulation of clouds and precipitation: 
         A particle-based and probabilistic microphysics model coupled with a non-hydrostatic model},
  author={Shima, Shin-ichiro and Kusano, Kanya and Kawano, Akio and Sugiyama, Tooru and Kawahara, Shintaro},
  journal=QJ,
  volume={135},
  number={642},
  pages={1307--1320},
  year={2009},
  publisher={Wiley Online Library}
}

@article {Xue-2017,
      author = "Lulin Xue and Jiwen Fan and Zachary J. Lebo and Wei Wu and Hugh Morrison and Wojciech W. Grabowski and Xia Chu and Istv{\'a}n Geresdi and Kirk North and Ronald Stenz and Yang Gao and Xiaofeng Lou and Aaron Bansemer and Andrew J. Heymsfield and Greg M. McFarquhar and Roy M. Rasmussen",
      title = "Idealized Simulations of a Squall Line from the MC3E Field Campaign Applying Three Bin Microphysics Schemes: {Dynamic} and Thermodynamic Structure",
      journal = MWR,
      year = "2017",
      publisher = "American Meteorological Society",
      address = "Boston MA, USA",
      volume = "145",
      number = "12",
      doi = "10.1175/MWR-D-16-0385.1",
      pages=      "4789 - 4812",
      url = "https://journals.ametsoc.org/view/journals/mwre/145/12/mwr-d-16-0385.1.xml"
}

@article{Seifert-2019,
author = {Seifert, Axel and Leinonen, Jussi and Siewert, Christoph and Kneifel, Stefan},
title = {The Geometry of Rimed Aggregate Snowflakes: A Modeling Study},
journal = JAMES,
volume = {11},
number = {3},
pages = {712-731},
doi = {10.1029/2018MS001519},
year = {2019}
}

@article{Brdar-Seifert-2018,
author = {Brdar, S. and Seifert, A.},
title = {McSnow: A Monte-Carlo Particle Model for Riming and Aggregation of Ice Particles 
         in a Multidimensional Microphysical Phase Space},
journal = JAMES,
volume = {10},
number = {1},
pages = {187-206},
doi = {10.1002/2017MS001167},
year = {2018}
}

@article{Zaengl-2015,
  title={The ICON (ICOsahedral Non-hydrostatic) modelling framework of DWD and MPI-M: Description of the non-hydrostatic dynamical core},
  author={Z{\"a}ngl, G{\"u}nther and Reinert, Daniel and R{\'\i}podas, Pilar and Baldauf, Michael},
  journal=QJ,
  volume={141},
  number={687},
  pages={563--579},
  year={2015},
  doi={10.1002/qj.2378},
  publisher={Wiley}
}

@article{Phillips-2008,
  title={An empirical parameterization of heterogeneous ice nucleation for multiple chemical species of aerosol},
  author={Phillips, Vaughan TJ and DeMott, Paul J and Andronache, Constantin},
  journal=JAS,
  volume={65},
  number={9},
  pages={2757--2783},
  year={2008},
  publisher={American Meteorological Society}
}

@article{Seifert-Beheng-2006,
  title={A two-moment cloud microphysics parameterization for mixed-phase clouds. {Part 1}: {Model} description},
  author={Seifert, Axel and Beheng, Klaus Dieter},
  journal=MAP,
  volume={92},
  number={1},
  pages={45--66},
  year={2006},
  doi={10.1007/s00703-005-0112-4},
  publisher={Springer}
}

@Article{Seifert-2012,
AUTHOR = {Seifert, A. and K\"ohler, C. and Beheng, K. D.},
TITLE = {Aerosol-cloud-precipitation effects over {Germany} as simulated by a convective-scale numerical weather prediction model},
JOURNAL = ACP,
VOLUME = {12},
YEAR = {2012},
NUMBER = {2},
PAGES = {709--725},
DOI = {10.5194/acp-12-709-2012}
}

@article{Morrison-2020,
  title={Confronting the challenge of modeling cloud and precipitation microphysics},
  author={Morrison, Hugh and van Lier-Walqui, Marcus and Fridlind, Ann M and Grabowski, Wojciech W
          and Harrington, Jerry Y and Hoose, Corinna and Korolev, Alexei and Kumjian, Matthew R and Milbrandt, Jason A
	  and Pawlowska, Hanna and others},
  journal=JAMES,
  volume={12},
  number={8},
  pages={e2019MS001689},
  year={2020},
  publisher={Wiley Online Library}
}

@article{Shima-2020,
  title={Predicting the morphology of ice particles in deep convection using the super-droplet method:
         development and evaluation of SCALE-SDM 0.2. 5-2.2. 0,-2.2. 1, and-2.2. 2},
  author={Shima, Shin-ichiro and Sato, Yousuke and Hashimoto, Akihiro and Misumi, Ryohei},
  journal=GMD,
  volume={13},
  number={9},
  pages={4107--4157},
  year={2020},
  doi = {10.5194/gmd-13-4107-2020},
  publisher={Copernicus Publications G{\"o}ttingen, Germany}
}

@article{Welss-2024,
  title={Explicit habit-prediction in the Lagrangian super-particle ice microphysics model McSnow},
  author={Welss, Jan-Niklas and Siewert, C and Seifert, A},
  journal=JAMES,
  volume={16},
  number={4},
  pages={e2023MS003805},
  year={2024},
  publisher={Wiley Online Library}
}

@article{Boehm-1992a,
title = "A general hydrodynamic theory for mixed-phase microphysics. Part I: drag and fall speed of hydrometeors",
journal = "Atmos. Res.",
volume = "27",
number = "4",
pages = "253 - 274",
year = "1992",
issn = "0169-8095",
doi = "10.1016/0169-8095(92)90035-9",
author = {Böhm, Johannes P.},
}

@article{Boehm-1992b,
title = "A general hydrodynamic theory for mixed-phase microphysics. Part II: collision kernels for coalescence",
journal = "Atmos. Res.",
volume = "27",
number = "4",
pages = "275 - 290",
year = "1992",
issn = "0169-8095",
doi = "10.1016/0169-8095(92)90036-A",
author = {Böhm, Johannes P.},
}

@article{Boehm-1992c,
title = "A general hydrodynamic theory for mixed-phase microphysics. Part III: Riming and aggregation",
journal = "Atmos. Res.",
volume = "28",
number = "2",
pages = "103 - 123",
year = "1992",
issn = "0169-8095",
doi = "10.1016/0169-8095(92)90023-4",
author = {Böhm, Johannes P.},
}

@article{Shaw-2003,
  title={Particle-turbulence interactions in atmospheric clouds},
  author={Shaw, Raymond A},
  journal={Annual Review of Fluid Mechanics},
  volume={35},
  number={1},
  pages={183--227},
  year={2003},
  publisher={Annual Reviews, Palo Alto, CA, USA}
}

@article{Grabowski-Wang-2013,
  title={Growth of cloud droplets in a turbulent environment},
  author={Grabowski, Wojciech W and Wang, Lian-Ping},
  journal={Annual review of fluid mechanics},
  volume={45},
  number={1},
  pages={293--324},
  year={2013},
  publisher={Annual Reviews}
}

@article{Onishi-Seifert-2016,
  title={Reynolds-number dependence of turbulence enhancement on collision growth},
  author={Onishi, Ryo and Seifert, Axel},
  journal=ACP,
  volume={16},
  number={19},
  pages={12441--12455},
  year={2016},
  doi = {10.5194/acp-16-12441-2016}
}

@article{Hallett-Mossop-1974,
  title={Production of secondary ice particles during the riming process},
  author={Hallett, J and Mossop, SC},
  journal={Nature},
  volume={249},
  number={5452},
  pages={26--28},
  year={1974},
  publisher={Nature Publishing Group UK London}
}

@article{Beheng-1987,
  title={Microphysical properties of glaciating cumulus clouds: Comparison of measurements with a numerical simulation},
  author={Beheng, KD},
  journal=QJ,
  volume={113},
  number={478},
  pages={1377--1382},
  year={1987},
  publisher={Wiley Online Library}
}

@article{Sun-2010,
  title={Mystery of ice multiplication in warm-based precipitating shallow cumulus clouds},
  author={Sun, Jiming and Ariya, Parisa A and Leighton, Henry G and Yau, MK},
  journal=GRL,
  volume={37},
  number={10},
  year={2010},
  publisher={Wiley Online Library}
}

@article{Mossop-1985,
  title={Microphysical properties of supercooled cumulus clouds in which an ice particle multiplication process operated},
  author={Mossop, SC},
  journal=QJ,
  volume={111},
  number={467},
  pages={183--198},
  year={1985},
  publisher={Wiley Online Library}
}

@article{Blyth-Latham-1997,
  title={A multi-thermal model of cumulus glaciation via the Hallett-Mossop process},
  author={Blyth, Alan M and Latham, John},
  journal=QJ,
  volume={123},
  number={541},
  pages={1185--1198},
  year={1997},
  publisher={Wiley Online Library}
}

@article{Blyth-Latham-1993,
  title={Development of ice and precipitation in New Mexican summertime cumulus clouds},
  author={Blyth, Alan M and Latham, John},
  journal=QJ,
  volume={119},
  number={509},
  pages={91--120},
  year={1993},
  publisher={Wiley Online Library}
}

@article{Harris-Hobbs-Cooper-1987,
  title={Field evidence supporting quantitative predictions of secondary ice production rates},
  author={Harris-Hobbs, Raymond L and Cooper, William A},
  journal=JAS,
  volume={44},
  number={7},
  pages={1071--1082},
  year={1987}
}

@article{Seidel-2024,
  title={Secondary ice production -- no evidence of efficient rime-splintering mechanism},
  author={Seidel, Johanna S and Kiselev, Alexei A and Keinert, Alice and Stratmann, Frank and Leisner, Thomas and Hartmann, Susan},
  journal=ACP,
  volume={24},
  number={9},
  pages={5247--5263},
  year={2024},
  publisher={Copernicus Publications G{\"o}ttingen, Germany}
}

@article{Qu-2025,
  title={The impacts of secondary ice production on the microphysics and dynamics of mid-latitude cold season convection},
  author={Qu, Zhipeng and Korolev, Alexei and Milbrandt, Jason A and Heckman, Ivan and Cholette, M{\'e}lissa and Nguyen, Cuong and Wolde, Mengistu},
  journal={Atmospheric Chemistry and Physics},
  volume={25},
  number={23},
  pages={17845--17868},
  year={2025},
  publisher={Copernicus Publications G{\"o}ttingen, Germany}
}

@article{Grzegorczyk-2025a,
  title={Investigating secondary ice production in a deep convective cloud with a 3D bin microphysics model: Part I-Sensitivity study of microphysical processes representations},
  author={Grzegorczyk, Pierre and Wobrock, Wolfram and Canzi, Antoine and Niquet, Laurence and Tridon, Fr{\'e}d{\'e}ric and Planche, C{\'e}line},
  journal=AR,
  volume={313},
  pages={107774},
  year={2025},
  publisher={Elsevier}
}

@article{Grzegorczyk-2025b,
  title={Investigating secondary ice production in a deep convective cloud with a 3D bin microphysics model: Part II-Effects on the cloud formation and development},
  author={Grzegorczyk, Pierre and Wobrock, Wolfram and Canzi, Antoine and Niquet, Laurence and Tridon, Fr{\'e}d{\'e}ric and Planche, C{\'e}line},
  journal=AR,
  volume={314},
  pages={107797},
  year={2025},
  publisher={Elsevier}
}

@article{Grzegorczyk2025c,
  title={Influence of secondary ice production on cloud and rain properties: analysis of the HYMEX IOP7a heavy-precipitation event},
  author={Grzegorczyk, Pierre and Wobrock, Wolfram and Dziduch, Aymeric and Planche, C{\'e}line},
  journal=ACP,
  volume={25},
  number={17},
  pages={10403--10420},
  year={2025},
  publisher={Copernicus Publications G{\"o}ttingen, Germany}
}

@article{Sun-2025,
  title={Influence of secondary ice formation on tropical deep convective clouds simulated by the Unified Model},
  author={Sun, Mengyu and Connolly, Paul J and Field, Paul R and Finney, Declan L and Blyth, Alan M},
  journal=ACP,
  volume={25},
  number={24},
  pages={18549--18569},
  year={2025},
  publisher={Copernicus Publications G{\"o}ttingen, Germany}
}

@article{Korolev-Leisner-2020,
  title={Review of experimental studies of secondary ice production},
  author={Korolev, Alexei and Leisner, Thomas},
  journal=ACP,
  volume={20},
  number={20},
  pages={11767--11797},
  year={2020},
  publisher={Copernicus GmbH}
}

@article{Field-2017,
  title={Secondary ice production: Current state of the science and recommendations for the future},
  author={Field, Paul R and Lawson, R Paul and Brown, Philip RA and Lloyd, Gary and Westbrook, Christopher and Moisseev, Dmitri and Miltenberger, Annette and Nenes, Athanasios and Blyth, Alan and Choularton, Thomas and others},
  journal={Meteorological Monographs},
  volume={58},
  pages={7--1},
  year={2017}
}

@article{Blyth-Latham-1998,
  title={Comments on cumulus glaciation papers by PV Hobbs and AL Rangno},
  author={Blyth, Alan M and Latham, John},
  journal=QJ,
  volume={124},
  pages={1007--1008},
  year={1998},
  publisher={ROYAL METEOROLOGICAL SOC}
}

@article{Seifert-2026,
      title={On the geometry of aggregate snowflakes},
      author={Axel Seifert and Christoph Siewert and Fabian Jakub and Leonie von Terzi and Stefan Kneifel},
      year={2026},
      journal={Tellus},
      volume={79},
      pages={87–110},
      doi={10.16993/tellus.4127},
}

@article{Bayley-2025b,
  title={CLEO: The Numerical Methods of a New Superdroplet Model including a Droplet Breakup Algorithm},
  author={Bayley, Clara JA and Naumann, Ann Kristin and Poydenot, Florian and Vogel, Raphaela and Stevens, Bjorn and Shima, Shin-Ichiro},
  journal={EGUsphere},
  volume={2025},
  pages={1--26},
  year={2025},
  publisher={Copernicus Publications G{\"o}ttingen, Germany}
}

@article{Chen-Lamb-1994,
  title={The theoretical basis for the parameterization of ice crystal habits: Growth by vapor deposition},
  author={Chen, Jen-Ping and Lamb, Dennis},
  journal=JAS,
  volume={51},
  number={9},
  pages={1206--1222},
  year={1994}
}

@article{Cober-List-1993,
  title={Measurements of the heat and mass transfer parameters characterizing conical graupel growth},
  author={Cober, Stewart G and List, Roland},
  journal=JAS,
  volume={50},
  number={11},
  pages={1591--1609},
  year={1993}
}

@article{Rasmussen-1984a,
  title={A wind tunnel and theoretical study of the melting behavior of atmospheric ice particles. {II:}
         {A} theoretical study for frozen drops of radius $<$ 500 $\mu$m},
  author={Rasmussen, Roy M and Levizzani, V and Pruppacher, HR},
  journal=JAS,
  volume={41},
  number={3},
  pages={374--380},
  year={1984}
}

@article{Rasmussen-1984b,
  title={A wind tunnel and theoretical study on the melting behavior of atmospheric ice particles: {III.}
         {Experiment} and theory for spherical ice particles of radius $>$ 500 $\mu$m},
  author={Rasmussen, Roy M and Levizzani, V and Pruppacher, HR},
  journal=JAS,
  volume={41},
  number={3},
  pages={381--388},
  year={1984}
}

@article{Rasmussen-Heymsfield-1987a,
  title={Melting and shedding of graupel and hail. {Part {I}: Model physics}},
  author={Rasmussen, Roy M and Heymsfield, Andrew J},
  journal=JAS,
  volume={44},
  number={19},
  pages={2754--2763},
  year={1987},
  publisher={American Meteorological Society}
}

@article{Rasmussen-Heymsfield-1987b,
  title={Melting and shedding of graupel and hail. {Part {II}: Sensitivity study}},
  author={Rasmussen, Roy M and Heymsfield, Andrew J},
  journal=JAS,
  volume={44},
  number={19},
  pages={2764--2782},
  year={1987},
  publisher={American Meteorological Society}
}

@article{Barklie-Gokhale-1959,
  title={The freezing of supercooled water drops},
  author={Barklie, RHD and Gokhale, NR},
  journal={McGill University, Stormy Weather Group, Scientific Report MW-30, Part},
  volume={3},
  pages={43--64},
  year={1959}
}

@article{Lauber-2018,
  title={Secondary ice formation during freezing of levitated droplets},
  author={Lauber, Annika and Kiselev, Alexei and Pander, Thomas and Handmann, Patricia and Leisner, Thomas},
  journal=JAS,
  volume={75},
  number={8},
  pages={2815--2826},
  year={2018}
}

@article{Keinert-2020,
  title={Secondary ice production upon freezing of freely falling drizzle droplets},
  author={Keinert, Alice and Spannagel, Dominik and Leisner, Thomas and Kiselev, Alexei},
  journal=JAS,
  volume={77},
  number={8},
  pages={2959--2967},
  year={2020}
}

@article{Kleinheins-2021,
  title={Thermal imaging of freezing drizzle droplets: pressure release events as a source of secondary ice particles},
  author={Kleinheins, Judith and Kiselev, Alexei and Keinert, Alice and Kind, Matthias and Leisner, Thomas},
  journal=JAS,
  volume={78},
  number={5},
  pages={1703--1713},
  year={2021}
}

@article{Korolev-2020,
  title={A new look at the environmental conditions favorable to secondary ice production},
  author={Korolev, Alexei and Heckman, Ivan and Wolde, Mengistu and Ackerman, Andrew S and Fridlind, Ann M
          and Ladino, Luis A and Lawson, R Paul and Milbrandt, Jason and Williams, Earle},
  journal=ACP,
  volume={20},
  number={3},
  pages={1391--1429},
  year={2020},
  publisher={Copernicus Publications G{\"o}ttingen, Germany}
}

@article{Pasquier-2022,
  title={Conditions favorable for secondary ice production in Arctic mixed-phase clouds},
  author={Pasquier, Julie Th{\'e}r{\`e}se and Henneberger, Jan and Ramelli, Fabiola and Lauber, Annika and David, Robert Oscar
          and Wieder, J{\"o}rg and Carlsen, Tim and Gierens, Rosa and Maturilli, Marion and Lohmann, Ulrike},
  journal=ACP,
  volume={22},
  number={23},
  pages={15579--15601},
  year={2022},
  publisher={Copernicus GmbH}
}

@article{Korolev-2022,
  title={Observation of secondary ice production in clouds at low temperatures},
  author={Korolev, Alexei and DeMott, Paul J and Heckman, Ivan and Wolde, Mengistu and Williams, Earle and Smalley, David J and Donovan, Michael F},
  journal=ACP,
  volume={22},
  number={19},
  pages={13103--13113},
  year={2022},
  publisher={Copernicus GmbH}
}

@article{Lachapelle-2024,
  title={Effect of secondary ice production processes on the simulation of ice pellets using the Predicted Particle Properties microphysics scheme},
  author={Lachapelle, Mathieu and Cholette, M{\'e}lissa and Th{\'e}riault, Julie M},
  journal=ACP,
  volume={24},
  number={19},
  pages={11285--11304},
  year={2024},
  publisher={Copernicus Publications G{\"o}ttingen, Germany}
}

@article{Cholette-2024,
  title={Secondary ice production improves simulations of freezing rain},
  author={Cholette, M{\'e}lissa and Milbrandt, Jason A and Morrison, Hugh and Kirk, Sabrina and Lalonde, Louis-{\'E}mile},
  journal={Geophysical research letters},
  volume={51},
  number={8},
  pages={e2024GL108490},
  year={2024},
  publisher={Wiley Online Library}
}

@article{Phillips-2017,
  title={Ice multiplication by breakup in ice--ice collisions. Part II: Numerical simulations},
  author={Phillips, Vaughan TJ and Yano, Jun-Ichi and Formenton, Marco and Ilotoviz, Eyal
  and Kanawade, Vijay and Kudzotsa, Innocent and Sun, Jiming and Bansemer, Aaron and Detwiler, Andrew G and Khain, Alexander and others},
  journal=JAS,
  volume={74},
  number={9},
  pages={2789--2811},
  year={2017}
}

@article{Han-2024,
  title={Secondary ice production in simulated deep convective clouds: A sensitivity study},
  author={Han, Cunbo and Hoose, Corinna and D{\"u}rlich, Viktoria},
  journal={Journal of the Atmospheric Sciences},
  volume={81},
  number={5},
  pages={903--921},
  year={2024},
  publisher={American Meteorological Society}
}

@article{Dedekind-2021,
  title={Sensitivity of precipitation formation to secondary ice production in winter orographic mixed-phase clouds},
  author={Dedekind, Zane and Lauber, Annika and Ferrachat, Sylvaine and Lohmann, Ulrike},
  journal=ACP,
  volume={21},
  number={19},
  pages={15115--15134},
  year={2021},
  publisher={Copernicus GmbH}
}

@article{Sullivan-2018,
  title={The effect of secondary ice production parameterization on the simulation of a cold frontal rainband},
  author={Sullivan, Sylvia C and Barthlott, Christian and Crosier, Jonathan and Zhukov, Ilya and Nenes, Athanasios and Hoose, Corinna},
  journal=ACP,
  volume={18},
  number={22},
  pages={16461--16480},
  year={2018},
  publisher={Copernicus Publications G{\"o}ttingen, Germany}
}

@article{Waman-2022,
  title={Dependencies of four mechanisms of secondary ice production on cloud-top temperature in a continental convective storm},
  author={Waman, Deepak and Patade, Sachin and Jadav, Arti and Deshmukh, Akash and Gupta, Ashok Kumar and Phillips, Vaughan TJ and Bansemer, Aaron and DeMott, Paul J},
  journal=JAS,
  volume={79},
  number={12},
  pages={3375--3404},
  year={2022}
}

@article{Waman-2026,
  title={Impacts of Secondary Ice Production on the Microphysics and Dynamics of Deep Convective Clouds in Different Environments},
  author={Waman, Deepak and Meusel, Julian and Keshtgar, Behrooz and Wallentin, Gabriella and Barthlott, Christian and Patade, Sachin and Shete, Sonali and Prabhakaran, Thara and Fievet, Romain and Finney, Declan and others},
  journal={EGUsphere},
  volume={2026},
  pages={1--55},
  year={2026},
  publisher={Copernicus Publications G{\"o}ttingen, Germany}
}

@article{Gautam-2024,
  title={Fragmentation in Collisions of Snow with Graupel/Hail: New Formulation from Field Observations},
  author={Gautam, Martanda and Waman, Deepak and Patade, Sachin and Deshmukh, Akash and Phillips, Vaughan and Jackowicz-Korczynski, Marcin and Paul, Freddy Pazhambilly and Smith, Paul and Bansemer, Aaron},
  journal=JAS,
  volume={81},
  number={12},
  pages={2149--2164},
  year={2024},
  publisher={American Meteorological Society}
}

@article{Grzegorczyk-2023,
  title={Fragmentation of ice particles: laboratory experiments on graupel--graupel and graupel--snowflake collisions},
  author={Grzegorczyk, Pierre and Yadav, Sudha and Zanger, Florian and Theis, Alexander and Mitra, Subir K and Borrmann, Stephan and Szak{\'a}ll, Mikl{\'o}s},
  journal=ACP,
  volume={23},
  number={20},
  pages={13505--13521},
  year={2023},
  publisher={Copernicus Publications G{\"o}ttingen, Germany}
}

@article{Yadav-2025,
  title={Measurement report: Influence of particle density on secondary ice production by graupel and frozen drop collisions},
  author={Yadav, Sudha and Metten, Lilly and Grzegorczyk, Pierre and Theis, Alexander and Mitra, Subir K and Szak{\'a}ll, Mikl{\'o}s},
  journal=ACP,
  volume={25},
  number={15},
  pages={8671--8682},
  year={2025},
  publisher={Copernicus Publications G{\"o}ttingen, Germany}
}

@article{Gao-2023,
  title={A numerical simulation study of secondary ice productions in a squall line case},
  author={Gao, Jie and Han, Xuqing and Chen, Yichen and Li, Shuangxu and Xue, Huiwen},
  journal={Atmosphere},
  volume={14},
  number={12},
  pages={1752},
  year={2023},
  publisher={MDPI}
}

@article{Pfeifer-2025,
  title={Efficient ice multiplication from freezing raindrop fragmentation},
  author={Pfeifer, Nils and Mom, Bernd and Moisseev, Dmitri and Hartmann, Susan and Meusel, Julian and Hoose, Corinna and Maahn, Maximilian},
  journal={Communications Earth \& Environment},
  volume={6},
  number={1},
  pages={942},
  year={2025},
  publisher={Nature Publishing Group UK London}
}

@article{Sheikh-2024,
  title={Effect of turbulence on the collision rate between settling ice crystals and droplets},
  author={Sheikh, MZ and Gustavsson, K and L{\'e}v{\^e}que, Emmanuel and Mehlig, B and Pumir, Alain and Naso, Aurore},
  journal={Journal of the Atmospheric Sciences},
  volume={81},
  number={5},
  pages={887--901},
  year={2024},
  publisher={American Meteorological Society}
}

@article{Siewert-2014b,
  title={Collision rates of small ellipsoids settling in turbulence},
  author={Siewert, C and Kunnen, RPJ and Schr{\"o}der, Wolfgang},
  journal={J.~of Fluid Mech.},
  volume={758},
  pages={686--701},
  year={2014},
  publisher={Cambridge University Press}
}

@article{Siewert-2014a,
  title={Orientation statistics and settling velocity of ellipsoids in decaying turbulence},
  author={Siewert, C and Kunnen, RPJ and Meinke, Meinke and Schr{\"o}der, Wolfgang},
  journal=AR,
  volume={142},
  pages={45--56},
  year={2014},
  publisher={Elsevier}
}

@article{Voth-2017,
  title={Anisotropic particles in turbulence},
  author={Voth, Greg A and Soldati, Alfredo},
  journal={Annual Review of Fluid Mechanics},
  volume={49},
  number={1},
  pages={249--276},
  year={2017},
  publisher={Annual Reviews}
}

@article{Ayala-Wang-2008,
  title={Effects of turbulence on the geometric collision rate of sedimenting droplets. Part 2. Theory and parameterization},
  author={Ayala, Orlando and Rosa, Bogdan and Wang, Lian-Ping},
  journal={New Journal of Physics},
  volume={10},
  number={7},
  pages={075016},
  year={2008}
}

@article{Phillips-2021,
  title={Comment on “Review of experimental studies of secondary ice production” by Korolev and Leisner (2020)},
  author={Phillips, Vaughan TJ and Yano, Jun-Ichi and Deshmukh, Akash and Waman, Deepak},
  journal=ACP,
  volume={21},
  number={15},
  pages={11941--11953},
  year={2021},
  publisher={Copernicus Publications G{\"o}ttingen, Germany}
}

@article{Weisman-Klemp-1982,
  title={The dependence of numerically simulated convective storms on vertical wind shear and buoyancy},
  author={Weisman, Morris L and Klemp, Joseph B},
  journal=MWR,
  volume={110},
  number={6},
  pages={504--520},
  year={1982}
}

@article{Rotunno-1988,
  title={A theory for strong, long-lived squall lines},
  author={Rotunno, Richard and Klemp, Joseph B and Weisman, Morris L},
  journal=JAS,
  volume={45},
  number={3},
  pages={463--485},
  year={1988}
}

@article{Segal-Khain-2006,
  title={Dependence of droplet concentration on aerosol conditions in different cloud types: 
         Application to droplet concentration parameterization of aerosol conditions},
  author={Segal, Y and Khain, A},
  journal=JGR,
  volume={111},
  number={D15},
  year={2006},
  publisher={Wiley Online Library}
}

@article{Bryan-2003,
  title={Resolution requirements for the simulation of deep moist convection},
  author={Bryan, George H and Wyngaard, John C and Fritsch, J Michael},
  journal=MWR,
  volume={131},
  number={10},
  pages={2394--2416},
  year={2003}
}

@article{Weisman-1997,
  title={The resolution dependence of explicitly modeled convective systems},
  author={Weisman, Morris L and Skamarock, William C and Klemp, Joseph B},
  journal=MWR,
  volume={125},
  number={4},
  pages={527--548},
  year={1997}
}

@article{Tijhuis-2025,
  title={Comparing the dynamics of idealized squall lines between NWP and LES models},
  author={Tijhuis, Mirjam and Seifert, Axel and de Lozar, Alberto and van Stratum, Bart JH and van Heerwaarden, Chiel C},
  journal={submitted to {\it Mon.~Wea.~Rev.}, preprint available at {\it arXiv},},
  url={https://arxiv.org/abs/2506.19435}, 
  year={2025}
}

@article{Phillips-2018,
  title={Secondary ice production by fragmentation of freezing drops: Formulation and theory},
  author={Phillips, Vaughan TJ and Patade, Sachin and Gutierrez, Julie and Bansemer, Aaron},
  journal=JAS,
  volume={75},
  number={9},
  pages={3031--3070},
  year={2018}
}

@article{deJong-2023,
  title={Breakups are complicated: an efficient representation of collisional breakup in the superdroplet method},
  author={de Jong, Emily and Mackay, John Ben and Bulenok, Oleksii and Jaruga, Anna and Arabas, Sylwester},
  journal=GMD,
  volume={16},
  number={14},
  pages={4193--4211},
  year={2023},
  publisher={Copernicus GmbH}
}

@Misc{Seifert-Siewert-2023,
  author = 	 {Axel Seifert and Christoph Siewert},
  title = 	 {Collisional breakup with constant super-droplet number},
  howpublished = {Interactive comment to "Breakups are complicated: an efficient representation of 
                     collisional breakup in the superdroplet method ", {\it Geosci.~Model~Dev.}},
  year = 	 2023,
  url = {https://doi.org/10.5194/egusphere-2022-1243-CC2},
  publisher={Copernicus GmbH}
}

@article{Heymsfield-1982,
  title={A comparative study of the rates of development of potential graupel and hail embryos in high plains storms},
  author={Heymsfield, Andrew J},
  journal=JAS,
  volume={39},
  number={12},
  pages={2867--2897},
  year={1982}
}

@article{Jensen-Harrington-2015,
  title={Modeling ice crystal aspect ratio evolution during riming: A single-particle growth model},
  author={Jensen, Anders A and Harrington, Jerry Y},
  journal=JAS,
  volume={72},
  number={7},
  pages={2569--2590},
  year={2015}
}

@article{Onishi-2015,
  title={Lagrangian tracking simulation of droplet growth in turbulence -- turbulence enhancement of autoconversion rate},
  author={Onishi, Ryo and Matsuda, Keigo and Takahashi, Keiko},
  journal=JAS,
  volume={72},
  number={7},
  pages={2591--2607},
  year={2015}
}

@article{Zhou-2001,
  title={Modelling turbulent collision of bidisperse inertial particles},
  author={Zhou, Yong and Wexler, Anthony S and Wang, Lian-Ping},
  journal=JFM,
  volume={433},
  pages={77--104},
  year={2001},
  publisher={Cambridge University Press}
}

@article{Wang-2000,
  title={Statistical mechanical description and modelling of turbulent collision of inertial particles},
  author={Wang, Lian-Ping and Wexler, Anthony S and Zhou, Yong},
  journal=JFM,
  volume={415},
  pages={117--153},
  year={2000},
  publisher={Cambridge University Press}
}

@article{Wang-2008,
  title={Turbulent collision efficiency of heavy particles relevant to cloud droplets},
  author={Wang, Lian-Ping and Ayala, Orlando and Rosa, Bogdan and Grabowski, Wojciech W},
  journal={New Journal of Physics},
  volume={10},
  number={7},
  pages={075013},
  year={2008}
}

@article{Wang-Grabowski-2009,
  title={The role of air turbulence in warm rain initiation},
  author={Wang, Lian-Ping and Grabowski, Wojciech W},
  journal=ASL,
  volume={10},
  number={1},
  pages={1--8},
  year={2009},
  publisher={Wiley Online Library}
}

@article{Kruis-1997,
  title={The collision rate of particles in turbulent flow},
  author={Kruis, FE and Kusters, KA},
  journal={Chem.~Eng.~Commun.},
  volume={158},
  number={1},
  pages={201--230},
  year={1997},
  publisher={Taylor \& Francis}
}

@article{Gustavsson-2017,
  title = {Statistical Model for the Orientation of Nonspherical Particles Settling in Turbulence},
  author = {Gustavsson, K. and Jucha, J. and Naso, A. and L\'ev\^eque, E. and Pumir, A. and Mehlig, B.},
  journal = {Phys. Rev. Lett.},
  volume = {119},
  issue = {25},
  pages = {254501},
  numpages = {6},
  year = {2017},
  month = {Dec},
  publisher = {American Physical Society},
  doi = {10.1103/PhysRevLett.119.254501},
  url = {https://link.aps.org/doi/10.1103/PhysRevLett.119.254501}
}

@misc{Pruppacher-Klett-1998,
  title={Microphysics of clouds and precipitation},
  author={Pruppacher, Hans R and Klett, James D},
  year={1998},
  publisher={Taylor \& Francis}
}

@article{Gavze-Khain-2022,
  title={Gravitational collision of small nonspherical particles: Swept volumes of prolate and oblate spheroids in calm air},
  author={Gavze, Ehud and Khain, Alexander},
  journal=JAS,
  volume={79},
  number={6},
  pages={1493--1514},
  year={2022}
}

@article{Luo-2026,
  title={Theoretical collision model of elongated particles in turbulence},
  author={Luo, Haoran and Huang, Jianda and Zhao, Lihao},
  journal=JFM,
  volume={1030},
  pages={A38},
  year={2026},
  publisher={Cambridge University Press}
}

@article{Zaichik-2009,
 author    = {Zaichik, L. I. and Fede, P. and Simonin, O. and Alipchenkov, V. M.},
 title     = {Statistical models for predicting the effect of bidisperse particle collisions on particle velocities and stresses in homogeneous anisotropic turbulent flows},
 journal   = {Int. J. Multiphase Flow},
 year      = {2009},
 volume    = {35},
 number    = {9},
 pages     = {868--878},
 doi       = {10.1016/j.ijmultiphaseflow.2009.05.003}
}

@article{Jaruga-2018,
  title={libcloudph++ 2.0: {Aqueous-phase} chemistry extension of the particle-based cloud microphysics scheme},
  author={Jaruga, Anna and Pawlowska, Hanna},
  journal=GMD,
  volume={11},
  number={9},
  pages={3623--3645},
  year={2018},
  publisher={Copernicus Publications G{\"o}ttingen, Germany}
}
\endgroup

\end{document}